\documentclass[conference]{IEEEtran}

\usepackage{array, calc, color, multicol}
\usepackage{mathtools}

\DeclarePairedDelimiter\floor{\lfloor}{\rfloor}
\usepackage{graphics, epstopdf, graphicx, epsfig, psfrag, epsf, subfigure}
\usepackage{amssymb, amsfonts, amsmath, times}
\usepackage{relsize}
\usepackage{multicol,balance, verbatim}
\usepackage{textcomp, mathrsfs, stmaryrd, setspace}
\usepackage{amssymb}
\usepackage{soul}
\usepackage[dvipsnames]{xcolor}
\usepackage{cite}
\usepackage{url}
\usepackage{amsmath,stackengine}
\stackMath

\newcommand{\ie}{{\em i.e., }}
\newcommand{\Ie}{{\em I.e., }}
\newcommand{\eg}{{\em e.g., }}

\newtheorem{theorem}{Theorem}
\newtheorem{lemma}[theorem]{Lemma}

\newtheorem{corollary}[theorem]{Corollary}
\newtheorem{definition}{Definition}
\newtheorem{example}{Example}

\DeclareGraphicsExtensions{.eps, .pdf, .png, .jpg}   

\newcommand{\Nset}{\mathcal{N}}

\newcommand{\Eset}{\mathcal{E}}

\newcommand{\yas}[1]{\textcolor{magenta}{#1}}

\usepackage{lettrine}
\usepackage[ruled,vlined]{algorithm2e}

\IEEEoverridecommandlockouts

\begin{document}


\title{Adaptive Gap Entangled Polynomial Coding for Multi-Party Computation at the Edge
}

\author{\IEEEauthorblockN{Elahe Vedadi}
\IEEEauthorblockA{
\textit{University of Illinois at Chicago}\\
evedad2@uic.edu}
\and
\IEEEauthorblockN{ Yasaman Keshtkarjahromi}
\IEEEauthorblockA{
\textit{Seagate Technology}\\
yasaman.keshtkarjahromi@seagate.com}
\and
\IEEEauthorblockN{Hulya Seferoglu}
\IEEEauthorblockA{
\textit{University of Illinois at Chicago}\\
hulya@uic.edu}
}

\maketitle

{$\hphantom{a}$}{}

\begin{abstract}
Multi-party computation (MPC) is promising for designing privacy-preserving machine learning algorithms at edge networks.  An emerging approach is coded-MPC (CMPC), which advocates the use of coded computation to improve the performance of MPC in terms of the required number of workers involved in computations. The current approach for designing CMPC algorithms is to merely combine efficient coded computation constructions with MPC. Instead, we propose a new construction; Adaptive Gap Entangled polynomial (AGE) codes, where the degrees of  polynomials used in computations are optimized for  MPC. We show that MPC with AGE codes (AGE-CMPC) performs better than existing CMPC algorithms in terms of the required number of workers as well as storage, communication and computation load. 

 \end{abstract}
 

\section{\label{sec:introduction}Introduction}
\IEEEPARstart{M}{assive} amount of data is generated at edge networks.  
For example, the data generated by IoT devices are expected to reach 73.1 ZB by 2025, growing from 18.3 ZB in 2019 \cite{IDCReport}. 
This vast data is expected to be processed in real-time in many time sensitive edge applications, which is extremely challenging if not impossible with existing centralized cloud due to limited bandwidth between an edge network and centralized cloud \cite{DemocratizingNetworkEdge, EdgeComputingVideo, EdgeEatCloud}.

\begin{figure}[t!]
\centering
{ \scalebox{.28}{\includegraphics{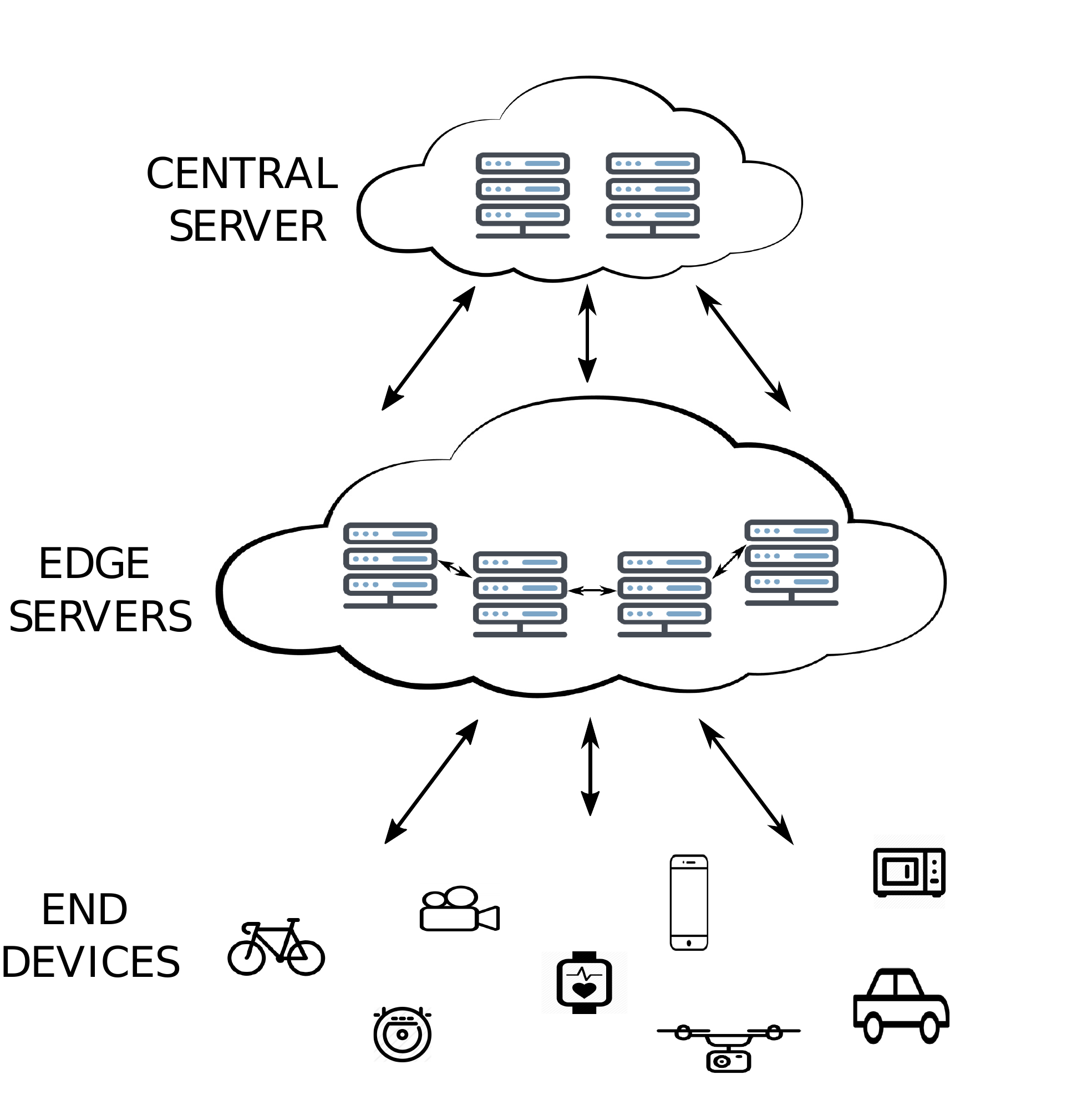}} }
\vspace{-10pt}
\caption{
An edge computing system. End devices generate and/or collect data, edge servers process data, and a central server 
collects the outcome of the processed data. 
}
\vspace{-10pt}
\label{fig:main_figure}
\end{figure}

We consider a distributed computing system at the edge, where data is generated and collected by end devices, Fig.~\ref{fig:main_figure}. 
Computationally intensive aspects are distributively processed by edge servers and a central server collects the outcome of the processed data.  In this context, it is crucial to design efficient computation mechanisms at edge servers by taking into account the limited resources, including the number of edge serves, computing power, storage, and communication cost, while preserving privacy of data at end devices. 

Multi-party computation (MPC) is a privacy-preserving distributed computing framework \cite{scalableMPC}. In MPC, several parties (end devices in Fig. \ref{fig:main_figure}) have private data and the goal is to compute a function of data collectively with the participation of all parties, while preserving privacy, \ie each party only knows its own information. MPC can be categorized into cryptographic solutions \cite{Yao}, \cite{GMW} and information-theoretic solutions \cite{BGW}. In this paper, our focus is on the information-theoretic MPC solution; BGW 
\cite{BGW} using Shamir's secret-sharing scheme \cite{ShamirSS} thanks to its lower computational complexity and quantum safe nature \cite{10.1007/3-540-48405-1_4}. Despite its potential, BGW does not take into account the limited resources at the edge. 

An emerging approach is coded-MPC (CMPC), which advocates the use of coded computation \cite{SpeedUp-journal, Tradeoff-journal} to improve the performance of BGW in terms of the required number of workers (edge servers in Fig. \ref{fig:main_figure}) involved in computationally intensive computations. For example, there may be two end devices in Fig.~\ref{fig:main_figure} possessing private matrices $A$ and $B$. The goal is to calculate $Y = A^TB$ with the help of edge servers, while preserving privacy. This multiplication is a computationally intensive task when the sizes of $A$ and $B$ are large.

The common approach for designing CMPC algorithms is to merely combine efficient coded computation constructions with MPC. This approach fails short of being efficient as it does not take into account the interaction between coded computation and MPC \cite{PolyDot-CMPC}. Indeed, CMPC mechanisms based on Shamir's secret shares create a polynomial for each matrix for matrix multiplication with two parts; \emph{coded} and \emph{secret} terms \cite{8613446, Zhu2021ImprovedCF, 9333639}. The multiplication of matrices, \ie the multiplication of these polynomials create cross terms of coded and secret terms, some of which are not used in the decoding process, so named \emph{garbage terms} \cite{PolyDot-CMPC}.

The garbage terms are crucial for the performance of CMPC. In fact, even if a construction is optimized for coded computation, it may not perform well in CMPC. It was shown in \cite{PolyDot-CMPC} that PolyDot coded MPC is better than entangled polynomial coded MPC (Entangled-CMPC) \cite{8613446} in terms of the required number of workers for a range of colluding workers. This result is surprising as it is known that entangled polynomial codes always outperform PolyDot codes in terms of the number of required workers for coded computation \cite{YuFundamentalLimits2018}. Motivated by this observation, we propose a new construction in this paper; Adaptive Gap Entangled (AGE) codes, where the degrees of  polynomials used in computations are optimized for  MPC. 
We show through analysis and simulations that MPC with AGE codes performs better than existing CMPC algorithms including PolyDot-CMPC \cite{PolyDot-CMPC}, Entangled-CMPC \cite{8613446}, SSMM \cite{Zhu2021ImprovedCF}, and GCSA-NA \cite{9333639} in terms of the required number of workers as well as storage, communication and computation load.

\section{Related Work} \label{sec:related}
Coded computation advocates higher reliability and smaller delay in distributed computation by introducing redundancy in the offloaded sub-tasks to the workers \cite{SpeedingUp}. Significant effort is being put on constructing codes for fast and distributed matrix-vector multiplication \cite{SpeedingUp}, \cite{FerdinandAnytime}, matrix-matrix multiplication \cite{YuPolynomial2017, LeeHighDim2017, PolyDotMatDot, YuFundamentalLimits2018}, dot product and convolution of two vectors \cite{ShortDot}, \cite{DuttaCodedConvolution2017}, gradient descent \cite{TandonGradientCoding2017, HalbawiImprovingDistGradient2018, RavivGradientCyclic2018}, distributed optimization \cite{KarakusRedundancy2019}, Fourier transform \cite{YuCodedFourier2017}, and linear transformations \cite{YangComputeLinear2017}. As compared to this line of works, we consider privacy-preserving computation at edge networks.

Privacy is studied in coded computation. In \cite{8382305, SecureCoded, GASP}, the problem of matrix-matrix multiplication is considered for the case that a master possesses the input data and would like to perform multiplication on the data with the help of parallel workers, while the data is kept confidential from the workers. In \cite{bitar_trans_PRAC} and \cite{BPR17}, privacy is addressed for the same system model of master-worker setup, but for matrix-vector multiplication. As compared to this line of work, we focus on the MPC system setup, where there are multiple sources each having private input data, and the goal is that a master learns the result of computation of matrix multiplication on the input data with the help of parallel workers. The input data should be kept confidential from workers and the master according to the  information-theoretic security. 

There is a line of work investigating CMPC. Lagrange Coded Computing (LCC) is designed  \cite{LCC1} in a coded computation setup for security and privacy. This work is extended for MPC setup \cite{LCC2}. 
The problem of limited memory at each party in MPC setup is addressed in \cite{PolynomCMPC}  by leveraging polynomial coded computation. This work is generalized using entangled polynomial codes for block-wise matrix multiplication \cite{8613446}. Secure multi-party batch matrix multiplication is considered in  \cite{9333639,Zhu2021ImprovedCF}, which modify the MPC system setup by employing the idea of noise alignment to reduce the communication load among workers. 
As compared to this line of work, we design AGE-CMPC by taking into account the interaction of coded computation and MPC for the limited edge resources. 

\section{System Model} \label{sec:system}

\textbf{Notations.} 
%
\emph{Set of polynomial degrees:} The set of powers of the terms in a given polynomial $f(x) = \sum_{i=0}^n a_ix^i$ with non-zero coefficients  is denoted by $\mathbf{P}(f(x))$, \ie $ \mathbf{P}(f(x)) = \{i \in \mathbb{Z}: 0 \leq i \leq n,\; a_i \neq 0\}$. 

\emph{Set definitions and operations:} We use the following standard notations for arbitrary sets $\mathbf{A}$ and $\mathbf{B}$, where the elements of $\mathbf{A}$, $\mathbf{B}$ are integers, \ie  $a, b \in \mathbb{Z}$; (i) $\mathbf{A} + \mathbf{B} = \{ a + b: a \in \mathbf{A},\; b \in \mathbf{B}\}$, and (ii) $\mathbf{A} + b = \{a+b: a \in \mathbf{A}\}$.   
The cardinality of $\mathbf{A}$ is shown by $|\mathbf{A}|$. The set of integers between $a$ and $b$ is represented by $\Omega_{a}^{b}$, \ie $\Omega_{a}^{b} = \{a, \ldots, b\}$. Furthermore, $k|m$ means that $m$ is divisible by $k$, \ie $\mod\{m,k\}=0$.
 
\emph{Matrix splitting:} 
If a matrix $A$ is divided into $s$ row-wise and $t$ column-wise partitions, it is represented as
\begin{align}\label{eq:blockwise-part}
A = \left[ {\begin{array}{ccc}
   A_{0,0}&
   \ldots & A_{0,t-1}\\   
   \vdots&\ddots&\vdots \\ A_{s-1,0} &
   \ldots&A_{s-1,t-1}
  \end{array} } \right],
  \end{align}
  {where for $A \in \mathbb{F}^{m \times m}$}, $A_{j,i} \in \mathbb{F}^{\frac{m}{s} \times \frac{m}{t}}$ for $j \in \Omega_{0}^{s-1}$ and $i \in \Omega_{0}^{t-1}$.

\textbf{Setup.} We consider a system setup with $E$ end devices, $N$ edge servers (workers), and a central server (master) as shown in Fig. \ref{fig:main_figure}. Each source  $e \in \Eset$, where $E = |\Eset|$, has private data $X_{e} \in \mathbb{F}^{\mu \times \nu}$, where $\mathbb{F}$ is a finite field. Each source is connected to all workers, and offloads its data to workers for privacy-preserving computation. Each worker $W_n, n \in \Nset$ ($|\Nset| = N$) is connected to other workers  as well as the master. 
The sources, workers, and the  master are all edge devices with limited available resources. 

\textbf{Application.} The goal is to calculate a function of per source data; $Y = \gamma(X_{1}, \ldots, X_{E})$, while the privacy of data $X_{1}, \ldots, X_{E}$ is preserved. While function $\gamma(.)$ could be any polynomial function in MPC setup, we focus on matrix multiplication as (i) we would like to present our ideas in a simple way, and (ii) matrix multiplication forms an essential building block of many signal processing and machine learning algorithms (gradient descent, classification, etc.) \cite{SpeedUp-journal}. In particular, we consider $Y=\gamma(A, B) = A^TB$, where $X_1 = A$, $X_2 = B$, $A \in \mathbb{F}^{m \times m} $, $B \in \mathbb{F}^{m \times m}$. We note that we use square matrices from two sources for easy exposition, and it is straightforward to extend our results for more general matrices and larger number of sources. 
 
\textbf{Attack Model.}
We assume a semi-honest system model, where the sources, the workers, and the master follow the defined protocols,  
but they are curious about the private data. We assume that $z$ among $N$ workers can collude to maximize the information that they can access. 
We design our AGE-CMPC mechanism against $z$ colluding workers to provide privacy-preserving computation. 


\textbf{Privacy Requirements.} We define the privacy requirements from the perspective of the sources, workers, and the master. 

\textit{Source perspective:} Source devices should not learn anything about the private data of any other sources. This requirement is satisfied in our system as there is no communication among the source devices. Also, the workers and the master do not send any information to sources.
    
\textit{Worker perspective:} Each worker should not learn anything about the private data $X_{1}, \ldots, X_{E}$ from the perspective of information-theoretic security. 
In particular, workers should not learn anything when they communicate with each other as well as when they receive data from the sources, \ie  $\tilde{H}(X_{1},\ldots,X_{E}|\underset{n \in \Nset_c}{\bigcup} (\{G_{n'}(\alpha_n), n'\in \{1,\ldots,N\}\},\underset{e \in \Eset}{\cup} F_e(\alpha_n)))  =\tilde{H}(X_{1}, X_{2},...,X_{E})$, 
    where $\tilde{H}$ denotes the Shannon entropy, $\alpha_n$ is a priori parameter associated by worker $W_n$, \eg we set $\alpha_n = n$, 
    %
    $G_{n'}(\alpha_n)$ is the data each worker $W_n$ receives from another worker $W_{n'}$, $F_e(\alpha_n)$ is the data received by each worker $W_n$ from source $e$ for $n \in \Nset_c$, and $\Nset_c$ is any subset of $\Nset$ satisfying $|\Nset_c| \leq z$. 
    
    
\textit{Master perspective}: The master node should not learn anything more than the final result $Y$, \ie $\tilde{H}(\chi_{1},\ldots, \chi_{E}|Y,\underset{n \in \Nset}{\bigcup}I(\alpha_n)) = \tilde{H}(\chi_{1},\ldots, \chi_{E}|Y)$,  
where $I(\alpha_n)$ is the data that the master receives from $W_n$.

\section{Adaptive Gap Entangled Polynomial Coding}
In this section, we introduce Adaptive Gap Entangled polynomial (AGE) codes and present our CMPC design with AGE codes; AGE-CMPC.
\subsection{AGE Codes} \label{sec:AGECodes}
We consider the generalized formulation  \cite{YuFundamentalLimits2018} for coded computation of matrices $A$ and $B$, where $A$ and $B$ are represented with the following polynomials.
\begin{align}\label{eq:generalEntangled}
    C_A(x) = & \sum\limits_{i=0}^{t-1}\sum\limits_{j=0}^{s-1}A_{i,j}x^{j\alpha+i\beta}, \nonumber \\
    C_B(x) = & \sum\limits_{k=0}^{s-1}\sum\limits_{l=0}^{t-1}B_{k,l}x^{(s-1-k)\alpha+\theta l} 
\end{align} where $\alpha, \beta, \theta \in \mathbb{N}$, $A_{i,j} \in A^T$ and $B_{k,l} \in B$. In this setup, instead of multiplying $Y=A^TB$, we can multiply $C_A(x)$ and $C_B(x)$, which can be decomposed to support distributed computing. Several codes that have been designed for coded computation can be considered as the special case of (\ref{eq:generalEntangled}) by considering different values of $(\alpha,\beta,\theta)$. 
%
%
For example, PolyDot codes \cite{PolyDotMatDot} correspond to $(\alpha,\beta,\theta)=(t,1,t(2s-1))$, while Generalized PolyDot codes \cite{Dutta2018AUC} and Entangled polynomial codes \cite{YuFundamentalLimits2018} follow $(\alpha,\beta,\theta)=(1,s,ts)$, where $t$ is the number of column-wise partitions and $s$ is the number of row-wise partitions of matrices $A$ and $B$.

We design our AGE codes by considering $(\alpha,\beta,\theta)=(1,s,ts+\lambda)$, where $\lambda$ is a parameter in the range of $0\leq \lambda\leq z$, which 
we optimize to achieve the minimum required number of workers for MPC. 
We note that Entangled polynomial codes \cite{YuFundamentalLimits2018} also follows (\ref{eq:generalEntangled}) for $\alpha=1$ and $\beta=s$, but as they are designed for coded computation, $\theta$ is optimized to achieve the minimum recovery threshold, so $\theta$ is set to $\theta=ts$. Instead, we set $\theta = ts+\lambda$ and optimize $\lambda$. Next, we prove the decodability of our AGE codes. 

\begin{theorem}\label{th:decodabilityofAGEcodes}
 AGE code guarantees the decodability of $Y=A^TB$ from the polynomial $C_Y(x)=C_A(x)C_B(x)$.
\end{theorem}

{\em Proof:}
 The proof is provided in Appendix A. \hfill $\Box$

\subsection{AGE-CMPC} \label{sec:AGE-CMPC}

\textbf{Phase 1 - Sources Share Data with Workers.} In the first phase, sources split their matrices $A$ and $B$ into $s\geq 1$ row-wise and $t \geq 1$ column-wise partitions\footnote{
We note that in AGE-CMPC, we exclude the case of no partitioning, \ie $s=t=1$; This case is considered in BGW, where coding is not required, and thus is excluded from our Coded MPC setup.} as in (\ref{eq:blockwise-part}), 
where $s|m$  and $t|m$ are satisfied. 
Assuming $A_{i,j} \in A^T$ and $B_{k,l} \in B$, where $i,l \in \Omega_{0}^{t-1}$, $j,k \in \Omega_{0}^{s-1}$, the sources create polynomials $F_A(x)$ and $F_B(x)$, which comprise coded and secret terms; \ie $F_{i'}(x)=C_{i'}(x)+S_{i'}(x),\; i' \in \{A,B\}$, where $C_{i'}(x)$'s are the coded terms defined by AGE codes and $S_{i'}(x)$'s are the secret terms which are defined by our AGE-CMPC design, which we explain next. 

Let $\mathbf{P}(C_A(x))$ and $\mathbf{P}(C_B(x))$ be the set of all powers in the polynomials $C_A(x)$ and $C_B(x)$, with non-zero coefficients. 
\begin{align}\label{eq:AGE-p(CA)-th}
    \mathbf{P}(C_{A}(x)) = & \{j+si: i \in \Omega_{0}^{t-1},\; 
    j \in \Omega_{0}^{s-1}\} \nonumber \\
    = & \{0,\ldots,ts-1\},
\end{align}
\begin{align}\label{eq:AGE-p(CB)-th}
    \mathbf{P}(C_B(x)) = & \{(s-1-k)+l(ts+\lambda): 
    k \in \Omega_{0}^{s-1},\;l \in \Omega_{0}^{t-1}\},
\end{align}
%
where, $s,t \in \mathbb{N}$ and $\lambda \in \Omega_{0}^{z}$. $S_A(x)$ and $S_B(x)$ are defined such that $\mathbf{P}(C_A(x)S_B(x))$, $\mathbf{P}(S_A(x)C_B(x))$, and $\mathbf{P}(S_A(x)S_B(x))$ do not have common terms with the important powers of $\mathbf{P}(C_A(x)C_B(x))$, which are equal to $(s-1)+si+(ts+\lambda)l$ for $i,l \in \Omega_{0}^{t-1}$. 
%
%
%
The reason is that $\{(s-1-k+j)\alpha+i\beta+\theta l:\;i,l \in \Omega_{0}^{t-1},\; j,k \in \Omega_{0}^{s-1}\;s,t \in \mathbb{N}\}$ is the set of powers of polynomial $C_A(x)C_B(x)$. The components of the desired product $Y=A^TB$ are equal to $Y_{i,l}=\sum_{j=0}^{s-1}A_{ij}B_{jl}$, for $i,l \in \Omega_{0}^{t-1}$, that are the summation of the coefficients of the terms with $j=k \in \Omega_{0}^{s-1}$. Therefore $\{(s-1)\alpha+i\beta+\theta l:\;i,l \in \Omega_{0}^{t-1},\;s,t \in \mathbb{N}\}$ is the set of important powers of $C_A(x)C_B(x)$, and for successful recovery of $Y$, these components should not have any overlap with the other components, \ie garbage terms.  
In other words, the following conditions should be satisfied: 
\begin{align}\label{eq:non_eq-AGE-conditions}
    & \text{C1: } (s-1)+si+(ts+\lambda)l \not\in \mathbf{P}(S_B(x))+\mathbf{P}(C_A(x)), \nonumber \\
   & \text{C2: } (s-1)+si+(ts+\lambda)l \not\in \mathbf{P}(S_A(x))+\mathbf{P}(C_B(x)), \nonumber \\
    & \text{C3: } (s-1)+si+(ts+\lambda)l \not\in \mathbf{P}(S_A(x))+\mathbf{P}(S_B(x)).
\end{align}

Our strategy for determining $\mathbf{P}(S_A(x))$ and $\mathbf{P}(S_B(x))$ is as follows. First, we set the elements of $\mathbf{P}(S_{B}(x))$ as $z$ consecutive elements starting from the maximum important power plus one, \ie $s-1+s(t-1)+(ts+\lambda)(t-1)$ plus one; $\mathbf{P}(S_B(x))=\{ts+(ts+\lambda)(t-1),\dots,ts+(ts+\lambda)(t-1)+z-1\}$ or equivalently:
$\mathbf{P}(S_B(x))= \{ts+\theta(t-1)+r, r\in \Omega_0^{z-1}\}$. 
We note that the elements of $\mathbf{P}(C_A(x))$ and $\mathbf{P}(S_A(x))$ are powers of polynomials, so they are non-negative. Therefore, by starting the elements of $\mathbf{P}(S_B(x))$ from the maximum important power plus one, C1 and C3 are satisfied. Then, we find all elements of the subset of $\mathbf{P}(S_A(x))$, starting from the minimum possible element, that satisfies C2 in (\ref{eq:non_eq-AGE-conditions}). Using this rule, we can determine $S_A(x)$ and $S_B(x)$ as
\begin{align}\label{eq:S-A}
   S_A(x)= \bigg\{ \begin{array}{cc}
   S_{A_1}(x) 
     &  z > \lambda, \text{ and } t \neq 1\\
   S_{A_2}(x) & z = \lambda, \text{ or } t=1
\end{array}\end{align} where $S_{A_1}(x) = \sum_{w=0}^{\lambda-1} \sum_{l=0}^{q-1}   \bar{A}_{(w+\theta l)}x^{ts+\theta l+w}  +  \sum_{u=0}^{z-1-q\lambda}   \bar{A}_{(u+\lambda+\theta(q-1))}  x^{ts+\theta q+u}$, $S_{A_2}(x)$ $=$ $\sum_{u=0}^{z-1}$ $\bar{A}_{u}x^{ts+u}$, and $\bar{A}_{(w+\theta l)}$, $\bar{A}_{(u+\lambda+\theta(q-1))}$, and $\bar{A}_{u}$ are chosen independently and uniformly at random in $\mathbb{F}^{\frac{m}{t} \times \frac{m}{s}}$. 
\begin{align}\label{eq:S-B}
   S_{B}(x)=\sum_{r=0}^{z-1} \bar{B}_rx^{ts+\theta(t-1)+r},
\end{align} where $\bar{B}_r$ is chosen independently and uniformly at random in $\mathbb{F}^{\frac{m}{s} \times \frac{m}{t}}$. 
\begin{theorem}\label{th:FA-FB-AGE-thrm}
The polynomials $S_{A}(x)$ and $S_{B}(x)$ defined in (\ref{eq:S-A}) and (\ref{eq:S-B}) satisfy the conditions in (\ref{eq:non_eq-AGE-conditions}). 
\end{theorem}
{\em Proof:}
The proof is provided in Appendix B. \hfill $\Box$

In phase 1, source 1 shares $F_{A}(\alpha_n)$ and source 2 shares $F_B(\alpha_n)$ with each worker $W_n$. Due to using $z$ random terms in constructing $F_{A}(x)$ and $F_B(x)$, no information about $A$ and $B$ is revealed to any workers.

\textbf{Phase 2 - Workers Compute and Communicate.} The second phase consists of workers processing data received from the sources and sharing the results with each other. In this phase, each worker $W_n$ calculates $H(\alpha_n) = F_{A}(\alpha_n)F_B(\alpha_n)$, where ${H}(x)$ is defined as: 
\begin{align}\label{eq:HxPolyDotCMPC}
    {H}(x) = \sum_{n=0}^{\deg(F_A(x))+\deg(F_B(x))} {H}_nx^n = F_{A}(x)F_B(x),
\end{align}
where ${H}_u = \sum_{j=0}^{s-1}A_{i,j}B_{j,l}$ are the coefficients that are required for calculating $A^TB$, \ie $u = si+(s-1)+\theta l$ for $i,l \in \Omega_{0}^{t-1}$. 
Each worker $W_n$ has the knowledge of one point from ${H}(x)$ through calculation of $H(\alpha_n)=F_{A}(\alpha_n)F_B(\alpha_n)$. By applying Lagrange interpolation on (\ref{eq:HxPolyDotCMPC}), there exist $r_n^{(i,l)}$'s such that
\begin{align}\label{lagrangeinterpolpolydot}
    H_u=\sum_{j=0}^{s-1} A_{ij}B_{jl} = \sum_{n=1}^{N} r_n^{(i,l)} {H}(\alpha_n).
\end{align} Thus, each worker $W_n$ multiplies $r_n^{(i,l)}$'s with $H(\alpha_n)$ and shares them with the other workers, securely. In particular, for each worker $W_n$, there are $t^2$ coefficients of $r_n^{(i,l)}$. Therefore, each worker $W_n$ creates a polynomial $G_n(x)$ with the first $t^2$ terms allocated to multiplication of $r_n^{(i,l)}$ with $H(\alpha_n)$ and the last $z$ terms allocated to random coefficients to keep $H(\alpha_n)$ confidential from $z$ colluding workers:
\begin{align} \label{eq:FnPolyDotCMPC}
    G_n(x) = \sum_{i=0}^{t-1}\sum_{l=0}^{t-1}r_n^{(i,l)}{H}(\alpha_n)x^{i+tl}+ \sum_{w=0}^{z-1} R^{(n)}_wx^{t^2+w},
\end{align}
where $R^{(n)}_w, w \in \Omega_{0}^{z-1}$ are chosen independently and uniformly at random from $\mathbb{F}^{\frac{m}{t} \times \frac{m}{t}}$.
Each worker $W_n$ sends $G_n(\alpha_{n'})$ to all other workers $W_{n'}$. After all the data exchanges, each worker $W_{n'}$ has the knowledge of $G_n(\alpha_{n'})$, which sums them up and sends it to the master in the last phase. The following equation represents the polynomial that is equal to the summation of $G_n(x)$:
\begin{align}
    I(x) = \sum_{n=1}^{N} G_n(x),
\end{align}
which can be equivalently written as:
\begin{align}\label{eq:FxPolyDotCMPC}
    I(x)= & \sum_{i=0}^{t-1}\sum_{l=0}^{t-1}\sum_{n=1}^{N}r_{n}^{(i,l)}{H}(\alpha_n)x^{i+tl}+ \sum_{w=0}^{z-1}\sum_{n=1}^{N} R^{(n)}_wx^{t^2+w} \nonumber \\
    = &  \sum_{i=0}^{t-1}\sum_{l=0}^{t-1}\sum_{j=0}^{s-1}A_{ij}B_{jl}x^{i+tl} + \sum_{w=0}^{z-1}\sum_{n=1}^{N} R^{(n)}_wx^{t^2+w}.
\end{align}
\textbf{Phase 3 - Master Node Reconstructs $Y = A^TB$.} As seen in (\ref{eq:FxPolyDotCMPC}), the coefficients for the first $t^2$ terms of $I(x)$ represent the components of the matrix $Y=A^TB$. On the other hand, the degree of $I(x)$ is $t^2+z-1$, therefore, the master can reconstruct $I(x)$ and extract $Y=A^TB$ after receiving $I(\alpha_n)$ from $t^2+z$ workers. 

\begin{theorem} \label{th:N_AGE}
The total number of workers required to compute $Y = A^TB$ using AGE-CMPC, when there exist $z$ colluding workers and each worker can work on at most $\frac{1}{st}$ 
fraction of data from each source due to the computation or storage constraints, is expressed as
\begin{align}\label{eq:eq:N-AGE-CMPC-optimization}
    N_{\text{AGE-CMPC}} =\begin{cases} \displaystyle\min_{\lambda} \Gamma(\lambda) & t \neq 1\\
    2s+2z-1 & t=1
    \end{cases}
\end{align} 
where $\Gamma(\lambda)$ is defined as
\begin{align}\label{eq:N-AGE-CMPC}
\Gamma(\lambda)=\begin{cases}
   \Upsilon_1(\lambda), & z>ts-s,\; \lambda=0\\
   \Upsilon_2(\lambda), & z \leq ts-s,\; \lambda=0\\
   \Upsilon_3(\lambda), & \lambda=z\\
   \Upsilon_4(\lambda), & z>ts,\; 0 < \lambda < z\\
   \Upsilon_5(\lambda), & z\leq ts,\; 0< \lambda < z,\; ts<\lambda+s-1\\
   \Upsilon_6(\lambda), & \lambda+s-1<z\leq ts,\;0 < \lambda < z,\; q\lambda \geq s \\
   \Upsilon_7(\lambda), & \lambda+s-1<z\leq ts,\;0 < \lambda < z,\; q\lambda < s\\
   \Upsilon_8(\lambda), & z \leq \lambda+s-1\leq ts,\;0 < \lambda < z,\; q\lambda \geq s\\
   \Upsilon_9(\lambda), & z \leq \lambda+s-1 \leq ts,\;0 < \lambda < z,\; q\lambda < s,
\end{cases}
\end{align}
and $\Upsilon_1(0)=2st^2+2z-1$, $\Upsilon_2(0)=st^2+3st-2s+t(z-1)+1$, $\Upsilon_3(z)=2ts+(ts+z)(t-1)+2z-1$, $\Upsilon_4(\lambda)=(q+2)ts+\theta(t-1)+2z-1$, $\Upsilon_5(\lambda)=3ts+\theta(t-1)+2z-1$, $\Upsilon_6(\lambda)=2ts+\theta(t-1)+(q+2)z-q-1$, $\Upsilon_7(\lambda)= \theta(t+1)+q(z-1)-2\lambda +z+ts+\min\{0, z+s(1-t)-\lambda q-1\}$, $\Upsilon_8(\lambda)=2ts+\theta(t-1)+3z+(\lambda+s-1)q-\lambda-s-1$, $\Upsilon_9(\lambda)=\theta(t+1)+q(s-1)-3\lambda+3z-1\nonumber+\min\{0,ts-z+1+\lambda q-s\}$, $s \geq 1$, $t \geq 2$, $s|m$, $t|m$ are satisfied, $q=\min\{\floor{\frac{z-1}{\lambda}},t-1\}$ and $\theta = ts+\lambda$.
\end{theorem}
{\em Proof:}
The proof is provided in Appendix C. \hfill $\Box$

\begin{example}\label{ex:AGECMPC}
This example shows the operation of AGE-CMPC in a scenario when $s=t=z=2$. In particular, there are two sources (Source A and Source B) that have matrices $A$ and $B$. The sources partition the matrices to $st=4$ sub-matrices; \ie $s=2$ row-wise and $t=2$ column-wise partitions. These sub-matrices will be multiplied with the help of a number of workers, where $z=2$ workers are adversaries.   

First, we solve the optimization problem in (\ref{eq:eq:N-AGE-CMPC-optimization}) to determine $\lambda$ that optimizes the required number of workers $N_{\text{AGE-CMPC}}$. The solution of (\ref{eq:eq:N-AGE-CMPC-optimization})  becomes $N_{\text{AGE-CMPC}} = 17$ and $\lambda^* = 2$ 
when $s=t=z=2$. This means that 17 workers are required by AGE-CMPC to guarantee privacy. We note that the required number of workers by Entangled-CMPC  \cite{8613446} is $N_{\text{Entangled-CMPC}}=19$. As seen, AGE-CMPC reduces the required number of workers as compared to Entangled-CMPC. 

Now that $\lambda$ is fixed to $\lambda = \lambda^* = 2$, we follow the phases in Section \ref{sec:AGE-CMPC}. The sources determine the coded terms $C_A(x)$ and $C_B(x)$ by the AGE codes described in Section \ref{sec:AGECodes} when $\lambda = 2$. We obtain  $C_A(x) = A_{00} + A_{01}x + A_{10}x^2 + A_{11}x^3$ and $C_B(x) = B_{00}x + B_{10} + B_{01}x^7 + B_{11}x^6$. Then, the degrees of the secret terms are determined as $S_A(x)$ and $S_B(x)$ that satisfy the conditions in (\ref{eq:non_eq-AGE-conditions}). We obtain $S_A(x) = \bar{A}_0x^4 + \bar{A}_1x^5$ and $S_B(x) = \bar{B}_0x^{10} + \bar{B}_1x^{11}$. Thus, Source A constructs $ F_{A}(x ) = A_{00} + A_{01}x + A_{10}x^2 + A_{11}x^3 + \bar{A}_0x^4 + \bar{A}_1x^5$, and Source B constructs $F_B(x) = B_{00}x + B_{10} + B_{01}x^7 + B_{11}x^6 + \bar{B}_0x^{10} + \bar{B}_1x^{11}$.  Sources A and B send $F_A(\alpha_n)$ and $F_B(\alpha_n)$ to each worker $W_n$, for some distinct $\alpha_1, \ldots, \alpha_N$.

In the second phase, each worker $W_n$ calculates $H(\alpha_n) = F_A(\alpha_n)F_B(\alpha_n)$. Then, all $N_{\text{AGE-CMPC}}=17$ workers collaborate to apply Lagrange interpolation on the polynomial $H(x) = \sum_{n=0}^{16} H_nx^n = F_A(x)F_B(x)$ (through calculation of $H(\alpha_n)$) to determine $r_n^{(i,l)}, i,l = \{0, 1\}, n=\{1, ..., 17\}$, such that:
 \begin{align}
     & H_1 = A_{00}B_{00} + A_{01}B_{10},  H_3 = A_{10}B_{00} + A_{11}B_{10}, \nonumber \\
     & H_7 = A_{00}B_{01} + A_{01}B_{11},  H_9 = A_{10}B_{01} + A_{11}B_{11}. \nonumber
 \end{align}
 Next, each worker $W_n$ multiplies $r_n^{(i,j)}, i,j \in \{0,1\}$ with $H(\alpha_n)$ and creates the polynomial $G_n(x)$ as 
 \begin{align}
          G_n(x) = &
           r_n^{(0,0)}{H}(\alpha_n)+ r_n^{(1,0)}{H}(\alpha_n)x+r_n^{(0,1)}{H}(\alpha_n)x^2 \nonumber \\
           &+r_n^{(1,1)}{H}(\alpha_n)x^3
           +R^{(n)}_0x^4 + R^{(n)}_1x^5. \nonumber
 \end{align}
 Then, each worker $W_n$ sends  $G_n(\alpha_{n'})$ to $W_{n'}$. After all data exchanges, each worker $W_{n'}$ has the knowledge of $G_n(\alpha_{n'})$, and sends $\sum_{n=1}^{17} G_n(\alpha_{n'})$ to the master.
 
 In the last phase, the master reconstructs $I(x)$ once it receives $I(\alpha_n)=\sum_{n'=1}^{17} G_{n'}(\alpha_{n})$ from $t^2+z=6$ workers.
 \begin{align}
     I(x) = 
     &(A_{00}B_{00} + A_{01}B_{10})+ (A_{10}B_{00} + A_{11}B_{10}) x + \nonumber  \\
     &(A_{00}B_{01} + A_{01}B_{11}) x^2+ (A_{10}B_{01} + A_{11}B_{11})x^3+ \nonumber \\
     &\sum_{n=1}^{17} R^{(n)}_0x^4+\sum_{n=1}^{17} R^{(n)}_1x^5 \nonumber
  \end{align}
 After reconstructing $I(x)$ and determining all coefficients, $Y$ is calculated as
 \begin{align}
   Y = A^TB = \left[{\begin{array}{cc}
   A_{00}B_{00} + A_{01}B_{10} & A_{00}B_{01} + A_{01}B_{11}\\
   A_{10}B_{00} + A_{11}B_{10} & A_{10}B_{01} + A_{11}B_{11}
   \end{array} } \right], \nonumber
 \end{align} in a privacy-preserving manner.
\hfill $\Box$
\end{example}
\subsection{AGE-CMPC in Perspective}\label{sec:AGECMPCvsOthers}
In this section we compare AGE-CMPC with Entangled-CMPC \cite{8613446}, SSMM \cite{Zhu2021ImprovedCF}, GCSA-NA \cite{9333639} (considering batch size as one), and PolyDot-CMPC \cite{PolyDot-CMPC} in terms of the number of required workers. For this purpose, let us define $\lambda^*$ as the optimum solution of the optimization problem in (\ref{eq:eq:N-AGE-CMPC-optimization}).
\begin{lemma}\label{corol:AGE-Vs-Entang}
$N_{\text{AGE-CMPC}}$ is less than the number of workers required by Entangled-CMPC \cite{8613446}, $N_{\text{Entangled-CMPC}}$, when $0 < \lambda^* \leq z$. For the case of $\lambda^*=0$, $N_{\text{AGE-CMPC}}=N_{\text{Entangled-CMPC}}$.
\end{lemma}

{\em Proof:}
 The proof is provided in Appendix D.A.\hfill $\Box$
\begin{lemma}\label{corol:AGE-Vs-SSMM} $N_{\text{AGE-CMPC}}$ is less than the number of workers required by SSMM \cite{Zhu2021ImprovedCF}, $N_{\text{SSMM}}$, when $0 \leq \lambda^* < z$. For the case of $\lambda^*=z$, $N_{\text{AGE-CMPC}}=N_{\text{SSMM}}$.
\end{lemma}

{\em Proof:}
 The proof is provided in Appendix D.B. \hfill $\Box$
\begin{lemma}\label{corol:AGE-Vs-GCSANA}
$N_{\text{AGE-CMPC}}$ is less than the number of workers required by GCSA-NA (for one matrix multiplication) \cite{9333639}, $N_{\text{GCSA-NA}}$, when $0 < \lambda^* \leq z$. For the case of $\lambda^*=0$, $N_{\text{AGE-CMPC}} \leq N_{\text{GCSA-NA}}$.
\end{lemma}

{\em Proof:}
 The proof is provided in Appendix D.C. \hfill $\Box$
\begin{lemma}\label{corol:AGE-Vs-polydot-1}
$N_{\text{AGE-CMPC}}$ is always less than or equal to the number of workers required by PolyDot-CMPC \cite{PolyDot-CMPC}. 
\end{lemma}

{\em Proof:}
 The proof is provided in Appendix D.D. \hfill $\Box$

\section{Computation, Storage, Communication and Privacy Analysis}\label{sec:recoverythreshold,comp,com,S analysis} 

In this section, we provide a theoretical analysis for the computation, storage, and communication overhead of AGE-CMPC, and discuss its privacy guarantee.


\subsection{Computation Overhead} 

We define the computation overhead as the total number of scalar multiplications performed by each worker. We neglect additions as the computation complexity of addition is negligible compared to multiplication. 
\begin{corollary} \label{cor:computation_all_methods}
The total computation overhead per worker to compute $Y=A^TB$ using AGE-CMPC 
is equal expressed as
\begin{align} \label{eq:AGE_computation_cost}
    \xi_{\text{AGE-CMPC}} =  \frac{m^3}{st^2}+m^2+N_{\text{AGE-CMPC}}(t^2+z-1)\frac{m^2}{t^2},
\end{align} where $m$ is the number of rows/columns of matrices $A$ and $B$, $s$ and $t$ are the number of row-wise and column-wise partitions, respectively, and $z$ is the number of colluding workers.
\end{corollary} 
{\em Proof:} First, each worker computes $H(\alpha_n)=F_A(\alpha_n)F_B(\alpha_n)$. 
In AGE-CMPC, $F_A(\alpha_n) \in \mathbb{F}^{ \frac{m}{t} \times \frac{m}{s}}$ and $F_B(\alpha_n) \in \mathbb{F}^{\frac{m}{s} \times \frac{m}{t}}$, so $\frac{m^3}{st^2}$ scalar multiplications are computed. 
After computing $H(\alpha_n)$, each worker $W_n$ needs to compute polynomial $G_n(x)$ for $N$ different points; $\alpha_{n'}, n'=1, \ldots, N$ following (\ref{eq:FnPolyDotCMPC}). For this purpose, worker $W_n$ first  multiplies $r_n^{i,j}$ for $i,j \in \{0,\ldots,t-1\}$ with $H(\alpha_n)\in \mathbb{F}^{\frac{m}{t} \times \frac{m}{t}}$. This requires $t^2\frac{m^2}{t^2}=m^2$ scalar multiplications. Then,  $r_n^{i,j}H(\alpha_n)$ is multiplied with $\alpha_{n'}^{i+tj}$ for all $N$ workers. This requires $N(t^2-1)\frac{m^2}{t^2}$ scalar multiplications. To calculate the second part of $G_n(x), n=1,\ldots,N$, $W_n$ multiplies $\alpha_{n'}^{t^2+w}$ with random matrices $R^{(n)}_w \in \mathbb{F}^{\frac{m}{t} \times \frac{m}{t}}$, for $w = \{0,\ldots,z-1\}$. This requires $Nz\frac{m^2}{t^2}$ scalar multiplications. In total, each worker $W_n$ computes $m^2+N(t^2+z-1)\frac{m^2}{t^2}$ scalar multiplications to obtain $G_n(\alpha_{n'})$'s. By adding the number of scalar multiplications required for computing $H(\alpha_n)$, the computation overhead of AGE-CMPC becomes $\frac{m^3}{st^2}+m^2+N(t^2+z-1)\frac{m^2}{t^2}$. This concludes the proof. \hfill $\Box$

\subsection{Storage Overhead}
We define the storage overhead as the total number of scalar parameters that should be stored in all phases of coded MPC at each worker.\footnote{We note that it is possible to delete some of the data after each phase once they are not needed for future steps, but we do not consider deleting data for easy exposition.} These parameters include the received parameters from the other workers as well as those that are computed and stored to be used in the next computations.
\begin{corollary} \label{cor:Storage_all_methods}
The total storage overhead per worker to compute $Y=A^TB$ using AGE-CMPC is expressed as
\begin{align} \label{eq:AGE_storage}
    \sigma_{\text{AGE-CMPC}} = (2N_{\text{AGE-CMPC}}+z+1)\frac{m^2}{t^2}+\frac{2m^2}{st}+t^2, 
\end{align} where $m$ is the number of rows/columns of matrices $A$ and $B$, $s$ and $t$ are the number of row-wise and column-wise partitions, respectively, and $z$ is the number of colluding workers.
\end{corollary}
{\em Proof:} Storage overhead consists of the following components. In the first phase, each worker $W_n$ receives $F_A(\alpha_n)$ and $F_B(\alpha_n)$ each with the size of $\frac{m}{t} \times \frac{m}{s}$ from the sources. This requires storing $SC_1 = \frac{2m^2}{st}$ scalar parameters.

In the second phase, each worker $W_n$ stores $H(\alpha_n)$ with the size of $\frac{m^2}{t^2}$ computed by multiplying $F_A(\alpha_n)$ with $F_B(\alpha_n)$. This requires storing $SC_2 = \frac{m^2}{t^2}$ scalar parameters.

Next, each worker $W_n$ creates the polynomial $G_n(x)$ to calculate different points of it. This requires to store the coefficients for each term of this polynomial. For this purpose, in AGE-CMPC, according to (\ref{eq:FnPolyDotCMPC}), the random variables $r_n^{i,j}, 0 \leq i,j \leq t-1$ with the total number of $t^2$ scalar parameters are stored. In addition, the random matrices $R^{(n)}_w \in \mathbb{F}^{\frac{m}{t} \times \frac{m}{t}}$ for $w \in \{0,\ldots,z-1\}$ are stored. In total, this requires storing $SC_3 = t^2+z\frac{m^2}{t^2}$ scalar parameters.

After creating $G_n(x)$, worker $W_n$ needs to compute it at points $\alpha_n',\; n' \in \Nset\setminus {n}$ to send them to the other workers as well as at point $\alpha_n$ required for calculating $I(\alpha_n)$. Also, worker $W_n$ receives $G_{n'}(\alpha_n)$ from the other workers, which will be stored in its storage. In AGE-CMPC, $G_{n'}(\alpha_n) \in \mathbb{F}^{\frac{m}{t} \times \frac{m}{t}}$, so in total, this requires storing $SC_4 = (2N-1)\frac{m^2}{t^2}$ scalar parameters.

Finally, worker $W_n$ needs to compute and store $I(\alpha_n)$. In AGE-CMPC, $I(\alpha_n) \in \mathbb{F}^{\frac{m}{t} \times \frac{m}{t}}$, so this requires storing $SC_5 = \frac{m^2}{t^2}$ scalar parameters.

We can derive (\ref{eq:AGE_storage}) by adding the storage components; \ie $\sum_{i=1}^{N}SC_i$. This concludes the proof. \hfill $\Box$

\subsection{Communication Overhead} 
We define the communication overhead as the total number of scalar parameters that are exchanged among all workers in phase 2. Note that there are other data transmissions; from sources to workers in phase 1, and from workers to the master in phase 3. We do not include these communications in the communication overhead calculation as they are negligible as compared to the data exchange among workers in phase 2.
\begin{corollary}
\label{cor:communication_worker_worker_all_methods}
Communication overhead to compute $Y=A^TB$, using AGE-CMPC is expressed as
\begin{align} \label{eq:AGE_comm}
    \zeta_{\text{AGE-CMPC}} = N_{\text{AGE-CMPC}}(N_{\text{AGE-CMPC}}-1)\frac{m^2}{t^2},
\end{align} where $m$ is the number of rows/columns of matrices $A$ and $B$, and $s$ and $t$ are the number of row-wise and column-wise partitions, respectively.
\end{corollary}
{\em Proof:} In phase 2 of AGE-CMPC, each worker $W_n, n \in \Nset$ sends $G_n(\alpha_{n'})$ to worker $W_{n'}, n' \in \Nset\setminus{n}$.  
In AGE-CMPC, $G_n(\alpha_{n'}) \in \mathbb{F}^{\frac{m}{t} \times \frac{m}{t}}$. Therefore, the communication overhead among workers for AGE-CMPC is equal to $N(N-1)\frac{m^2}{t^2}$. This concludes the proof. \hfill $\Box$


\subsection{Privacy Analysis}
AGE-CMPC satisfies the privacy requirements stated in Section \ref{sec:system}. The proof directly follows from the proof of Theorem 3 in \cite{PolynomCMPC}.


\begin{figure}[t!]
\centering
\subfigure{ 
\scalebox{.17}{ \includegraphics{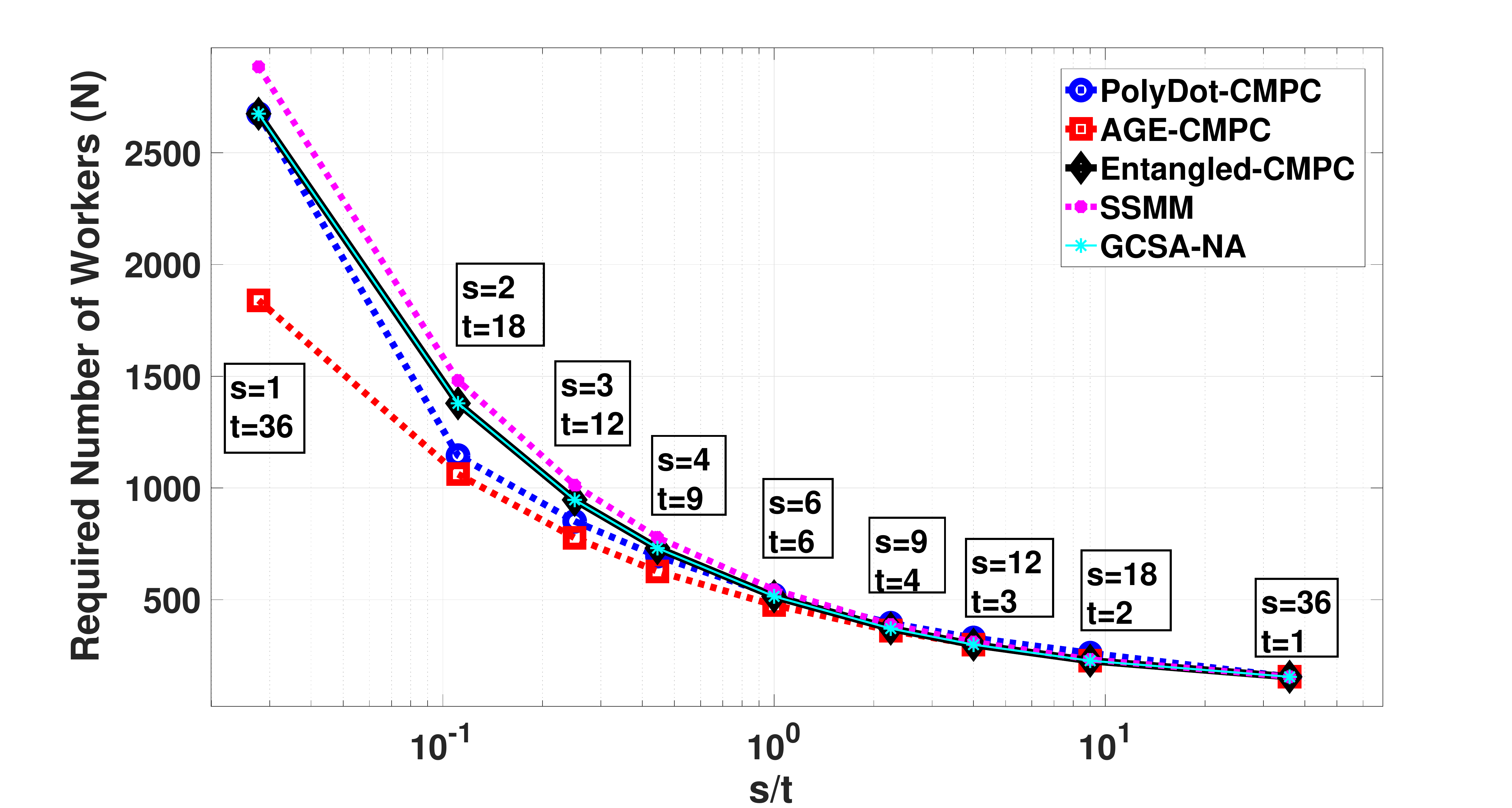}} } 
\vspace{-5pt}
\caption{Required number of workers.}
\vspace{-10pt}
\label{fig:N-AGE}
\end{figure}

\section{Performance Evaluation} \label{sec:simulation} 
We evaluate the performance of our algorithm, and compare with the baselines, (i) PolyDot-CMPC \cite{PolyDot-CMPC}, (ii) Entangled-CMPC \cite{8613446}, (iii) SSMM \cite{Zhu2021ImprovedCF}, and (iv) GCSA-NA \cite{9333639}. The system model parameters are considered as follows: the size of each matrix $A$ and $B$ is $m\times m=36000 \times 36000$, the number of colluding workers is $z=42$, the number of partitions of matrices $A$ and $B$ is $st=36$. 

Fig.~\ref{fig:N-AGE} shows the required number of workers needed to compute the multiplication of $Y=A^TB$ versus $s/t$, the number of row partitions over the number of column partitions. As seen, the required number of workers of AGE-CMPC is less than or equal to the other baselines, which confirms our Lemmas \ref{corol:AGE-Vs-Entang}, \ref{corol:AGE-Vs-SSMM}, \ref{corol:AGE-Vs-GCSANA}, and \ref{corol:AGE-Vs-polydot-1}. Moreover, by decreasing the number of column partitions $t$, the required number of workers of AGE-CMPC gets closer to the baselines, and for $t \leq 3$, it is equal to Entangled-CMPC. The reason is that the parameter $\theta=ts+\lambda$ in the power of $C_B(x)$ in (\ref{eq:generalEntangled}) is the coefficient of $l \in \Omega_{0}^{t-1}$, which is related to the number of column partitions, $t$, and it affects the number of gaps among powers of $C_B(x)$ when we have column partitions. Thus, decreasing $t$ results in decreasing the effect of $\theta$ in the number of gaps among powers of polynomial $C_A(x)C_B(x)$, \ie the garbage terms, hence the required number of workers. Finally, we can see through this figure that the required number of workers of all methods have direct relationship with the number of column partitions $t$.
%
%

\begin{figure}[t!]
\centering
\subfigure[Storage]{ \scalebox{.17}{\includegraphics{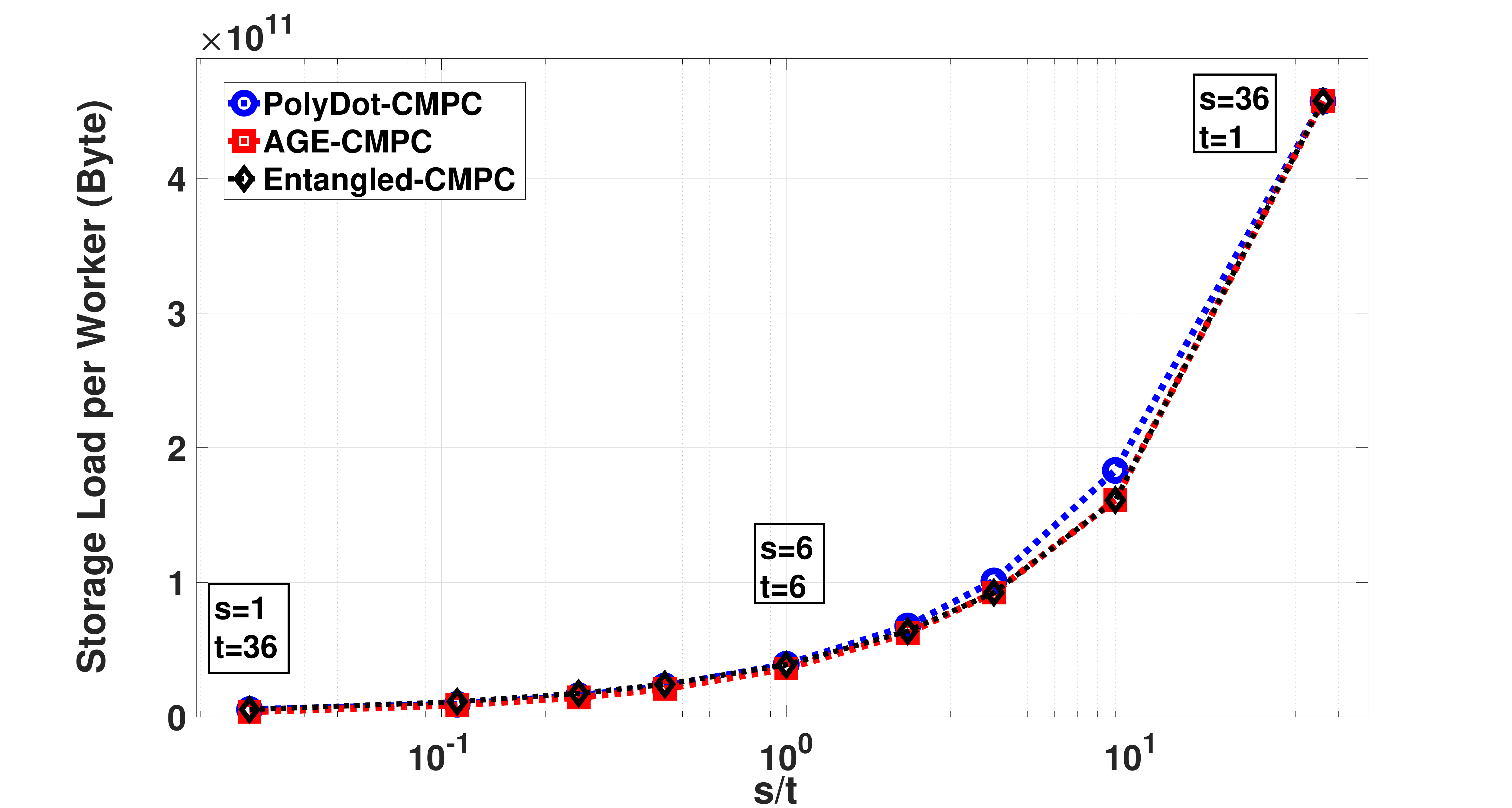}} } 
\subfigure[Computation]{ \scalebox{.17}{\includegraphics{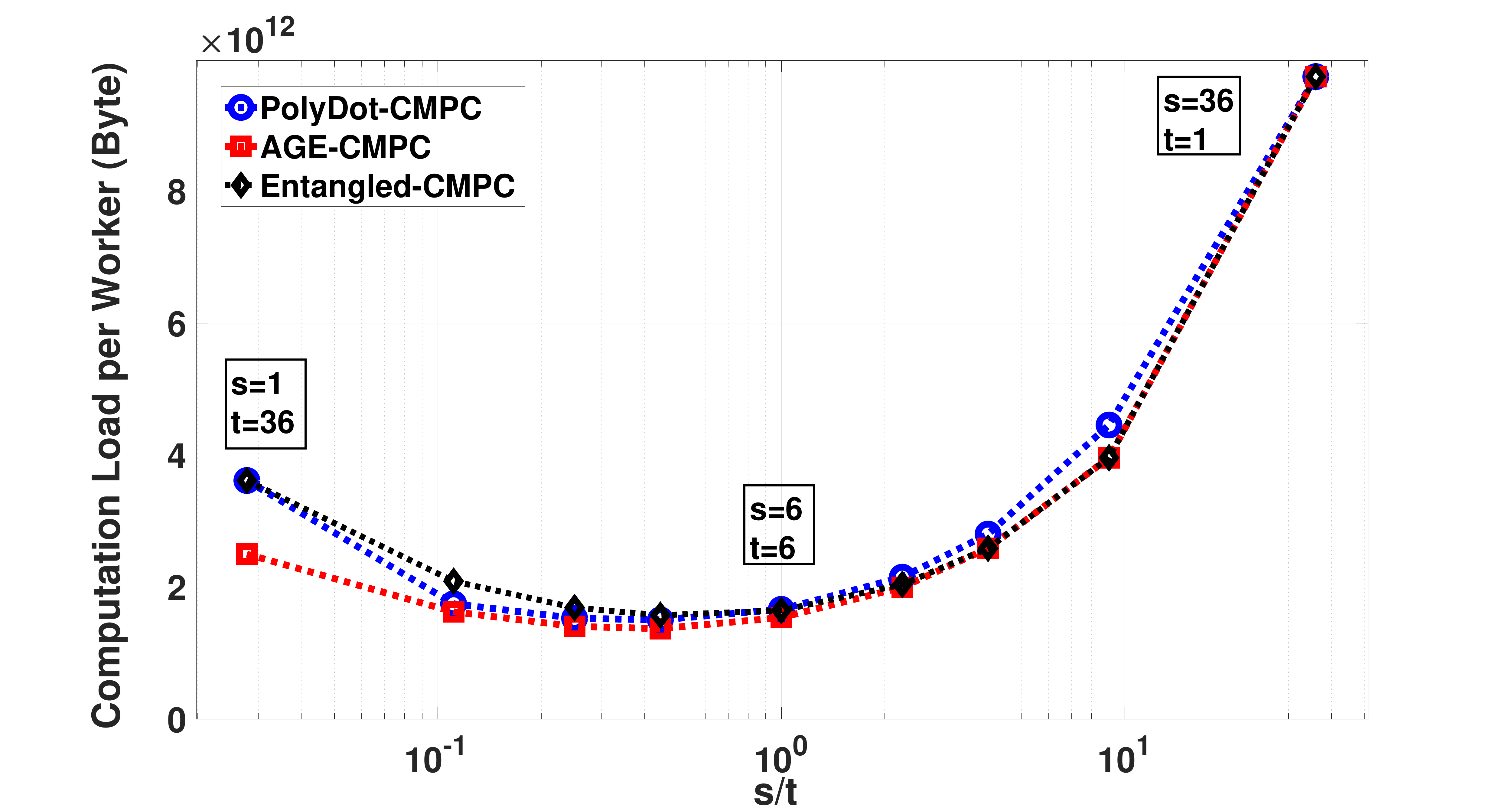}} } 
\subfigure[Communication]{ \scalebox{.17}{\includegraphics{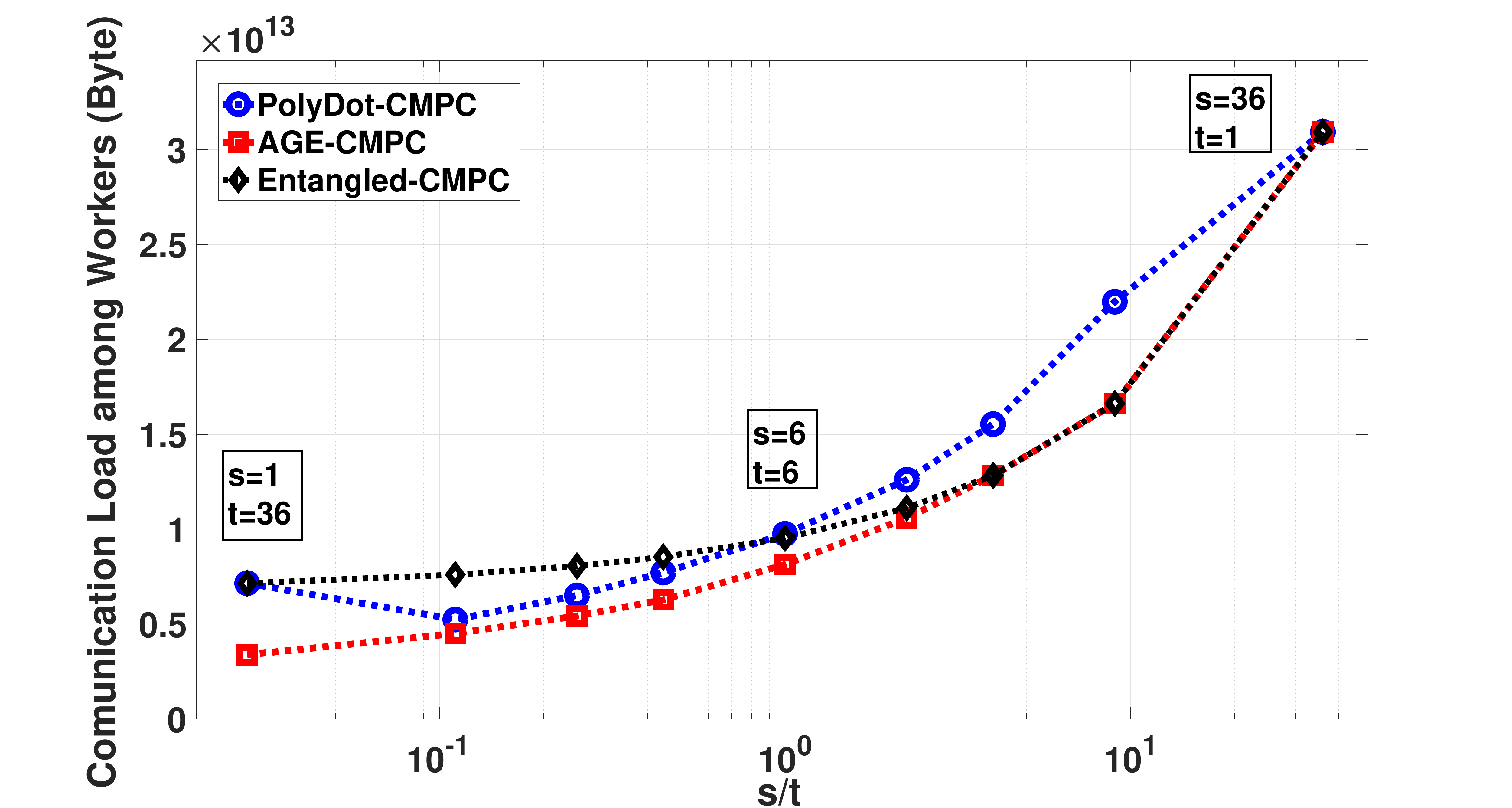}} } 
\caption{(a) Storage,  (b) computation, and (c) communication loads.}
\vspace{-15pt}
\label{fig:storage-comm-comp}
\end{figure}

Fig.~\ref{fig:storage-comm-comp}(a) shows the storage cost per worker, where the size of each stored scalar is 1 Byte, versus $s/t$. AGE-CMPC reduces the storage load per worker as compared to baselines. 
The reason is that there is a direct relationship between the required number of workers and storage load per worker in CMPC setup when we fix $s$ and $t$. As we described in phase 2, each worker $n$ needs to compute polynomial $G_n(\alpha_{n'})$ and send it to workers $n'$ for $n' \in \{0,\dots,N\}\setminus \{n\}$, also it needs to receive $G_{n'}(\alpha_n)$ from worker $n'$. Therefore, the smaller required number of workers of AGE-CMPC results in the smaller storage load per worker as compared to PolyDot-CMPC and Entangled-CMPC. On the other hand, for a fixed $st=36$, if we change $s$ and $t$ from $s=1, t=36$ to $s=36, t=1$, one can see the existing trade-off between the storage and the required number of workers by comparing Fig.~\ref{fig:N-AGE} with Fig.~\ref{fig:storage-comm-comp}(a). The reason is that the required number of workers has a direct relation with the number of column partitions $t$, while the storage load per worker has an inverse relation with $t$. Therefore, based on the resource limitations of the system, one can choose an appropriate $s$ and $t$ to exploit this trade-off.

Fig.~\ref{fig:storage-comm-comp}(b) shows the computation cost per worker versus $s/t$. Similar to the discussion in Fig.~\ref{fig:storage-comm-comp}(a), for the fixed amounts of $s$ and $t$, required number of workers has a direct relation with computation load per worker, \ie larger amounts of workers results in larger computation load per worker. Therefore, computation load per worker of AGE-CMPC is less than or equal to the other methods. However, as seen in Fig.~\ref{fig:storage-comm-comp}(b), computation load per worker does not have a monotonic behavior by decreasing $t$. In other words, for fixed $st=36$, if we decrease $t$ from $36$ to $9$, computation load per worker will decrease, and if we decrease $t$ from $9$ to $1$, computation load per worker will increase. The reason is that computation load per worker has a direct relationship with both storage load per worker and the required number of workers. 


Fig.~\ref{fig:storage-comm-comp}(c) shows the communication overhead versus $s/t$. The communication overhead consists of; (i) from sources to workers in phase 1, (ii) among workers in phase 2, and (iii) from workers to the master in phase 3. We consider the communication cost among workers as it is the dominating communication cost in this system.
We assume that each scalar that is transmitted among workers is 1 Byte. Similar to our discussions for storage and computation loads, for the fixed $s$ and $t$, the required number of workers has a direct relationship with the communication load among workers, \ie larger number of workers results in larger communication load among workers. Therefore, the communication load among workers of AGE-CMPC is less than or equal to the other methods. 

\section{Conclusion} \label{sec:conc}
We have investigated coded privacy-preserving computation using Shamir's secret sharing. We have designed a novel coded computation method; AGE codes that is customized for coded privacy-preserving computations. We also designed a coded privacy-preserving computation mechanism; AGE coded MPC (AGE-CMPC) by employing  AGE codes. We designed our algorithm such that it takes advantage of the ``garbage terms''. Also, we have analyzed AGE-CMPC in terms of the required number of workers as well as its computation, storage, and communication overhead, and shown that AGE-CMPC provides significant improvement. 

\bibliographystyle{IEEEtran}

\bibliography{refs}

\begin{thebibliography}{10}
\providecommand{\url}[1]{#1}
\csname url@samestyle\endcsname
\providecommand{\newblock}{\relax}
\providecommand{\bibinfo}[2]{#2}
\providecommand{\BIBentrySTDinterwordspacing}{\spaceskip=0pt\relax}
\providecommand{\BIBentryALTinterwordstretchfactor}{4}
\providecommand{\BIBentryALTinterwordspacing}{\spaceskip=\fontdimen2\font plus
\BIBentryALTinterwordstretchfactor\fontdimen3\font minus
  \fontdimen4\font\relax}
\providecommand{\BIBforeignlanguage}[2]{{%
\expandafter\ifx\csname l@#1\endcsname\relax
\typeout{** WARNING: IEEEtran.bst: No hyphenation pattern has been}%
\typeout{** loaded for the language `#1'. Using the pattern for}%
\typeout{** the default language instead.}%
\else
\language=\csname l@#1\endcsname
\fi
#2}}
\providecommand{\BIBdecl}{\relax}
\BIBdecl

\bibitem{IDCReport}
R.~Swearingen, ``Idc report 2020: Iot growth demands rethink of long-term
  storage strategies, says idc,'' 2020.

\bibitem{DemocratizingNetworkEdge}
\BIBentryALTinterwordspacing
L.~Peterson, T.~Anderson, S.~Katti, N.~McKeown, G.~Parulkar, J.~Rexford,
  M.~Satyanarayanan, O.~Sunay, and A.~Vahdat, ``Democratizing the network
  edge,'' \emph{SIGCOMM Comput. Commun. Rev.}, vol.~49, no.~2, pp. 31--36, May
  2019. [Online]. Available: \url{http://doi.acm.org/10.1145/3336937.3336942}
\BIBentrySTDinterwordspacing

\bibitem{EdgeComputingVideo}
\BIBentryALTinterwordspacing
P.~Levine and A.~Horowitz, ``Return to the edge and the end of cloud
  computing,'' 2017. [Online]. Available:
  \url{https://www.youtube.com/watch?v=-QRXQTSZxdQ}
\BIBentrySTDinterwordspacing

\bibitem{EdgeEatCloud}
G.~M. Research, ``The edge will eat the cloud,'' 2017.

\bibitem{scalableMPC}
J.~Saia and M.~Zamani, ``Recent results in scalable multi-party computation,''
  in \emph{SOFSEM 2015: Theory and Practice of Computer Science}, G.~F.
  Italiano, T.~Margaria-Steffen, J.~Pokorn{\'y}, J.-J. Quisquater, and
  R.~Wattenhofer, Eds.\hskip 1em plus 0.5em minus 0.4em\relax Berlin,
  Heidelberg: Springer Berlin Heidelberg, 2015, pp. 24--44.

\bibitem{Yao}
A.~C.-C. Yao, ``How to generate and exchange secrets,'' in \emph{27th Annual
  Symposium on Foundations of Computer Science (sfcs 1986)}, 1986, pp.
  162--167.

\bibitem{GMW}
S.~M. O.~Goldreich and A.~Wigderson, ``How to play any mental game,'' in
  \emph{Proc. of the 19th STOC}, 1987, pp. 218--229.

\bibitem{BGW}
M.~Ben-Or, S.~Goldwasser, and A.~Wigderson, ``Completeness theorems for
  non-cryptographic fault-tolerant distributed computation,'' in
  \emph{Providing Sound Foundations for Cryptography: On the Work of Shafi
  Goldwasser and Silvio Micali}, 2019, pp. 351--371.

\bibitem{ShamirSS}
A.~Shamir, ``How to share a secret,'' \emph{Communications of the ACM},
  vol.~22, no.~11, pp. 612--613, 1979.

\bibitem{10.1007/3-540-48405-1_4}
U.~Maurer, ``Information-theoretic cryptography,'' in \emph{Advances in
  Cryptology --- CRYPTO' 99}, M.~Wiener, Ed.\hskip 1em plus 0.5em minus
  0.4em\relax Berlin, Heidelberg: Springer Berlin Heidelberg, 1999, pp. 47--65.

\bibitem{SpeedUp-journal}
K.~{Lee}, M.~{Lam}, R.~{Pedarsani}, D.~{Papailiopoulos}, and K.~{Ramchandran},
  ``Speeding up distributed machine learning using codes,'' \emph{IEEE
  Transactions on Information Theory}, vol.~64, no.~3, March 2018.

\bibitem{Tradeoff-journal}
S.~{Li}, M.~A. {Maddah-Ali}, Q.~{Yu}, and A.~S. {Avestimehr}, ``A fundamental
  tradeoff between computation and communication in distributed computing,''
  \emph{IEEE Transactions on Information Theory}, vol.~64, no.~1, pp. 109--128,
  Jan 2018.

\bibitem{PolyDot-CMPC}
\BIBentryALTinterwordspacing
E.~Vedadi, Y.~Keshtkarjahromi, and H.~Seferoglu, ``Polydot coded privacy
  preserving multi-party computation at the edge,'' in \emph{IEEE Signal
  Processing Advances in Wireless Communications (SPAWC) (invited paper)},
  2022. [Online]. Available: \url{https://https://nrl.ece.uic.edu/}
\BIBentrySTDinterwordspacing

\bibitem{8613446}
H.~A. Nodehi, S.~R.~H. Najarkolaei, and M.~A. Maddah-Ali, ``Entangled
  polynomial coding in limited-sharing multi-party computation,'' in \emph{2018
  IEEE Information Theory Workshop (ITW)}, 2018, pp. 1--5.

\bibitem{Zhu2021ImprovedCF}
J.~Zhu, Q.~Yan, and X.~Tang, ``Improved constructions for secure multi-party
  batch matrix multiplication,'' \emph{IEEE Transactions on Communications},
  vol.~69, pp. 7673--7690, 2021.

\bibitem{9333639}
Z.~Chen, Z.~Jia, Z.~Wang, and S.~A. Jafar, ``Gcsa codes with noise alignment
  for secure coded multi-party batch matrix multiplication,'' \emph{IEEE
  Journal on Selected Areas in Information Theory}, vol.~2, no.~1, pp.
  306--316, 2021.

\bibitem{YuFundamentalLimits2018}
Q.~Yu, M.~A. Maddah-Ali, and A.~S. Avestimehr, ``Straggler mitigation in
  distributed matrix multiplication: Fundamental limits and optimal coding,''
  \emph{IEEE Transactions on Information Theory}, vol.~66, no.~3, pp.
  1920--1933, 2020.

\bibitem{SpeedingUp}
K.~Lee, M.~Lam, R.~Pedarsani, D.~Papailiopoulos, and K.~Ramchandran, ``Speeding
  up distributed machine learning using codes,'' \emph{IEEE Transactions on
  Information Theory}, vol.~64, no.~3, pp. 1514--1529, 2018.

\bibitem{FerdinandAnytime}
N.~S. Ferdinand and S.~C. Draper, ``Anytime coding for distributed
  computation,'' in \emph{2016 54th Annual Allerton Conference on
  Communication, Control, and Computing (Allerton)}, 2016, pp. 954--960.

\bibitem{YuPolynomial2017}
Q.~Yu, M.~A. Maddah-Ali, and S.~Avestimehr, ``Polynomial codes: an optimal
  design for high-dimensional coded matrix multiplication,'' in \emph{NIPS},
  2017, pp. 4406--4416.

\bibitem{LeeHighDim2017}
K.~Lee, C.~Suh, and K.~Ramchandran, ``High-dimensional coded matrix
  multiplication,'' in \emph{2017 IEEE International Symposium on Information
  Theory (ISIT)}, 2017, pp. 2418--2422.

\bibitem{PolyDotMatDot}
M.~Fahim, H.~Jeong, F.~Haddadpour, S.~Dutta, V.~Cadambe, and P.~Grover, ``On
  the optimal recovery threshold of coded matrix multiplication,'' in
  \emph{2017 55th Annual Allerton Conference on Communication, Control, and
  Computing (Allerton)}.\hskip 1em plus 0.5em minus 0.4em\relax IEEE, 2017, pp.
  1264--1270.

\bibitem{ShortDot}
S.~Dutta, V.~Cadambe, and P.~Grover, ``“short-dot”: Computing large linear
  transforms distributedly using coded short dot products,'' \emph{IEEE
  Transactions on Information Theory}, vol.~65, no.~10, pp. 6171--6193, 2019.

\bibitem{DuttaCodedConvolution2017}
------, ``Coded convolution for parallel and distributed computing within a
  deadline,'' in \emph{2017 IEEE International Symposium on Information Theory
  (ISIT)}, 2017, pp. 2403--2407.

\bibitem{TandonGradientCoding2017}
\BIBentryALTinterwordspacing
R.~Tandon, Q.~Lei, A.~G. Dimakis, and N.~Karampatziakis, ``Gradient coding:
  Avoiding stragglers in distributed learning,'' in \emph{Proceedings of the
  34th International Conference on Machine Learning}, ser. Proceedings of
  Machine Learning Research, D.~Precup and Y.~W. Teh, Eds., vol.~70.\hskip 1em
  plus 0.5em minus 0.4em\relax PMLR, 06--11 Aug 2017, pp. 3368--3376. [Online].
  Available: \url{http://proceedings.mlr.press/v70/tandon17a.html}
\BIBentrySTDinterwordspacing

\bibitem{HalbawiImprovingDistGradient2018}
W.~Halbawi, N.~Azizan, F.~Salehi, and B.~Hassibi, ``Improving distributed
  gradient descent using reed-solomon codes,'' in \emph{2018 IEEE International
  Symposium on Information Theory (ISIT)}, 2018, pp. 2027--2031.

\bibitem{RavivGradientCyclic2018}
N.~Raviv, I.~Tamo, R.~Tandon, and A.~G. Dimakis, ``Gradient coding from cyclic
  mds codes and expander graphs,'' \emph{IEEE Transactions on Information
  Theory}, vol.~66, no.~12, pp. 7475--7489, 2020.

\bibitem{KarakusRedundancy2019}
\BIBentryALTinterwordspacing
C.~Karakus, Y.~Sun, S.~Diggavi, and W.~Yin, ``Redundancy techniques for
  straggler mitigation in distributed optimization and learning,''
  \emph{Journal of Machine Learning Research}, vol.~20, no.~72, pp. 1--47,
  2019. [Online]. Available: \url{http://jmlr.org/papers/v20/18-148.html}
\BIBentrySTDinterwordspacing

\bibitem{YuCodedFourier2017}
Q.~Yu, M.~A. Maddah-Ali, and A.~S. Avestimehr, ``Coded fourier transform,'' in
  \emph{2017 55th Annual Allerton Conference on Communication, Control, and
  Computing (Allerton)}, 2017, pp. 494--501.

\bibitem{YangComputeLinear2017}
Y.~Yang, P.~Grover, and S.~Kar, ``Computing linear transformations with
  unreliable components,'' \emph{IEEE Transactions on Information Theory},
  vol.~63, no.~6, pp. 3729--3756, 2017.

\bibitem{8382305}
H.~Yang and J.~Lee, ``Secure distributed computing with straggling servers
  using polynomial codes,'' \emph{IEEE Transactions on Information Forensics
  and Security}, vol.~14, no.~1, pp. 141--150, Jan 2019.

\bibitem{SecureCoded}
J.~Kakar, S.~Ebadifar, and A.~Sezgin, ``On the capacity and
  straggler-robustness of distributed secure matrix multiplication,''
  \emph{IEEE Access}, vol.~7, pp. 45\,783--45\,799, 2019.

\bibitem{GASP}
R.~G.~L. D’Oliveira, S.~El~Rouayheb, and D.~Karpuk, ``Gasp codes for secure
  distributed matrix multiplication,'' \emph{IEEE Transactions on Information
  Theory}, vol.~66, no.~7, pp. 4038--4050, 2020.

\bibitem{bitar_trans_PRAC}
R.~Bitar, Y.~Xing, Y.~Keshtkarjahromi, V.~Dasari, S.~El~Rouayheb, and
  H.~Seferoglu, ``Private and rateless adaptive coded matrix-vector
  multiplication,'' \emph{EURASIP Journal on Wireless Communications and
  Networking}, 2021.

\bibitem{BPR17}
R.~Bitar, P.~Parag, and S.~El~Rouayheb, ``Minimizing latency for secure
  distributed computing,'' in \emph{Information Theory (ISIT), 2017 IEEE
  International Symposium on}.\hskip 1em plus 0.5em minus 0.4em\relax IEEE,
  2017, pp. 2900--2904.

\bibitem{LCC1}
Q.~Yu, N.~Raviv, J.~So, and A.~S. Avestimehr, ``Lagrange coded computing:
  Optimal design for resiliency, security and privacy,'' \emph{arXiv preprint,
  arXiv:1806.00939}, 2018.

\bibitem{LCC2}
Q.~Yu, N.~Raviv, and A.~S. Avestimehr, ``Coding for private and secure
  multiparty computing,'' in \emph{2018 IEEE Information Theory Workshop
  (ITW)}, 2018, pp. 1--5.

\bibitem{PolynomCMPC}
H.~Akbari-Nodehi and M.~A. Maddah-Ali, ``Secure coded multi-party computation
  for massive matrix operations,'' \emph{IEEE Transactions on Information
  Theory}, vol.~67, no.~4, pp. 2379--2398, 2021.

\bibitem{Dutta2018AUC}
S.~Dutta, Z.~Bai, H.~Jeong, T.~M. Low, and P.~Grover, ``A unified coded deep
  neural network training strategy based on generalized polydot codes,''
  \emph{2018 IEEE International Symposium on Information Theory (ISIT)}, pp.
  1585--1589, 2018.

\end{thebibliography}

\section*{Appendix A: Proof of Theorem \ref{th:decodabilityofAGEcodes}} 
For AGE codes with $\{\alpha, \beta,\theta\}=\{1,s,ts+\lambda\}$, (\ref{eq:generalEntangled}) is reduced to:
\begin{align}\label{eq:AGECodes}
    C_A(x) = & \sum\limits_{i=0}^{t-1}\sum\limits_{j=0}^{s-1}A_{i,j}x^{j+is}, \nonumber \\
    C_B(x) = & \sum\limits_{k=0}^{s-1}\sum\limits_{l=0}^{t-1}B_{k,l}x^{(s-1-k)+(ts+\lambda)l},
\end{align}

To prove the decodability of AGE codes, we need to prove that the polynomial $C_Y(x)=C_A(x)C_B(x)=\sum\limits_{i=0}^{t-1}\sum\limits_{j=0}^{s-1} \sum\limits_{k=0}^{s-1}\sum\limits_{l=0}^{t-1} A_{i,j} B_{k,l}x^{j+is+(s-1-k)+(ts+\lambda)l}$ consists of $t$ distinct terms with coefficients $Y_{i,l} = \sum\limits_{j=0}^{s-1}A_{i,j}B_{j,l}, 0\leq i, l \leq t-1$. For this purpose, we define two sets of (i) $\mathbf{P}_1=\{s-1+is+(ts+\lambda)l, 0 \leq i,l \leq t-1\}$, representing the potential set of powers of the terms in $C_Y(x)$ with coefficients $Y_{i,l}$ (resulting from $j=k$), and (ii) $\mathbf{P}_2=\{j+is+(s-1-k)+(ts+\lambda)l, 0 \leq i,l \leq t-1, 0\leq k,j \leq s-1, j \neq k\}$, the set of powers of the remaining terms in $C_Y(x)$. Then, we prove that (i) $\mathbf{P}_1$ consists of $t^2$ distinct elements, and (ii) $\mathbf{P}_1$ and $\mathbf{P}_2$ do not have any overlap.

{\em{(i) Proving that $\mathbf{P}_1$ consists of $t^2$ distinct elements}}: From the definition of $\mathbf{P}_1$, it is equal to:
\begin{align}
    \mathbf{P}_1=\bigcup\limits_{l=0}^{t-1} \bigcup\limits_{i=0}^{t-1} \{s-1+is+(ts+\lambda)l\}.
\end{align}
For a given $l$, each subset of $\bigcup\limits_{i=0}^{t-1} \{s-1+is+(ts+\lambda)l\}$ consists of $t$ distinct elements. In addition, for two different values of $l=l_1$ and $l=l_2$ ($l_1 \neq l_2$), there is no overlap between $\bigcup\limits_{i=0}^{t-1} \{s-1+is+(ts+\lambda)l_1\}$ and $\bigcup\limits_{i=0}^{t-1} \{s-1+is+(ts+\lambda)l_2\}$. The reason is that for $0 \leq l_1 < l_2 \leq t-1$\footnote{Note that the assumption of $l_1<l_2$ does not result in loss of generality.}, the largest element of $\bigcup\limits_{i=0}^{t-1} \{s-1+is+(ts+\lambda)l_1\}$, \ie $ts-1+(ts+\lambda)l_1$ is less than the smallest element of $\bigcup\limits_{i=0}^{t-1} \{s-1+is+(ts+\lambda)l_2\}$, \ie $s-1+(ts+\lambda)l_2$:
\begin{align}
0 < s+\lambda 
\Rightarrow &ts-1 < s-1+ts+\lambda  \nonumber \\
\Rightarrow &ts-1+(ts+\lambda)l_1 < s-1+(ts+\lambda)(l_1+1) \nonumber \\
\Rightarrow &ts-1+(ts+\lambda)l_1 < s-1+(ts+\lambda)l_2.
\end{align}
Therefore, $\mathbf{P}_1$ consists of $t^2$ distinct elements.

{\em (ii) Proving that $\mathbf{P}_1$ and $\mathbf{P}_2$ have no overlap:} From the definition of $\mathbf{P}_1$ and $\mathbf{P}_2$, we have:
\begin{align}
    \mathbf{P}_1=&\bigcup\limits_{l_1=0}^{t-1} \bigcup\limits_{i_1=0}^{t-1} \mathbf{P}_1(l_1,i_1)\nonumber\\
    =&\bigcup\limits_{l_1=0}^{t-1} \bigcup\limits_{i_1=0}^{t-1} \{s-1+i_1s+(ts+\lambda)l_1\}
\end{align}
and
\begin{align}
    \mathbf{P}_2=&\bigcup\limits_{l_2=0}^{t-1} \bigcup\limits_{i_2=0}^{t-1}\mathbf{P}_2(l_2,i_2)\nonumber\\ =&\bigcup\limits_{l_2=0}^{t-1} \bigcup\limits_{i_2=0}^{t-1}\bigcup\limits_{\substack{j'=-(s-1)\\j'\neq 0}}^{s-1}\{j'+i_2s+s-1+(ts+\lambda)l_2\}.
\end{align}
To prove $\mathbf{P}_1 \cap \mathbf{P}_2 = \emptyset$, we consider the following five cases; (a) $l_1=l_2,i_1=i_2$, (b) $l_1=l_2,i_1<i_2$, (c) $l_1=l_2,i_1>i_2$, (d) $l_1>l_2$, (e) $l_1<l_2$. We prove that $\mathbf{P}_1(i_1,l_1) \cap \mathbf{P}_2(i_2,l_2)=\emptyset$ holds for each case.

(a) $l_1=l_2,i_1=i_2$: 
For this case, $\mathbf{P}_1(l_1,i_1)$ consists of the only element of $s-1+i_1s+(ts+\lambda)l_1$ which is not a member of $\mathbf{P}_2(l_2,i_2)$ as  $j'\neq0$. Therefore, $\mathbf{P}_1(i_1,l_1) \cap \mathbf{P}_2(i_2,l_2)=\emptyset$ for this case.

(b) $l_1=l_2,i_1<i_2$:
For this case, the smallest element of $\mathbf{P}_2(l_2,i_2)$ is always greater than $\mathbf{P}_1(l_1,i_1)=s-1+i_1s+(ts+\lambda)l_1$, as shown below:
\begin{align}
s-1+i_1s+(ts+&\lambda)l_1 < (i_1+1)s+(ts+\lambda)l_1\nonumber\\
&\leq i_2s+(ts+\lambda)l_2 \nonumber \\
&=-(s-1)+i_2s+(s-1)+(ts+\lambda)l_2
\end{align}
Therefore, $\mathbf{P}_1(i_1,l_1) \cap \mathbf{P}_2(i_2,l_2)=\emptyset$ holds for this case.

(c) $l_1=l_2,i_1>i_2$:
For this case, the largest element of $\mathbf{P}_2(l_2,i_2)$ is always less than $\mathbf{P}_1(l_1,i_1)=s-1+i_1s+(ts+\lambda)l_1$, as shown below:
\begin{align}
s-1+i_1s+(ts+\lambda)l_1 &> s-2+i_1s+(ts+\lambda)l_1\nonumber\\
&\geq s-2+(i_2+1)s+(ts+\lambda)l_2 \nonumber \\
&=2s-2+i_2s+(ts+\lambda)l_2
\end{align}
Therefore, $\mathbf{P}_1(i_1,l_1) \cap \mathbf{P}_2(i_2,l_2)=\emptyset$ holds for this case.

(d) $l_1>l_2$:
For this case, the smallest element of $\mathbf{P}_1(l_1,i_1)$, \ie $\mathbf{P}_1(l_1,0)$ is always greater than the largest element of $\mathbf{P}_2(l_2,i_2)$ \ie $\mathbf{P}_2(l_2, t-1), j'=s-1$, as shown below:
\begin{align}
s-1+(ts+\lambda)l_1 &\ge s-1+(ts+\lambda)(l_2+1)\nonumber\\
&= s-1+ts+\lambda+(ts+\lambda)l_2 \nonumber \\
&>s-1+(ts+\lambda)l_2
\end{align}
Therefore, $\mathbf{P}_1(i_1,l_1) \cap \mathbf{P}_2(i_2,l_2)=\emptyset$ holds for this case.

(e) $l_1<l_2$: 
For this case, the largest element of $\mathbf{P}_1(l_1,i_1)$, \ie $\mathbf{P}_1(l_1,t-1)$ is always less than the smallest element of $\mathbf{P}_2(l_2,i_2)$ \ie $\mathbf{P}_2(l_2, 0), j'=-(s-1)$, as shown below:
\begin{align}
(t-1)s+(ts+\lambda)l_1 &< (t-1)s+(ts+\lambda)(l_2-1)\nonumber\\
&= (t-1)s-ts-\lambda+(ts+\lambda)l_2 \nonumber \\
&=-s-\lambda+(ts+\lambda)l_2\nonumber\\
&< (ts+\lambda)l_2
\end{align}
Therefore, $\mathbf{P}_1(i_1,l_1) \cap \mathbf{P}_2(i_2,l_2)=\emptyset$ holds for this case.

This completes the proof of Theorem \ref{th:decodabilityofAGEcodes}.

\section*{Appendix B: Proof of Theorem \ref{th:FA-FB-AGE-thrm}} 
We first show that $\mathbf{P}(S_B(x))=\{ts+(ts+\lambda)(t-1),\ldots,ts+(ts+\lambda)(t-1)+z-1\}$ in (\ref{eq:S-B}) satisfies C1 in (\ref{eq:non_eq-AGE-conditions}). Then, we fix $\mathbf{P}(S_{B}(x))$ in C3 of (\ref{eq:non_eq-AGE-conditions}), and find $\mathbf{P}(S_A(x))$ that satisfies C2 and C3. Next, we explain these steps in details.

\emph{Showing that $\mathbf{P}(S_B(x))=\{ts+(ts+\lambda)(t-1),\ldots,ts+(ts+\lambda)(t-1)+z-1\}$ in (\ref{eq:S-B}) satisfies C1 in (\ref{eq:non_eq-AGE-conditions}).}
%
The largest element of the left side of C1 is equal to $(s-1)+(t-1)s+(ts+\lambda)(t-1)=ts+(ts+\lambda)(t-1)-1$ and the smallest element of the right side of C1 is equal to the smallest element of $\mathbf{P}(S_B(x))$, \ie $ts+(ts+\lambda)(t-1)$ plus the smallest element of $\mathbf{P}(C_A(x))$, \ie $0$. As $ts+(ts+\lambda)(t-1)-1$ is less than $ts+(ts+\lambda)(t-1)$, C1 is satisfied.

\emph{Fixing $\mathbf{P}(S_{B}(x))$ in C3 of (\ref{eq:non_eq-AGE-conditions}), and find $\mathbf{P}(S_A(x))$ that satisfies C2 and C3.}
C3 is satisfied for any choice of $\mathbf{P}(S_A(x))$ with non-negative elements. The reason is that the largest element of the left side of C3 is less than the smallest element of $\mathbf{P}(S_B(x))$. Next, we find $\mathbf{P}(S_A(x))$ with the smallest elements that satisfies C2, so (\ref{eq:non_eq-AGE-conditions}) is equal to
\begin{align}\label{eq:non_eq1-AGE_1}
    & (s-1)+si+(ts+\lambda)l \not\in \mathbf{P}(S_A(x)) \nonumber \\
    & +\{(s-1-k)+l'(ts+\lambda)\},
\end{align}
where $0\leq k \leq s-1,\; 0 \leq i,l,l' \leq t-1, 0 \leq \lambda \leq z$. The above equation is equivalent to:
\begin{align}\label{eq:non_eq1-AGE_2}
    \beta'+\theta l'' \not\in \mathbf{P}(S_A(x)),
\end{align}
for $l''=(l-l')$, $\theta = ts+\lambda$ and $\beta' = si+k$. The range of variable $\beta'$ is $\{si+k, 0\leq i \leq t-1, 0 \leq k \leq s-1\} = \bigcup\limits_{i=0}^{t-1} \{si,\ldots,si+s-1\} = \{0,\ldots,ts-1\}$. Therefore, we have
\begin{align}\label{eq:non_eq1-AGE_3}
    \mathbf{P}(S_A(x)) \not\in \bigcup\limits_{l=-(t-1)}^{t-1} \{\theta l,\ldots,\theta l+ts-1\},
\end{align}
Using the complement of the above intervals and the fact that the elements of $\mathbf{P}(S_A(x))$ is non-negative, we have
\begin{align}\label{eq:non_eq1-AGE_4}
    \mathbf{P}(S_A(x)) \in &\bigcup\limits_{l=0}^{t-2} \{ts+\theta l,\ldots,(l+1)\theta-1\}\nonumber\\
    &\cup \{ts+\theta (t-1), \ldots, +\infty\}, t>1
\end{align}
\begin{align}\label{eq:non_eq1-AGE_4t=1}
    \mathbf{P}(S_A(x)) \in 
    & \{s, \ldots, +\infty\}, t=1
\end{align}
Note that the required number of powers with non-zero coefficients for the secret term $S_A(x)$ is $z$, \ie
\begin{equation}
    |\mathbf{P}(S_A(x))| = z.
\end{equation}
Since our goal is to make the degree of polynomial $F_A(x)$ as small as possible, we choose the $z$ smallest powers from the sets in (\ref{eq:non_eq1-AGE_4}) to form $\mathbf{P}(S_A(x))$. 
Note that in (\ref{eq:non_eq1-AGE_4}), there are $t-1$ finite sets and one infinite set, where each finite set contains $\lambda=\theta-ts$ elements. Therefore, based on the value of $z$, we use the first interval and as many remaining intervals as required for $z > \lambda$, and the first interval only for $z = \lambda$ (Note that $0\leq \lambda \leq z$).
\begin{lemma}\label{lem:P(SA)-z large-AGE}
If $z > \lambda$ and $t \neq 1$, the set of all powers of polynomial $S_A(x)$ with non-zero coefficients is defined as 
\begin{align}\label{eq:finite_P(SA)_set_representation-z large-AGE}
    \mathbf{P}(S_A(x)) = &\Big(\bigcup\limits_{l=0}^{q-1} \{ts+\theta l,\ldots,(l+1)\theta-1\}\Big) \nonumber\\
    & \cup \{ts+q\theta,\ldots,ts+q\theta+z-1-q(\theta-ts)\}\\
    =&\{ts+\theta l+w, l\in\Omega_0^{q-1}, w\in \Omega_0^{\lambda-1}\} \nonumber \\
    &\cup \{ts+\theta q+u, u\in\Omega_0^{z-1-q\lambda}\}\label{eq:psa12-AGE}.
\end{align}
\end{lemma}
{\em Proof:}
For the case of $z > \lambda$, the number of elements in the first interval of (\ref{eq:non_eq1-AGE_4}), which is equal to $\lambda$, is not sufficient for selecting $z$ powers. Therefore, more than one interval is used. We show the number of selected intervals with $q+1$, where $q \ge 1$ is defined as $q=\min\{\floor{\frac{z-1}{\lambda}},t-1\}$. With this definition, the first $q$ intervals of (\ref{eq:non_eq1-AGE_4}) are selected in full. In other words, in total, we select $q\lambda$ elements to form the first $q$ intervals in (\ref{eq:finite_P(SA)_set_representation-z large-AGE}). The remaining $z-q\lambda$ elements are selected from the $(q+1)^\text{st}$ interval of (\ref{eq:non_eq1-AGE_4}) to form the last interval of   (\ref{eq:finite_P(SA)_set_representation-z large-AGE}). We can derive (\ref{eq:psa12-AGE}) from  (\ref{eq:finite_P(SA)_set_representation-z large-AGE}) by replacing $\theta$ with its equivalent value, $ts+\lambda$.
\hfill $\Box$

\begin{lemma}\label{lem:P(SA)-z small-AGE}
If $z = \lambda$ and $t \ne 1$, the set of all powers of polynomial $S_A(x)$ with non-zero coefficients is defined as the following:
\begin{align}\label{eq:finiteP(SA)-second-scenario-AGE}
\mathbf{P}(S_A(x)) = \{ts,\dots,ts+z-1\}, \nonumber\\
= \{ts+u, u\in \Omega_0^{z-1}\}.
\end{align}
\end{lemma}
{\em Proof:}
In this scenario since $z = \lambda$, the first interval of (\ref{eq:non_eq1-AGE_4}) is sufficient to select all $z$ elements of $\mathbf{P}(S_A(x))$. Therefore, $z$ elements are selected from the first interval of (\ref{eq:non_eq1-AGE_4}), as shown in (\ref{eq:finiteP(SA)-second-scenario-AGE}).\hfill $\Box$

\begin{lemma}\label{lem:P(SA)-t=1}
If $t=1$ , the set of all powers of polynomial $S_A(x)$ with non-zero coefficients is defined as the following:
\begin{align}\label{eq:finiteP(SA)-third-scenario-AGE}
\mathbf{P}(S_A(x)) = \{s,\dots,s+z-1\}, \nonumber\\
= \{s+u, u\in \Omega_0^{z-1}\}.
\end{align}
\end{lemma}
{\em Proof:}
In this scenario, z smallest elements are selected from (\ref{eq:non_eq1-AGE_4t=1}) 
as shown in (\ref{eq:finiteP(SA)-third-scenario-AGE}).\hfill $\Box$

This completes the proof of Theorem (\ref{th:FA-FB-AGE-thrm}).

\section*{Appendix C: Proof of Theorem \ref{th:N_AGE}}
To prove this theorem, we first consider the case that $t=1$. Then, we consider that case thats $t\neq 1$.
\begin{lemma}\label{lemma:t=1NAGECMPC} $N_{\text{AGE-CMPC}}=2s+2z-1$ when $t=1$. s
\end{lemma}
{\em Proof}: $F_A(x)$ and $F_B(x)$ are expressed as in the following for $t=1$ using 
(\ref{eq:AGE-p(CA)-th}), (\ref{eq:AGE-p(CB)-th}), (\ref{eq:S-A}) and (\ref{eq:S-B}). 
\begin{align}\label{eq:FA AGE-CMPC-t=1}
    F_{A}(x) = & \sum_{j=0}^{s-1} A_{j}x^{j}
    + \sum_{u=0}^{z-1}\bar{A}_{u}x^{s+u},
\end{align}
\begin{align}\label{eq:FB AGE-CMPC-t=1}
    F_{B}(x) = & \sum_{k=0}^{s-1} B_{k}x^{s-1-k}
    + \sum_{r=0}^{z-1}\bar{B}_{r}x^{s+r}. 
\end{align}
$F_A(x)$ and $F_B(x)$ are equal to the secret shares of Entangled-CMPC \cite{8613446}, for $t=1$. Thus, in this case, AGE-CMPC and Entangled-CMPC are equivalent, so we have $N_{\text{AGE-CMPC}}=N_{\text{Entangled-CMPC}}=2s+2z-1$ \cite{8613446}. 
This completes the proof. \hfill $\Box$
%
%
%

Now, we consider $t\neq 1$. The required number of workers is equal to the number of terms in $H(x)=F_A(x)F_B(x)$ with non-zero coefficients. The set of all powers of polynomial $H(x)$ with non-zero coefficients, shown by $\mathbf{P}({H}(x))$, is expressed as
 \begin{align}\label{eq:PHx-AGE}
 \mathbf{P}({H}(x)) = \mathbf{D}_1 \cup  \mathbf{D}_2\cup \mathbf{D}_3 \cup \mathbf{D}_4,
 \end{align}
 where
  \begin{align}
     & \mathbf{D}_1 = \mathbf{P}(C_A(x))+\mathbf{P}(C_B(x))
 \end{align}
\begin{align}
    & \mathbf{D}_2  =\mathbf{P}(C_A(x))+\mathbf{P}(S_B(x))
 \end{align}
   \begin{align}\label{eq:d3definition-AGE}
    & \mathbf{D}_3=\mathbf{P}(S_A(x))+\mathbf{P}(C_B(x))
 \end{align}
   \begin{align}\label{eq:d4definition-AGE}
    & \mathbf{D}_4=\mathbf{P}(S_A(x))+\mathbf{P}(S_B(x))
 \end{align}
 
Using (\ref{eq:AGE-p(CA)-th}) and (\ref{eq:AGE-p(CB)-th}), $\mathbf{D}_1$ is calculated as:
 \begin{align}
      \mathbf{D}_1 = & \mathbf{P}(C_{A}(x))+\mathbf{P}(C_B(x)) \nonumber \\
     = &  \{j+si
     : 0 \leq i \leq t-1,\; 0 \leq j \leq s-1,\} \nonumber \\ 
     &+ \{s-1-k+\theta l 
     : 0 \leq l \leq t-1,\; 0 \leq k \leq s-1\} \nonumber \\
     = &  \{j+si+s-1-k+\theta l:0 \leq i,l \leq t-1,\; \nonumber \\
     & 0 \leq j,k \leq s-1,\}\nonumber \\
     = & \bigcup\limits_{i=0}^{t-1}\{is,\ldots,(i+2)s-2\} + \{\theta l : 0\leq l\leq t-1\} \nonumber\\
     = & \{0,\ldots,ts+s-2\} + \{\theta l : 0\leq l\leq t-1\} \label{eq:d1AGE1}\\
     = & \bigcup\limits_{l=0}^{t-1}\{\theta l,\ldots,ts+s-2+\theta l\}, \label{eq:d1-AGE}
      \end{align}
where (\ref{eq:d1AGE1}) comes from the fact that the largest element of each $i^{\text{th}}$ subset of $\bigcup\limits_{i=0}^{t-1}\{is,\ldots,(i+2)s-2\}$ plus one, \ie $(i+2)s-2+1$ is greater than or equal to the smallest element of the $(i+1)^{\text{st}}$ subset, \ie $(i+1)s$ as $s\ge 1$. 
Using (\ref{eq:AGE-p(CA)-th}) and (\ref{eq:S-B}), $\mathbf{D}_2$ is calculated as:
 \begin{align}\label{eq:d2-AGE}
      \mathbf{D}_2 = & \mathbf{P}(C_{A}(x))+\mathbf{P}(S_B(x)) \nonumber \\
     = &  \{j+si
     : 0 \leq i \leq t-1,\; 0 \leq j \leq s-1,\} \nonumber \\ 
     &+ \{ts+\theta(t-1)+r: 
      0 \leq r \leq z-1\} \nonumber \\
     = &  \{j+si+ts+\theta(t-1)+r:0 \leq i \leq t-1,\; \nonumber \\
     & 0 \leq j\leq s-1,\; 0 \leq r \leq z-1\}\nonumber \\
     = & \bigcup\limits_{i=0}^{t-1} \{is, \ldots, (i+1)s+z-2\} +ts+\theta(t-1)\nonumber \\
     = & \{ts+\theta(t-1),\ldots,2ts+\theta(t-1)+z-2\},
      \end{align}       
where the last equality comes from the fact that there is no gap between the subsets of $\bigcup\limits_{i=0}^{t-1} \{is, \ldots, (i+1)s+z-2\}$. The reason is that the largest element of the $i^{\text{th}}$ subset, \ie $(i+1)s+z-2$ plus one is larger than or equal to the smallest element of the $(i+1)^{\text{st}}$ subset, \ie $(i+1)s$ as $z\ge 1$.

In the following, we consider different regions for the values of $z$ and $\lambda$ and calculate $|\mathbf{P}({H}(x))|$ through calculation of $\mathbf{D}_3$ and $\mathbf{D}_4$. In addition, we use the following lemma, whichhelps us to calculate $\mathbf{P}({H}(x))$ without requiring to consider all of the terms of $\mathbf{D}_3$  in some cases.
\begin{lemma}\label{lemma:UpperBoundPHx-AGE} The following inequality holds. 
\begin{align}\label{eq:UpperBoundPHx-AGE}
    |\mathbf{P}({H}(x))|\leq& \deg(S_A(x))+\deg(S_B(x))+1\nonumber\\
    &=\max\{\mathbf{D}_4\}+1
\end{align}
\end{lemma}
{\em Proof:} 
$|\mathbf{P}({H}(x))|$, which is equal to the number of terms in $H(x)$ with non-zero coefficients, is less than or equal to the number of all terms, which is equal to $\deg(H(x))+1$. Thus, 
\begin{align}\label{eq:phxUpperBound-AGE}
    |\mathbf{P}({H}(x))|\leq &\deg(H(x))+1 \nonumber \\ =&\deg((C_A(x)+S_A(x))(C_B(x)+S_B(x)))+1 \nonumber \\
    =&\max\{\deg(C_A(x)),\deg(S_A(x)) \} \nonumber \\
    &+\max\{\deg(S_B(x)), \deg(C_B(x))\}+1.
\end{align}
From (\ref{eq:AGE-p(CA)-th}), $\deg(C_A(x))=ts-1$. On the other hand, from (\ref{eq:S-A}), $\deg(S_A(x))\ge ts$. Therefore, $\max\{\deg(C_A(x)),\deg(S_A(x))\} = \deg(S_A(x))$. Moreover, From (\ref{eq:AGE-p(CB)-th}), $\deg(C_B(x))=s-1+\theta(t-1)$, and from (\ref{eq:S-B}), $\deg(S_B(x))\ge ts+\theta(t-1)$. Therefore, $\max\{\deg(C_B(x)),\deg(S_B(x))\} = \deg(S_B(x))$,
which results in the first inequality of (\ref{eq:UpperBoundPHx-AGE}).

On the other hand, from (\ref{eq:d4definition-AGE}), $\max\{\mathbf{D}_4\}=\max\{\mathbf{P}(S_A(x))\}+\max\{\mathbf{P}(S_B(x))\}=\deg(S_A(x))+\deg(S_B(x))$.

This completes the proof. \hfill $\Box$

\begin{lemma}\label{lemma:non-zero-coeff-AGE-lambda=0 and z greater ts-t}
For $z > ts-s, t\neq 1$ and $\lambda=0$, we have
\begin{align}\label{eq:non-zero-coef-AGE-lambda=0 z greater ts-s}
    |\mathbf{P}({H}(x))|= \Upsilon_1(0)= 2st^2+2z-1
\end{align}
\end{lemma}
{\em Proof:} By replacing $\lambda$ with $0$ in AGE-CMPC formulations, the scheme is equivalent to Entangled-CMPC in \cite{8613446}. Therefor, the proof of this lemma can be derived directly from the proof of Theorem 1 in \cite{8613446}.\hfill $\Box$

\begin{lemma}\label{lemma:non-zero-coeff-AGE-lambda=0 and z less ts-t}
For $z \leq ts-s, t\neq 1$ and $\lambda=0$, we have
\begin{align}
    |\mathbf{P}({H}(x))|&=\Upsilon_2(0)= st^2+3st-2s+t(z-1)+1
\end{align}
\end{lemma}
{\em Proof:} For this case, AGE-CMPC is equivalent to Entangled-CMPC. Therefore, the proof of this lemma can be derived directly from the proof of Theorem 1 in \cite{8613446}.\hfill $\Box$

\begin{lemma}\label{lemma:non-zero-coeff-AGE-lambda=z}
 For $\lambda=z, t\neq 1$, we have
\begin{align}
    |\mathbf{P}({H}(x))|&=\Upsilon_3(z)= 2ts+\theta(t-1)+2z-1\nonumber\\
    &=(ts+z)(1+t)-1
\end{align}
\end{lemma}
 
 {\em Proof:} 
 To prove this lemma, we first calculate $\mathbf{D}_3$ from (\ref{eq:AGE-p(CB)-th}) and (\ref{eq:S-A}):
 \begin{align}\label{eq:d3-AGE- lambda=z}
      \mathbf{D}_3 = & \mathbf{P}(S_{A}(x))+\mathbf{P}(C_B(x)) \nonumber \\
      = & \{ts+u: 0 \leq u \leq z-1\} \nonumber \\
      + & \{s-1-k+\theta l:0 \leq l \leq t-1,\; 0 \leq k \leq s-1,\}\nonumber \\
     = & \{ts,\dots,ts+z+s-2\}+\{\theta l: 0 \leq l \leq t-1\} \nonumber \\
     = & \bigcup_{l=0}^{t-1} \{\theta l+ts,\dots,\theta l +ts+z+s-2\}
     .
\end{align}
From (\ref{eq:d1-AGE}) and (\ref{eq:d3-AGE- lambda=z}), we can calculate $\mathbf{D}_{13}=\mathbf{D}_1 \cup \mathbf{D}_3$ as:
\begin{align}
    \mathbf{D}_{13}=&\mathbf{D}_1 \cup \mathbf{D}_3\nonumber\\
    =&\bigcup\limits_{l=0}^{t-1}\{\theta l,\ldots,ts+s-2+\theta l\}\nonumber\\
    &\cup\bigcup_{l=0}^{t-1} \{\theta l+ts,\dots,\theta l +ts+z+s-2\}\nonumber\\
    =&\bigcup_{l=0}^{t-1} \{\theta l,\ldots,\theta l +ts+z+s-2\}\label{eq:d13AGE11}\\
    =&\{0,\ldots,\theta (t-1) +ts+z+s-2\}\label{eq:d13AGE12},
\end{align}
where (\ref{eq:d13AGE11}) comes from the fact that $\theta l < \theta l+ts \leq ts+s-2+\theta l < \theta l +ts+z+s-2$ and (\ref{eq:d13AGE12}) comes from the fact that there is no gap between each two consecutive subsets of $\bigcup_{l=0}^{t-1} \{\theta l,\ldots,\theta l +ts+z+s-2\}$ as $\theta l+ts+z+s-2+1 = \theta l+\theta+s-1 \ge \theta (l+1)$. 
Next, we calculate $\mathbf{D}_{123}=\mathbf{D}_{13} \cup \mathbf{D}_{2}$ from (\ref{eq:d13AGE12}) and (\ref{eq:d2-AGE})
\begin{align}\label{eq:d123AGE1}
    \mathbf{D}_{123}=&\mathbf{D}_{1} \cup \mathbf{D}_{3} \cup \mathbf{D}_{2}\nonumber\\
    =&\{0,\ldots,\theta (t-1) +ts+z+s-2\} \nonumber\\
    &\cup\{ts+\theta(t-1),\ldots,2ts+\theta(t-1)+z-2\}\nonumber\\
    =&\{0,\ldots,2ts+\theta(t-1) +z-2\},
\end{align}
where the last equality 
comes from the fact that $0 < ts+\theta(t-1) \leq \theta(t-1)+ts+z+s-2 < 2ts+\theta(t-1)+z-2$. Next, we first calculate $\mathbf{D}_4$, and then its union with $\mathbf{D}_{123}$.
From (\ref{eq:S-A}) and (\ref{eq:S-B}), we have
 \begin{align}\label{eq:d4-AGE- lambda=z}
      \mathbf{D}_4 = & \mathbf{P}(S_{A}(x))+\mathbf{P}(S_B(x)) \nonumber \\
      = & \{ts,\dots,ts+z-1\} \nonumber \\
      + & \{ts+\theta(t-1),\dots,ts+\theta(t-1)+z-1\} \nonumber \\
      = & \{2ts+\theta(t-1),\dots,2ts+\theta(t-1)+2z-2\}.
\end{align}
From (\ref{eq:PHx-AGE}), (\ref{eq:d123AGE1}) and (\ref{eq:d4-AGE- lambda=z}), we have
\begin{align}
    \mathbf{P}({H}(x)) = & \mathbf{D}_{123} \cup \mathbf{D}_4 \nonumber\\
    &\{0,\ldots,2ts+\theta(t-1) +z-2\} \cup\nonumber\\ &\{2ts+\theta(t-1),\dots,2ts+\theta(t-1)+2z-2\}\nonumber\\
    =&\{0,\ldots,2ts+\theta(t-1)+2z-2\}.
\end{align}
Therefore, $|\mathbf{P}({H}(x))|=2ts+\theta(t-1)+2z-2+1$. 
This completes the proof. \hfill $\Box$ 

For the remaining regions of the values of $z$ and $\lambda$, where $\lambda < z$, we use the following lemma to calculate $\mathbf{P}({H}(x))$. 

\begin{lemma}
For $\lambda < z$, we have
\begin{align}\label{eq:d1cupd2cupd3AGE4}
    \mathbf{D}_1 \cup  \mathbf{D}_2\cup \mathbf{D}_3 = \mathbf{\widehat{D}}_{123'} \cup \mathbf{\widetilde{D}}'_3 \cup \mathbf{\widetilde{D}}''_3,
\end{align}
\end{lemma}
where $\mathbf{\widehat{D}}_{123'} =\{0,\dots,2ts+\theta(t-1)+z-2\}$, $\mathbf{\widetilde{D}}'_3=\bigcup_{l'=t-1}^{t+q-2} \{ts+\theta l',\dots,\theta(l'+1)+s-2\}$ and $\mathbf{\widetilde{D}}''_3=\{ts+(q+t-1)\theta,\dots,(q+1)ts+(t-1)\theta+s+z-2\}$.

{\em Proof:}
To prove this lemma, we first calculate and decompose $\mathbf{\widehat{D}}_{3}$ using (\ref{eq:S-A}) and (\ref{eq:AGE-p(CB)-th}):
\begin{align}
      \mathbf{D}_3 = & \mathbf{P}(S_{A}(x))+\mathbf{P}(C_B(x)) \nonumber \\
      = &(\{ts+\theta l+w, 0 \leq l \leq q-1, 0\leq w \leq \lambda-1\}\cup\nonumber \\
      &\{ts+\theta q+u, 0\leq u \leq z-1-q\lambda\})\nonumber\\
      +&\{s-1-k+l(ts+\lambda), 0 \leq k \leq s-1, 0 \leq l \leq t-1\}\nonumber\\
      =&\{ts+\theta l'+w', 0 \leq l' \leq t+q-2, \nonumber\\
      & \quad \quad \quad \quad \quad \quad 0 \leq w' \leq \lambda+s-2\}\nonumber\\
      &\cup \{ts+\theta l''+u', q \leq l'' \leq t+q-1, \nonumber\\
      & \quad \quad \quad \quad \quad \quad \quad 0 \leq u' \leq s+z-q\lambda-2\}\nonumber\\
      =&\{ts+\theta l'+w', 0 \leq l' \leq t+q-2, \nonumber\\
      & \quad \quad \quad \quad \quad \quad 0 \leq w' \leq \lambda+s-2\}\nonumber\\
      &\cup \{ts+\theta l''+u', q \leq l'' \leq t+q-2, \nonumber\\
      & \quad \quad \quad \quad \quad \quad \quad 0 \leq u' \leq s+z-q\lambda-2\}\nonumber\\
      &\cup \{ts+\theta l''+u', l'' = t+q-1, \nonumber\\
      & \quad \quad \quad \quad \quad \quad \quad 0 \leq u' \leq s+z-q\lambda-2\}\nonumber\\
      =&\{ts+\theta l'+w', 0 \leq l' \leq t+q-2, \nonumber\\
      & \quad \quad \quad \quad \quad \quad 0 \leq w' \leq \lambda+s-2\} \label{d3''hatSubsetd3'}\\
      &\cup \{ts+\theta (t+q-1)+u', 0 \leq u' \leq s+z-q\lambda-2\}\nonumber\\
      =&\{ts+\theta l'+w', 0 \leq l' \leq t-2, \nonumber\\
      & \quad \quad \quad \quad \quad \quad 0 \leq w' \leq \lambda+s-2\} \nonumber\\
      &\cup\{ts+\theta l'+w', t-1 \leq l' \leq t+q-2, \nonumber\\
      & \quad \quad \quad \quad \quad \quad 0 \leq w' \leq \lambda+s-2\} \nonumber\\
      &\cup \{ts+\theta (t+q-1)+u', 0 \leq u' \leq s+z-q\lambda-2\}\nonumber\\
      =&\mathbf{\widehat{D}}'_3\cup\mathbf{\widetilde{D}}'_3 \cup \mathbf{\widetilde{D}}''_3, \label{eq:d3-AGE-z grater than ts-1}
\end{align}
where
\begin{align}\label{eq:d'3hatAGE}
    \mathbf{\widehat{D}}'_3=\bigcup_{l'=0}^{t-2}\{ts+\theta l',\ldots,ts+\theta l'+\lambda+s-2\},
\end{align}
\begin{align}\label{eq:d'3tildeAGE}
    \mathbf{\widetilde{D}}'_3=\bigcup_{l'=t-1}^{t+q-2}\{ts+\theta l',\ldots,ts+\theta l'+\lambda+s-2\},
\end{align}
\begin{align}\label{eq:d''3tildeAGE}
    &\mathbf{\widetilde{D}}''_3=\nonumber\\
    &\{ts+\theta(t+q-1),\ldots,ts+\theta(t+q-1)+s+z-q\lambda-2\}\nonumber\\
    &=\nonumber\\
    &\{ts+\theta(t+q-1),\ldots,(q+1)ts+(t-1)\theta+s+z-2\},
\end{align}
and (\ref{d3''hatSubsetd3'}) comes from the fact that
\begin{align}
    z > \lambda &\Rightarrow z-1 \ge \lambda \nonumber\\
    & \Rightarrow q=\min\{\floor{\frac{z-1}{\lambda}},t-1\}=\floor{\frac{z-1}{\lambda}}\nonumber\\
    & \quad \quad  \& \quad q+1 > \frac{z-1}{\lambda} \nonumber\\
    & \Rightarrow \lambda+s-2 > s+z-q\lambda -3\nonumber \\
    & \Rightarrow \lambda+s-2 \ge s+z-q\lambda -2\nonumber
        \end{align}
    \begin{align}
    & \Rightarrow \{ts+\theta l''+u', q \leq l'' \leq t+q-2, \nonumber\\
      & \quad \quad \quad \quad \quad \quad \quad 0 \leq u' \leq s+z-q\lambda-2\}\nonumber\\
      &\subset \{ts+\theta l'+w', 0 \leq l' \leq t+q-2, \nonumber\\
      & \quad \quad \quad \quad \quad \quad 0 \leq w' \leq \lambda+s-2\}\nonumber\\
      .
\end{align}
Next, we calculate $\mathbf{\widehat{D}}_{123'} = \mathbf{D}_1 \cup \mathbf{\widehat{D}}'_3 \cup \mathbf{D}_2$ using (\ref{eq:d1-AGE}), (\ref{eq:d'3hatAGE}), and (\ref{eq:d2-AGE}):
\begin{align}
    \mathbf{\widehat{D}}_{123'} =& \mathbf{D}_1 \cup \mathbf{\widehat{D}}'_3 \cup \mathbf{D}_2\nonumber\\
    =&\bigcup\limits_{l=0}^{t-1}\{\theta l,\ldots,ts+s-2+\theta l\}\cup\nonumber\\
    &\bigcup_{l'=0}^{t-2}\{ts+\theta l',\ldots,ts+\theta l'+\lambda+s-2\}\cup\nonumber\\
    &\{ts+\theta(t-1),\ldots,2ts+\theta(t-1)+z-2\}\nonumber\\
    =&\bigcup\limits_{l=0}^{t-2}\{\theta l,\ldots,ts+s-2+\theta l\}\cup\nonumber\\
    &\{\theta (t-1),\ldots,ts+s-2+\theta (t-1)\}\cup\nonumber\\
    &\bigcup_{l'=0}^{t-2}\{ts+\theta l',\ldots,ts+\theta l'+\lambda+s-2\}\cup\nonumber\\
    &\{ts+\theta(t-1),\ldots,2ts+\theta(t-1)+z-2\}\nonumber\\
    =&\bigcup\limits_{l=0}^{t-2}\{\theta l,\ldots,ts+s-2+\theta l+\lambda\}\cup\nonumber\\
    &\{\theta (t-1),\ldots,ts+s-2+\theta (t-1)\}\cup\nonumber\\
    &\{ts+\theta(t-1),\ldots,2ts+\theta(t-1)+z-2\}\label{eq:dhat123'AGE1}\\
    =&\bigcup\limits_{l=0}^{t-2}\{\theta l,\ldots,ts+s-2+\theta l+\lambda\}\cup\nonumber\\
    &\{\theta (t-1),\ldots,2ts+\theta(t-1)+z-2\}\nonumber\\
    =&\{0,\ldots,ts+s-2+\theta(t-2)+\lambda\}\cup\nonumber\\
    &\{\theta (t-1),\ldots,2ts+\theta(t-1)+z-2\}\label{eq:dhat123'AGE2}\\
    =&\{0,\ldots,2ts+\theta(t-1)+z-2\}\label{eq:dhat123'AGEfinal},
\end{align}
where (\ref{eq:dhat123'AGE1}) comes from the fact that $s<ts+z$. Thus, $ts+s-2+\theta (t-1)< 2ts+\theta(t-1)+z-2$. We obtain (\ref{eq:dhat123'AGE2}) from the fact that $ts+s-2+\theta l+\lambda+1=s-1+\theta(l+1)\ge \theta(l+1)$ and the last equality comes from the fact that:
\begin{align}
    &\Rightarrow 0 \leq s-1 <2ts+z-2 \nonumber\\
    & \Rightarrow \theta(t-1) \leq s-1+\theta(t-1) < 2ts+z-2+\theta(t-1) \nonumber
    \end{align}
    \begin{align}
    & \Rightarrow \theta(t-1) \leq ts+s-1+\theta(t-2)+\lambda \nonumber\\
    &\quad \quad \quad \quad \quad \quad \quad \quad \quad \quad \quad< 2ts+\theta(t-1)+z-2.
\end{align}
We can derive (\ref{eq:d1cupd2cupd3AGE4}) from (\ref{eq:d3-AGE-z grater than ts-1}), (\ref{eq:d'3tildeAGE}), (\ref{eq:d''3tildeAGE}), and (\ref{eq:dhat123'AGEfinal}). 
This completes the proof. \hfill $\Box$ 

From (\ref{eq:S-A}) and (\ref{eq:S-B}), $\mathbf{D}_4$ for $z>\lambda$ is calculated as
 \begin{align}\label{eq:d4-AGE}
   \mathbf{D}_4 =& \mathbf{P}(S_{A}(x))+\mathbf{P}(S_B(x)) \nonumber \\
   = & \bigcup_{l=0}^{q-1} \{2ts+\theta(l+t-1),\dots,\nonumber\\
   &\quad \quad \quad \quad 2ts+\theta(l+t-1)+z-1+\lambda-1\} \nonumber \\
    &\cup  \{2ts+\theta(q+t-1),\dots, \nonumber \\
    & \quad \quad \quad 2ts+\theta(q+t-1)+2z-2-q(\theta-ts)\} \nonumber \\
   = & \bigcup_{l=0}^{q-1} \{2ts+\theta(l+t-1),\dots,ts+\theta(l+t)+z-2\} \nonumber \\
   &\cup \{2ts+\theta(q+t-1),\dots,\nonumber\\
   & \quad \quad \quad (q+2)ts+\theta(t-1)+2z-2\},
 \end{align}
where for $z>ts$. The above equation is a continuous set as there exist no gaps between each of its two consecutive subsets
. The reason is that, for $z>ts$, the greatest element of each subset plus one, \ie $ts+\theta(l+t)+z-1$, is greater than or equal to the smallest element of it's consecutive subset, \ie $2ts+\theta(l+t)$ for $l=\{0,\dots,q-1\}$. This is shown as
 \begin{align}
     & z > ts \Rightarrow ts+z-1 \geq 2ts \nonumber \\
     &  \Rightarrow ts+z-1+\theta(l+t) \geq 2ts+\theta(l+t).
 \end{align}
Therefore, for $z>\max\{ts,\lambda\}$, $\mathbf{D}_4$ is equal to:
\begin{align}\label{eq:d4-z greater ts-AGE}
    \mathbf{D}_4 = & 
    \{2ts+\theta(t-1),\dots,(q+2)ts+\theta(t-1)+2z-2\}.
\end{align}

\begin{lemma}\label{lemma:non-zero-coeff-AGE-z greater than ts}
For $z > ts, t\neq 1$ and $0 < \lambda < z$:
\begin{equation}
    |\mathbf{P}({H}(x))|= \Upsilon_4(\lambda)= (q+2)ts+\theta(t-1)+2z-1
\end{equation}
\end{lemma}
{\em Proof:} 
To prove this lemma, we calculate $\mathbf{P}({H}(x)) = \mathbf{D}_1 \cup  \mathbf{D}_2\cup \mathbf{D}_3 \cup \mathbf{D}_4$ using (\ref{eq:d1cupd2cupd3AGE4}) and (\ref{eq:d4-z greater ts-AGE}):
\begin{align}
    &\mathbf{P}({H}(x)) = \mathbf{D}_1 \cup  \mathbf{D}_2\cup \mathbf{D}_3 \cup \mathbf{D}_4\nonumber\\
    =&\mathbf{\widehat{D}}_{123'} \cup \mathbf{\widetilde{D}}'_3 \cup \mathbf{\widetilde{D}}''_3\cup\mathbf{D}_4\nonumber\\
    =&\{0,\dots,2ts+\theta(t-1)+z-2\}\nonumber\\
    &\cup\{2ts+\theta(t-1),\dots,(q+2)ts+\theta(t-1)+2z-2\}\nonumber\\
    &\cup \mathbf{\widetilde{D}}'_3 \cup \mathbf{\widetilde{D}}''_3\nonumber\\
    =&\{0,\dots,(q+2)ts+\theta(t-1)+2z-2\}\nonumber\\
    &\cup \mathbf{\widetilde{D}}'_3 \cup \mathbf{\widetilde{D}}''_3
\end{align}
From the above equation, $|\mathbf{P}({H}(x))| \ge (q+2)ts+\theta(t-1)+2z-1$.
On the other hand, from (\ref{eq:UpperBoundPHx-AGE}), $|\mathbf{P}({H}(x))| \leq \max\{\mathbf{D}_4\}+1=(q+2)ts+\theta(t-1)+2z-2+1$. Therefore, $|\mathbf{P}({H}(x))| = (q+2)ts+\theta(t-1)+2z-1$.  
This completes the proof.
\hfill $\Box$

\begin{lemma}\label{lemma:non-zero-coeff-AGE-z less than ts: 1}
For $z\leq ts<\lambda+s-1, t\neq 1$ and $0< \lambda<z$, we have
\begin{equation}
    |\mathbf{P}({H}(x))|= \Upsilon_5(\lambda)= 3ts+\theta(t-1)+2z-1
\end{equation}
\end{lemma}
{\em Proof:} 
For the conditions of this lemma, \ie $ ts-s+2 \leq \lambda \leq z-1$ and $z-1 \leq ts-1$, the range of variation of $\frac{z-1}{\lambda}$ and thus the value of $q$ is calculated as follows:
\begin{align}
     &1 \leq \frac{z-1}{\lambda} \leq \frac{ts-1}{ts-s+2}\nonumber \\
      \Rightarrow &1 \leq \frac{z-1}{\lambda} \leq \frac{ts-1}{ts-s+2} < 2 \label{eq:qforlemma40} \\
     \Rightarrow &q = \floor{\frac{z-1}{\lambda}} = 1,
\end{align}
where (\ref{eq:qforlemma40}) comes from the fact that $s(t-2)\ge 0$ and thus $s(t-2)+5>0$. By replacing $q=1$ in (\ref{eq:d1cupd2cupd3AGE4}) and (\ref{eq:d4-AGE}), we calculate $\mathbf{P}({H}(x)) = \mathbf{D}_1 \cup  \mathbf{D}_2\cup \mathbf{D}_3 \cup \mathbf{D}_4$ as
\begin{align}
    \mathbf{P}({H}(x)) =& \mathbf{D}_1 \cup  \mathbf{D}_2\cup \mathbf{D}_3 \cup \mathbf{D}_4\nonumber\\
    =&\mathbf{\widehat{D}}_{123'} \cup \mathbf{\widetilde{D}}'_3 \cup \mathbf{\widetilde{D}}''_3\cup\mathbf{D}_4\nonumber\\
    =&\{0,\dots,2ts+\theta(t-1)+z-2\}\nonumber\\
    &\cup \{ts+\theta(t-1),\dots,\theta t+s-2\} \nonumber\\
    &\cup \{ts+t\theta,\dots,2ts+(t-1)\theta+s+z-2\}\nonumber\\
    &\cup \{2ts+\theta(t-1),\dots,ts+\theta t+z-2\} \nonumber \\
   &\cup \{2ts+\theta t,\dots,3ts+\theta(t-1)+2z-2\}\nonumber\\
   =&\{0,\dots,2ts+\theta(t-1)+z-2\}\nonumber\\
   &\cup \{ts+t\theta,\dots,2ts+(t-1)\theta+s+z-2\}\nonumber\\
    &\cup \{2ts+\theta(t-1),\dots,ts+\theta t+z-2\} \nonumber \\
   &\cup \{2ts+\theta t,\dots,3ts+\theta(t-1)+2z-2\}\label{eq:d1cupd2cupd3cup4AGE39a}\\
   =&\{0,\dots,2ts+(t-1)\theta+s+z-2\}\nonumber\\
    &\cup \{2ts+\theta(t-1),\dots,ts+\theta t+z-2\} \nonumber \\
   &\cup \{2ts+\theta t,\dots,3ts+\theta(t-1)+2z-2\}\label{eq:d1cupd2cupd3cup4AGE39b}\\
   =&\{0,\dots,ts+\theta t+z-2\}\nonumber\\
    &\cup \{2ts+\theta(t-1),\ldots,3ts+\theta(t-1)+2z-2\}\label{eq:d1cupd2cupd3cup4AGE39c} \\
    =&\{0,\dots,ts+\theta (t-1)+\theta+z-2\}\nonumber\\ 
    &\cup \{2ts+\theta(t-1),\ldots,3ts+\theta(t-1)+2z-2\}\nonumber\\
     =&\{0,\dots,2ts+\theta (t-1)+\lambda+z-2\}\nonumber\\ 
    &\cup \{2ts+\theta(t-1),\ldots,3ts+\theta(t-1)+2z-2\}\nonumber\\
    =&\{0,\dots,3ts+\theta(t-1)+2z-2\}\label{eq:d1cupd2cupd3cup4AGE39final}
\end{align}
where (\ref{eq:d1cupd2cupd3cup4AGE39a}) comes from 
\begin{align}
    & \lambda < z \nonumber \\
    \Rightarrow & s(1-t)<0 < z-\lambda \nonumber \\
    \Rightarrow & s-2 < ts-\lambda z+z-2 \nonumber \\
    \Rightarrow & \theta t+s-2 < 2ts+\theta(t-1)+z-2\nonumber\\
    \Rightarrow & \{ts+\theta(t-1),\dots,\theta t+s-2\} \nonumber\\
    &\subset \{0,\dots,2ts+\theta(t-1)+z-2\},
\end{align} and
(\ref{eq:d1cupd2cupd3cup4AGE39b}) comes from 
\begin{align}
    & \lambda < z \nonumber \\
    \Rightarrow & \lambda \leq z-1 \nonumber \\
    \Rightarrow & ts+t\theta \leq 2ts+\theta(t-1)+z-2+1,
\end{align} and
(\ref{eq:d1cupd2cupd3cup4AGE39c}) comes from 
\begin{align}
    & ts-s+1 < \lambda \nonumber \\
    \Rightarrow & s(t-1)+1 < \lambda \nonumber \\
    \Rightarrow & s(2-1)+1 \leq s(t-1)+1 < \lambda \nonumber \\
    \Rightarrow & s < \lambda \nonumber \\
    \Rightarrow & 2ts+(t-1)\theta+s+z-2 < ts+\theta t +z-2,
\end{align}
and (\ref{eq:d1cupd2cupd3cup4AGE39final}) comes from the fact that $\lambda+z-2\ge 0$ (because $\lambda \ge 1$ and $z \ge 1$) and $\lambda < z$ (and thus $\lambda < z+ts$ and $2ts+\theta(t-1)+\lambda+z-2<3ts+\theta(t-1)+2z-2$). 
From (\ref{eq:d1cupd2cupd3cup4AGE39final}), $|\mathbf{P}({H}(x))|=3ts+\theta(t-1)+2z-1$. This completes the proof. \hfill $\Box$

In order to calculate $\mathbf{P}({H}(x))=\mathbf{\widehat{D}}_{123'} \cup \mathbf{\widetilde{D}}'_3 \cup \mathbf{\widetilde{D}}''_3\cup\mathbf{D}_4$ for the remaining regions of the values of $z$ and $\lambda$, \ie $z>\lambda>0$, we first calculate $\mathbf{\widehat{D}}_{123'}\cup\mathbf{D}_4$ using (\ref{eq:dhat123'AGEfinal}) and (\ref{eq:d4-AGE}) as follows

\begin{align}
    &\mathbf{P}({H}(x)) = \mathbf{D}_1 \cup  \mathbf{D}_2\cup \mathbf{D}_3 \cup \mathbf{D}_4\nonumber\\
    =&\mathbf{\widehat{D}}_{123'} \cup \mathbf{\widetilde{D}}'_3 \cup \mathbf{\widetilde{D}}''_3\cup\mathbf{D}_4\nonumber\\
    =&\{0,\ldots,2ts+\theta(t-1)+z-2\}\nonumber\\
    &\cup\bigcup_{l=0}^{q-1} \{2ts+\theta(l+t-1),\dots,ts+\theta(l+t)+z-2\} \nonumber \\
   &\cup \{2ts+\theta(q+t-1),\dots ,(q+2)ts+\theta(t-1)+2z-2\}\nonumber\\
    &\cup \mathbf{\widetilde{D}}'_3 \cup \mathbf{\widetilde{D}}''_3\nonumber\\
    =&\{0,\ldots,2ts+\theta(t-1)+z-2\}\nonumber\\
    &\cup\{2ts+\theta(t-1),\dots,ts+\theta t+z-2\}\nonumber\\
    &\cup\bigcup_{l=1}^{q-1} \{2ts+\theta(l+t-1),\dots,ts+\theta(l+t)+z-2\} \nonumber \\
   &\cup \{2ts+\theta(q+t-1),\dots ,(q+2)ts+\theta(t-1)+2z-2\}\nonumber\\
    &\cup \mathbf{\widetilde{D}}'_3 \cup \mathbf{\widetilde{D}}''_3\nonumber\\
    =&\{0,\ldots,ts+\theta t+z-2\}\nonumber\\
    &\cup\bigcup_{l=1}^{q-1} \{2ts+\theta(l+t-1),\dots,ts+\theta(l+t)+z-2\} \nonumber \\
   &\cup \{2ts+\theta(q+t-1),\dots ,(q+2)ts+\theta(t-1)+2z-2\}\nonumber\\
    &\cup \mathbf{\widetilde{D}}'_3 \cup \mathbf{\widetilde{D}}''_3 \label{eq:d1cupd2cupd3cupd3AGELastFour1}\\
    =&(\mathbf{\widehat{D}}_{123'4}\cup\mathbf{\widetilde{D}}'_3)\cup(\mathbf{\widehat{D}}_{123'4}\cup\mathbf{\widetilde{D}}''_3)\quad \quad \text{for    } 0<\lambda<z \label{eq:phxforthe4lastcases},
\end{align}
where (\ref{eq:d1cupd2cupd3cupd3AGELastFour1}) comes from the fact that $\lambda>0$ and thus $ts+\theta t+z-2 > 2ts+\theta(t-1)+z-2$ and $\mathbf{\widehat{D}}_{123'4}$ is equal to
\begin{align}\label{eq:d1234}
    &\mathbf{\widehat{D}}_{123'4} =\{0,\ldots,ts+\theta t+z-2\}\nonumber\\
    &\cup\bigcup_{l=1}^{q-1} \{2ts+\theta(l+t-1),\dots,ts+\theta(l+t)+z-2\} \nonumber \\
   &\cup \{2ts+\theta(q+t-1),\dots ,(q+2)ts+\theta(t-1)+2z-2\}.
\end{align}
Next, we calculate $\mathbf{\widehat{D}}_{123'4}\cup\mathbf{\widetilde{D}}'_3$ and $\mathbf{\widehat{D}}_{123'4}\cup\mathbf{\widetilde{D}}''_3$ for different regions of values of $z$ and $\lambda$.

\begin{lemma}\label{lemma:d1234cupd'3AGE}
\begin{align}
\mathbf{\widehat{D}}_{123'4}\cup\mathbf{\widetilde{D}}'_3 =\begin{cases}
\mathbf{D}_{123'4(a)}, & z > \lambda+s-1 \\
   \mathbf{D}_{123'4(b)}, & z \leq \lambda+s-1,
\end{cases}
\end{align}
\begin{align}
    &\mathbf{D}_{123'4(a)} = \{0,\ldots,ts+\theta t+z-2\}\nonumber\\
    &\cup\bigcup_{l=t}^{t+q-2} \{2ts+\theta l,\dots,ts+\theta(l+1)+z-2\} \nonumber \\
   &\cup \{2ts+\theta(q+t-1),\dots ,(q+2)ts+\theta(t-1)+2z-2\},
\end{align}
\begin{align}
    &\mathbf{D}_{123'4(b)}= \{0,\ldots,\theta(t+1)+s-2\}\nonumber\\
    &\cup\bigcup_{l=t}^{t+q-3} \{2ts+\theta  l,\dots,\theta(l+2)+s-2\} \nonumber\\
    &\cup \{2ts+\theta (t+q-2),\dots,ts+\theta(t+q-1)+z-2\} \nonumber \\
   &\cup \{2ts+\theta(q+t-1),\dots ,(q+2)ts+\theta(t-1)+2z-2\}
\end{align}
\end{lemma}
{\em Proof:}
From (\ref{eq:d1234}) and (\ref{eq:d1cupd2cupd3AGE4}), we have:
\begin{align}
&\mathbf{\widehat{D}}_{123'4}\cup\mathbf{\widetilde{D}}'_3 = \{0,\ldots,ts+\theta t+z-2\}\nonumber\\
    &\cup\bigcup_{l=1}^{q-1} \{2ts+\theta(l+t-1),\dots,ts+\theta(l+t)+z-2\} \nonumber \\
   &\cup \{2ts+\theta(q+t-1),\dots ,(q+2)ts+\theta(t-1)+2z-2\}\nonumber\\
   &\cup \bigcup_{l'=t-1}^{t+q-2}\{ts+\theta l',\ldots,\theta (l'+1)+s-2\}\nonumber\\
   = &\{0,\ldots,ts+\theta t+z-2\}\nonumber
         \end{align}
   \begin{align}
    &\cup\bigcup_{l=1}^{q-1} \{2ts+\theta(l+t-1),\dots,ts+\theta(l+t)+z-2\} \nonumber \\
   &\cup \{2ts+\theta(q+t-1),\dots ,(q+2)ts+\theta(t-1)+2z-2\}\nonumber\\
   &\cup \{ts+\theta (t-1),\ldots,\theta t+s-2\}\nonumber\\
   &\cup \bigcup_{l'=t}^{t+q-2}\{ts+\theta l',\ldots,\theta (l'+1)+s-2\}\nonumber\\
   = &\{0,\ldots,ts+\theta t+z-2\}\nonumber\\
    &\cup\bigcup_{l=1}^{q-1} \{2ts+\theta(l+t-1),\dots,ts+\theta(l+t)+z-2\} \nonumber
       \end{align}
   \begin{align}
   &\cup \{2ts+\theta(q+t-1),\dots ,(q+2)ts+\theta(t-1)+2z-2\}\nonumber\\
   &\cup \bigcup_{l'=t}^{t+q-2}\{ts+\theta l',\ldots,\theta (l'+1)+s-2\}\label{eq:d1cupd2cupd3cupd4cupdthilde3a}\\
   = &\{0,\ldots,ts+\theta t+z-2\}\nonumber
      \end{align}
   \begin{align}
    &\cup\bigcup_{l=t}^{t+q-2} \{2ts+\theta l,\dots,ts+\theta(l+1)+z-2\} \nonumber
    \end{align}
    \begin{align}
   &\cup \{2ts+\theta(q+t-1),\dots ,(q+2)ts+\theta(t-1)+2z-2\}\nonumber\\
   &\cup \bigcup_{l'=t}^{t+q-2}\{ts+\theta l',\ldots,\theta (l'+1)+s-2\}\nonumber\\
   = &\{0,\ldots,ts+\theta t+z-2\}\nonumber\\
    &\cup\bigcup_{l=t}^{t+q-2} \{2ts+\theta l,\dots,ts+\theta(l+1)+z-2\} \nonumber \\
   &\cup \{2ts+\theta(q+t-1),\dots ,(q+2)ts+\theta(t-1)+2z-2\}\nonumber\\
   &\cup \{ts+\theta t,\ldots,\theta (t+1)+s-2\}\nonumber\\
   &\cup \bigcup_{l'=t}^{t+q-3}\{ts+\theta (l'+1),\ldots,\theta (l'+2)+s-2\}
   \label{eq:d1cupd2cupd3cupd4cupdthilde3b}
\end{align}
where (\ref{eq:d1cupd2cupd3cupd4cupdthilde3a}) comes from the fact that $s<ts+z$ and thus $\theta t+s-2<ts+\theta t+z-2$; this results in $\{ts+\theta (t-1),\ldots,\theta t+s-2\} \subset \{0,\ldots,ts+\theta t+z-2\}$. 

Now, we consider the two cases;  Case 1: $z > \lambda+s-1$, and Case 2: $z \leq \lambda+s-1$, and simplify (\ref{eq:d1cupd2cupd3cupd4cupdthilde3b}) for each case. 

\begin{figure*}
		\centering
		\includegraphics[width=14cm]{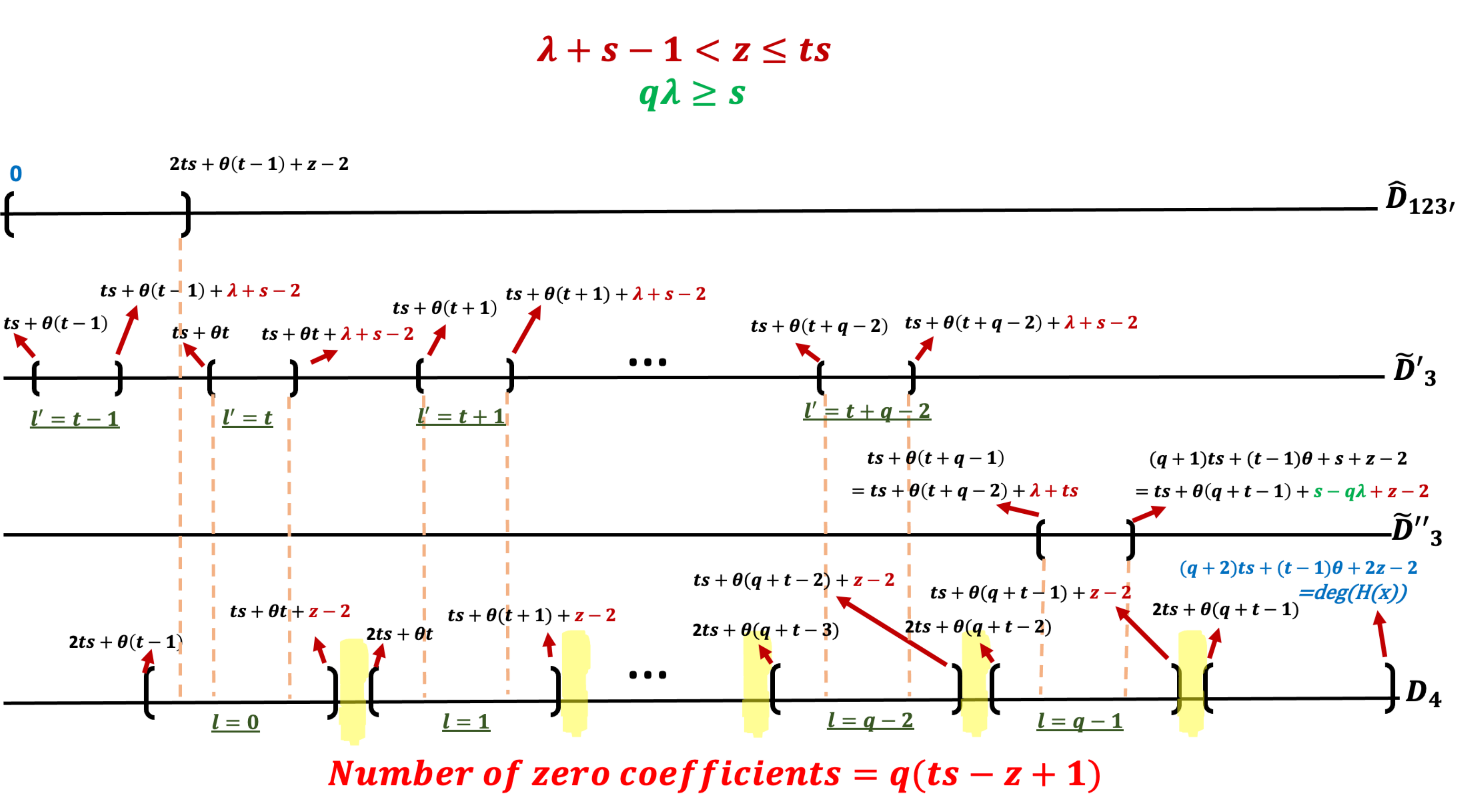}
	\caption{Illustration of $ \mathbf{\widehat{D}}_{123'} \cup \mathbf{\widetilde{D}}'_3 \cup \mathbf{\widetilde{D}}''_3 \cup \mathbf{D}_4$ for $\lambda+s-1 < z \leq ts$ and $q\lambda \geq s$.
	}
\label{fig:non-zero-coeff-AGE-z less than ts: 2}
\vspace{-10pt}
\end{figure*}

\begin{figure*}
		\centering
		\includegraphics[width=14cm]{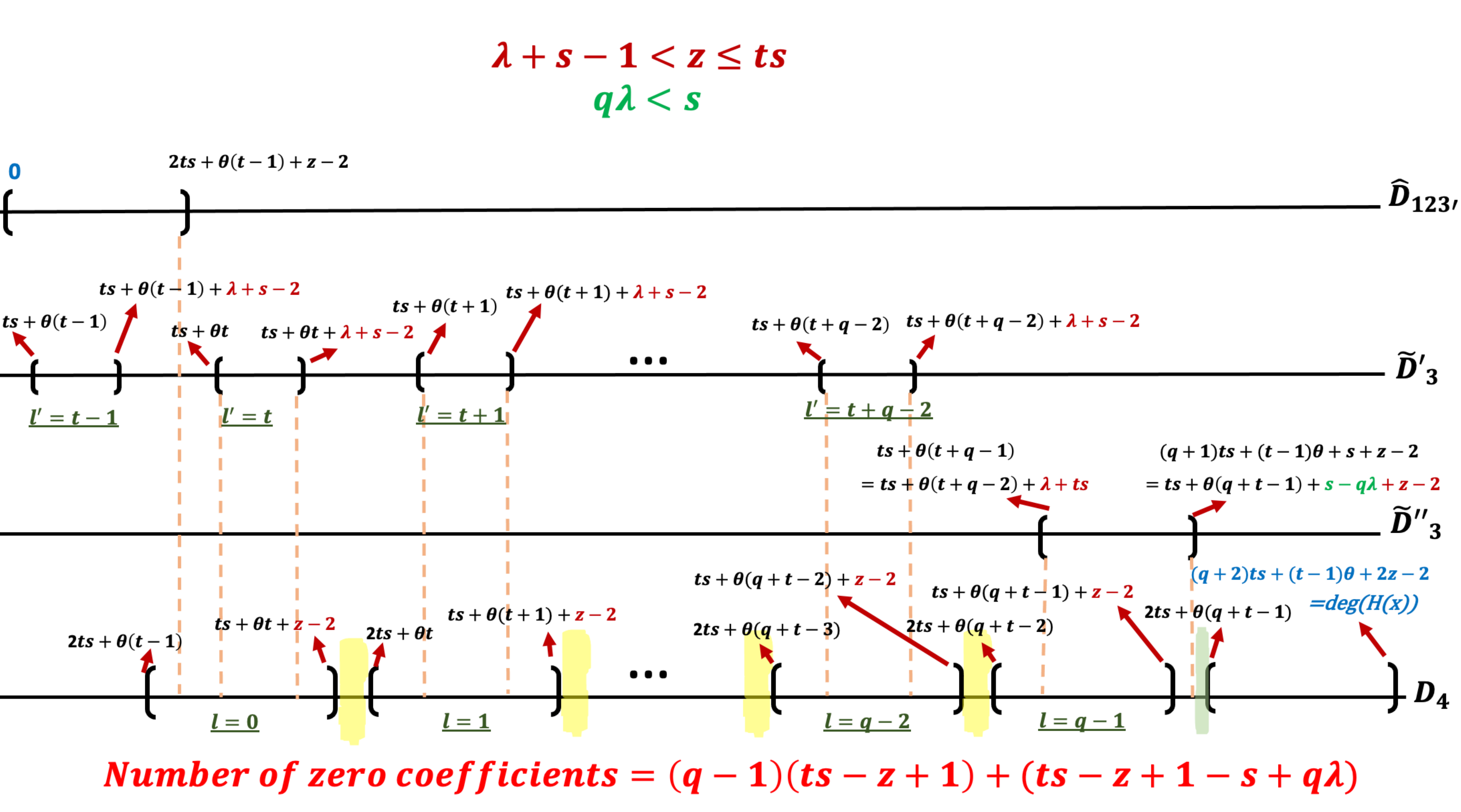}
	\caption{Illustration of $ \mathbf{\widehat{D}}_{123'} \cup \mathbf{\widetilde{D}}'_3 \cup \mathbf{\widetilde{D}}''_3 \cup \mathbf{D}_4$ for $\lambda+s-1 < z \leq ts$ and $q\lambda < s$.
	}
\label{fig:non-zero-coeff-AGE-z less than ts: 3}
\vspace{-10pt}
\end{figure*}

Case 1: $z > \lambda+s-1$. For this case, $\bigcup_{l'=t}^{t+q-3} \{ts+\theta (l'+1),\dots,\theta(l'+2)+s-2\}$ is a subset of $\bigcup_{l=t}^{t+q-2} \{2ts+\theta l,\dots,ts+\theta(l+1)+z-2\}$. This is formulated in the following and demonstrated in Fig. \ref{fig:non-zero-coeff-AGE-z less than ts: 2} and \ref{fig:non-zero-coeff-AGE-z less than ts: 3}.
\begin{align}\label{eq:subsetPrrofb}
    & 0 < \lambda \nonumber \\
    \Rightarrow & ts < \theta \nonumber \\
    \Rightarrow & 2ts+\theta l < ts+\theta(l+1),
\end{align}
and 
\begin{align}\label{eq:subsetPrrofc}
    & \lambda+s-1 < z \nonumber \\
    \Rightarrow & \lambda+s \leq z \nonumber \\
    \Rightarrow & \theta+s-2 \leq ts+z-2\nonumber\\
    \Rightarrow & \theta(l+2)+s-2 \leq ts+\theta(l+1)+z-2.
\end{align}

On the other hand, $\{ts+\theta t,\ldots,\theta (t+1)+s-2\}$ is a subset of $\{0,\ldots,ts+\theta t+z-2\}$. This is expressed in the following and demonstrated in Fig. \ref{fig:non-zero-coeff-AGE-z less than ts: 2} and \ref{fig:non-zero-coeff-AGE-z less than ts: 3}.
\begin{align}\label{eq:subsetProofa}
    & \lambda+s-1 < z \nonumber \\
    \Rightarrow & s+\lambda \leq z \nonumber \\
    \Rightarrow & ts+z \leq \theta+s\nonumber\\
    \Rightarrow & \theta (t+1)+s-2 \leq ts+\theta t+z-2,
\end{align}
Therefore, for the case of $z>\lambda+s-1$, (\ref{eq:d1cupd2cupd3cupd4cupdthilde3b}) is simplified as
\begin{align}
    &\mathbf{\widehat{D}}_{123'4}\cup\mathbf{\widetilde{D}}'_3 = \{0,\ldots,ts+\theta t+z-2\}\nonumber\\
    &\cup\bigcup_{l=t}^{t+q-2} \{2ts+\theta l,\dots,ts+\theta(l+1)+z-2\} \nonumber \\
   &\cup \{2ts+\theta(q+t-1),\dots ,(q+2)ts+\theta(t-1)+2z-2\}
\end{align}

\begin{figure*}
		\centering
		\includegraphics[width=14cm]{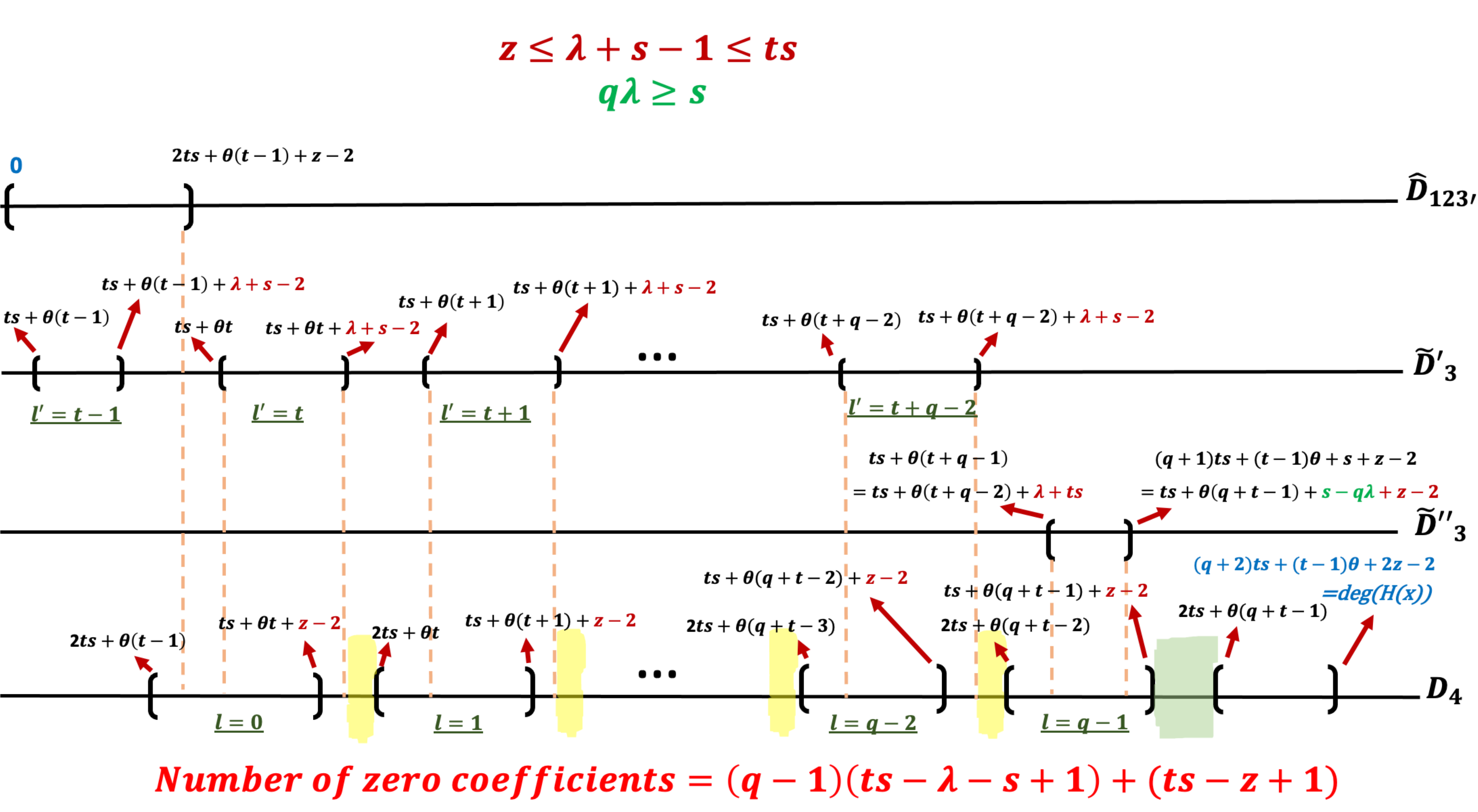}
	\caption{Illustration of $ \mathbf{\widehat{D}}_{123'} \cup \mathbf{\widetilde{D}}'_3 \cup \mathbf{\widetilde{D}}''_3 \cup \mathbf{D}_4$ for $z \leq \lambda+s-1 \leq ts$ and $q\lambda \geq s$.
	}
\label{fig:non-zero-coeff-AGE-z less than ts: 4}
\vspace{-10pt}
\end{figure*}

\begin{figure*}
		\centering
		\includegraphics[width=14cm]{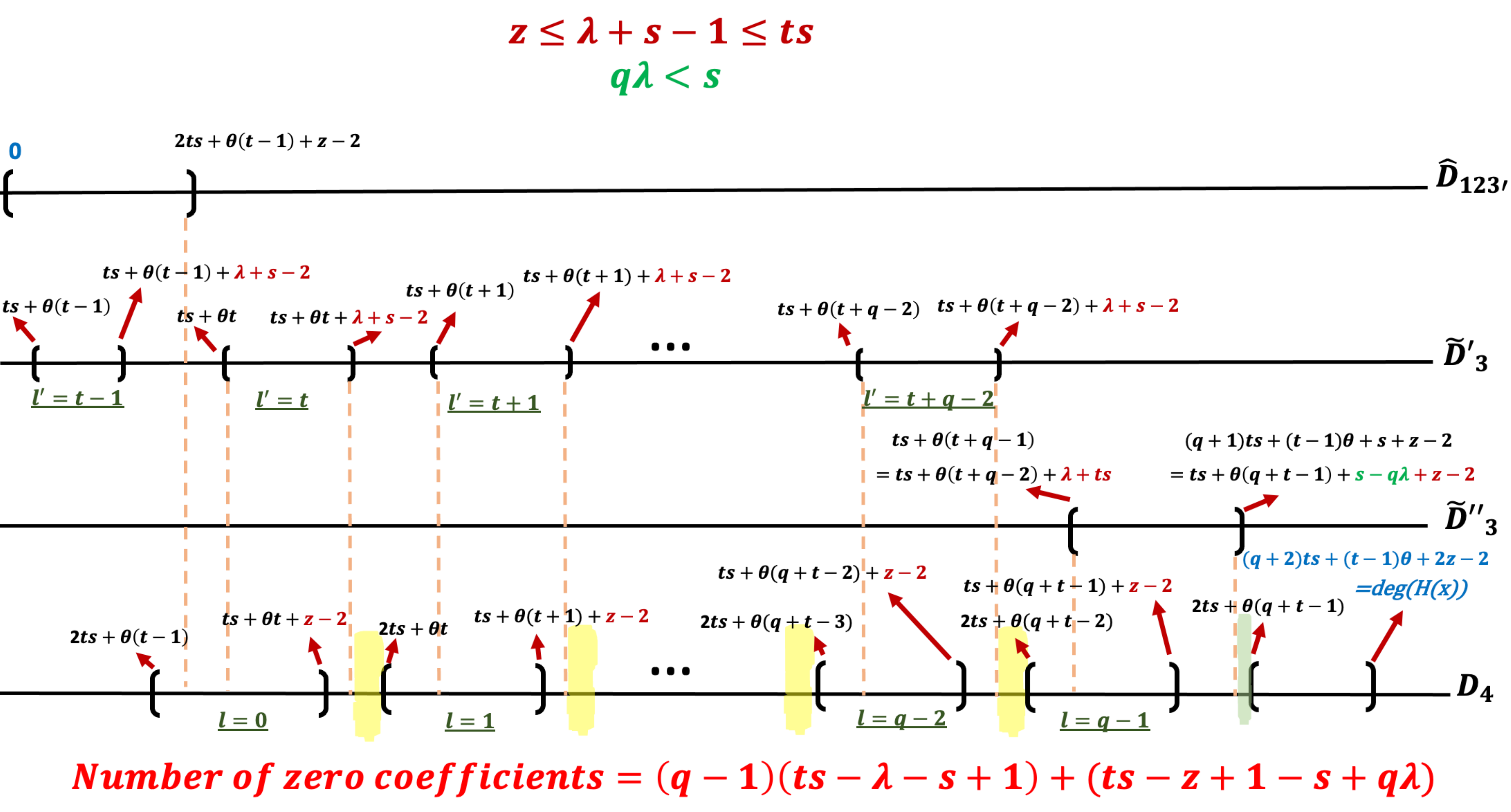}
	\caption{Illustration of $ \mathbf{\widehat{D}}_{123'} \cup \mathbf{\widetilde{D}}'_3 \cup \mathbf{\widetilde{D}}''_3 \cup \mathbf{D}_4$ for $z \leq \lambda+s-1 \leq ts$ and $q\lambda < s$.
	}
\label{fig::non-zero-coeff-AGE-z less than ts: 5}
\end{figure*}

Case 2: $z \leq \lambda+s-1$.
For this case, the union of $\{ts+\theta t,\ldots,\theta (t+1)+s-2\}$ and $\{0,\ldots,ts+\theta t+z-2\}$ is equal to $\{0,\ldots,\theta (t+1)+s-2\}$. This can be derived from (\ref{eq:subsetProofa}) and demonstrated in Fig. \ref{fig:non-zero-coeff-AGE-z less than ts: 4} and \ref{fig::non-zero-coeff-AGE-z less than ts: 5}.
On the other hand, the union of $\bigcup_{l'=t}^{t+q-3} \{ts+\theta (l'+1),\dots,\theta(l'+2)+s-2\}$ and $\bigcup_{l=t}^{t+q-2} \{2ts+\theta l,\dots,ts+\theta(l+1)+z-2\}$ is equal to $\bigcup_{l=t}^{t+q-3} \{2ts+\theta  l,\dots,\theta(l+2)+s-2\} \cup \{2ts+\theta (t+q-2),\dots,ts+\theta(t+q-1)+z-2\}$. This can be derived  from (\ref{eq:subsetPrrofb}) and (\ref{eq:subsetPrrofc}) and demonstrated in Fig. \ref{fig:non-zero-coeff-AGE-z less than ts: 4} and \ref{fig::non-zero-coeff-AGE-z less than ts: 5}.
Therefore, for the case of $z\leq \lambda+s-1$, (\ref{eq:d1cupd2cupd3cupd4cupdthilde3b}) is simplified as
\begin{align}
    &\mathbf{\widehat{D}}_{123'4}\cup\mathbf{\widetilde{D}}'_3 = \{0,\ldots,\theta(t+1)+s-2\}\nonumber\\
    &\cup\bigcup_{l=t}^{t+q-3} \{2ts+\theta  l,\dots,\theta(l+2)+s-2\} \nonumber\\
    &\cup \{2ts+\theta (t+q-2),\dots,ts+\theta(t+q-1)+z-2\} \nonumber \\
   &\cup \{2ts+\theta(q+t-1),\dots ,(q+2)ts+\theta(t-1)+2z-2\}
\end{align}
This completes the proof. \hfill $\Box$

\begin{lemma}\label{lemma:d1234cupd''3AGE} The following equalities hold.
\begin{align}
\mathbf{\widehat{D}}_{123'4}\cup\mathbf{\widetilde{D}}''_3 =\begin{cases}
\mathbf{D}_{123''4(a)}, & q\lambda \ge s \\
   \mathbf{D}_{123''4(b)}, & q\lambda < s,
\end{cases}
\end{align}
\begin{align}
&\mathbf{D}_{123''4(a)}= \{0,\ldots,ts+\theta t+z-2\}\nonumber\\
    &\cup\bigcup_{l=1}^{q-2} \{2ts+\theta(l+t-1),\dots,ts+\theta(l+t)+z-2\} \nonumber \\
    &\cup \{2ts+\theta(q+t-2),\dots,ts+\theta(q+t-1)+z-2\}\nonumber\\
   &\cup \{2ts+\theta(q+t-1),\dots ,(q+2)ts+\theta(t-1)+2z-2\},
\end{align}
\begin{align}
&\mathbf{D}_{123''4(b)}=\{0,\ldots,ts+\theta t+z-2\}\nonumber\\
    &\cup\bigcup_{l=1}^{q-2} \{2ts+\theta(l+t-1),\dots,ts+\theta(l+t)+z-2\} \nonumber \\
    &\cup \{2ts+\theta(q+t-2),\dots,(q+1)ts+(t-1)\theta+s+z-2\}\nonumber\\
   &\cup \{2ts+\theta(q+t-1),\dots ,(q+2)ts+\theta(t-1)+2z-2\}
\end{align}
\end{lemma}
{\em Proof:}
From (\ref{eq:d1234}) and (\ref{eq:d1cupd2cupd3AGE4}), we have
\begin{align}\label{eq:d1cupd2cupd3''cupd4AGE}
&\mathbf{\widehat{D}}_{123'4}\cup\mathbf{\widetilde{D}}''_3 = \{0,\ldots,ts+\theta t+z-2\}\nonumber\\
    &\cup\bigcup_{l=1}^{q-1} \{2ts+\theta(l+t-1),\dots,ts+\theta(l+t)+z-2\} \nonumber\\
   &\cup \{2ts+\theta(q+t-1),\dots ,(q+2)ts+\theta(t-1)+2z-2\}\nonumber\\
   &\cup \{ts+(q+t-1)\theta,\dots,(q+1)ts+(t-1)\theta+s+z-2\}\nonumber\\
   =& \{0,\ldots,ts+\theta t+z-2\}\nonumber\\
    &\cup\bigcup_{l=1}^{q-2} \{2ts+\theta(l+t-1),\dots,ts+\theta(l+t)+z-2\} \nonumber \\
    &\cup \{2ts+\theta(q+t-2),\dots,ts+\theta(q+t-1)+z-2\}\nonumber\\
   &\cup \{2ts+\theta(q+t-1),\dots ,(q+2)ts+\theta(t-1)+2z-2\}\nonumber\\
   &\cup \{ts+(q+t-1)\theta,\dots,(q+1)ts+(t-1)\theta+s+z-2\}
\end{align}
To simplify the above equation, we consider the two cases; Case 1: $q\lambda \ge s$, and Case 2: $q\lambda < s$. 

Case 1: $q\lambda \ge s$. For this case, $\{ts+(q+t-1)\theta,\dots,(q+1)ts+(t-1)\theta+s+z-2\}$ is a subset of $\{2ts+\theta(q+t-2),\dots,ts+\theta(q+t-1)+z-2\}$. This is shown mathematically in the following and demonstrated in Fig. \ref{fig:non-zero-coeff-AGE-z less than ts: 2} and \ref{fig:non-zero-coeff-AGE-z less than ts: 4}:
\begin{align}\label{eq:subsetProofd}
    & q\lambda \ge s \nonumber \\
    \Rightarrow & \theta q+z-2 \ge qts+s+z-2 \nonumber \\
    \Rightarrow & ts+\theta(q+t-1)+z-2 \ge\nonumber\\
    &\quad \quad \quad (q+1)ts+(t+1)\theta+s+z-2,
\end{align}
and
\begin{align}\label{eq:subsetProofe}
    & ts < ts+\lambda \nonumber \\
    \Rightarrow & 2ts+\theta(q+t-2) < ts+(q+t-1)\theta.
\end{align}
Therefore, for the case of $q\lambda \ge s$, (\ref{eq:d1cupd2cupd3''cupd4AGE}) is simplified as:
\begin{align}
&\mathbf{\widehat{D}}_{123'4}\cup\mathbf{\widetilde{D}}''_3 = \{0,\ldots,ts+\theta t+z-2\}\nonumber\\
    &\cup\bigcup_{l=1}^{q-2} \{2ts+\theta(l+t-1),\dots,ts+\theta(l+t)+z-2\} \nonumber \\
    &\cup \{2ts+\theta(q+t-2),\dots,ts+\theta(q+t-1)+z-2\}\nonumber\\
   &\cup \{2ts+\theta(q+t-1),\dots ,(q+2)ts+\theta(t-1)+2z-2\}
\end{align}

Case 2: $q\lambda < s$. For this case, the union of $\{ts+(q+t-1)\theta,\dots,(q+1)ts+(t-1)\theta+s+z-2\}$ and $\{2ts+\theta(q+t-2),\dots,ts+\theta(q+t-1)+z-2\}$ is equal to $\{2ts+\theta(q+t-2),\dots,(q+1)ts+(t-1)\theta+s+z-2\}$. This can be derived mathematically from (\ref{eq:subsetProofd}) and (\ref{eq:subsetProofe}) and demonstrated in Fig. \ref{fig:non-zero-coeff-AGE-z less than ts: 3} and \ref{fig::non-zero-coeff-AGE-z less than ts: 5}. Therefore, for the case of $q\lambda < s$, (\ref{eq:d1cupd2cupd3''cupd4AGE}) is simplified as
\begin{align}
&\mathbf{\widehat{D}}_{123'4}\cup\mathbf{\widetilde{D}}''_3 = \{0,\ldots,ts+\theta t+z-2\}\nonumber\\
    &\cup\bigcup_{l=1}^{q-2} \{2ts+\theta(l+t-1),\dots,ts+\theta(l+t)+z-2\} \nonumber \\
    &\cup \{2ts+\theta(q+t-2),\dots,(q+1)ts+(t-1)\theta+s+z-2\}\nonumber\\
   &\cup \{2ts+\theta(q+t-1),\dots ,(q+2)ts+\theta(t-1)+2z-2\}
\end{align}
This completes the proof. \hfill $\Box$

\begin{lemma}\label{lemma:non-zero-coeff-AGE-z less than ts: 2}
For $\lambda+s-1 < z \leq ts, 0<\lambda <z, t\neq 1$ and $q\lambda \geq s$, we have
\begin{equation}
    |\mathbf{P}({H}(x))|=\Upsilon_6(\lambda)= 2ts+\theta(t-1)+(q+2)z-q-1
\end{equation}
\end{lemma}
{\em Proof:} 
From (\ref{eq:phxforthe4lastcases}) and Lemmas \ref{lemma:d1234cupd'3AGE} and {\ref{lemma:d1234cupd''3AGE}}, we have
\begin{align}
    &\mathbf{P}({H}(x))=(\mathbf{\widehat{D}}_{123'4}\cup\mathbf{\widetilde{D}}'_3)\cup(\mathbf{\widehat{D}}_{123'4}\cup\mathbf{\widetilde{D}}''_3)\nonumber\\
    =&\{0,\ldots,ts+\theta t+z-2\}\nonumber\\
    &\cup\bigcup_{l=t}^{t+q-2} \{2ts+\theta l,\dots,ts+\theta(l+1)+z-2\} \nonumber \\
   &\cup \{2ts+\theta(q+t-1),\dots ,(q+2)ts+\theta(t-1)+2z-2\}\nonumber
            \end{align}
    \begin{align}
   &\cup\{0,\ldots,ts+\theta t+z-2\}\nonumber\\
    &\cup\bigcup_{l=1}^{q-2} \{2ts+\theta(l+t-1),\dots,ts+\theta(l+t)+z-2\} \nonumber \\
    &\cup \{2ts+\theta(q+t-2),\dots,ts+\theta(q+t-1)+z-2\}\nonumber\\
   &\cup \{2ts+\theta(q+t-1),\dots ,(q+2)ts+\theta(t-1)+2z-2\}\nonumber\\
   =&\{0,\ldots,ts+\theta t+z-2\}\nonumber\\
    &\cup\bigcup_{l=t}^{t+q-2} \{2ts+\theta l,\dots,ts+\theta(l+1)+z-2\} \nonumber \\
   &\cup \{2ts+\theta(q+t-1),\dots ,(q+2)ts+\theta(t-1)+2z-2\}\nonumber\\
    &\cup \{2ts+\theta(q+t-2),\dots,ts+\theta(q+t-1)+z-2\}\nonumber\\
   =&\{0,\ldots,ts+\theta t+z-2\}\nonumber\\
    &\cup\bigcup_{l=t}^{t+q-2} \{2ts+\theta l,\dots,ts+\theta(l+1)+z-2\} \nonumber \\
   &\cup \{2ts+\theta(q+t-1),\dots ,(q+2)ts+\theta(t-1)+2z-2\}\label{eq:phxAGEtheLastFoura}
\end{align}
Next, we show that the subsets shown in (\ref{eq:phxAGEtheLastFoura}) do not have overlap. 
\begin{align}
    & z \leq ts \nonumber \\
    \Rightarrow & z-2 < ts \nonumber \\
    \Rightarrow & ts+\theta t +z-2 < 2ts+\theta t\nonumber\\
    & \& \quad ts+\theta(l+1)+z-2 < 2ts+\theta(l+1).
\end{align}
Therefore, by calculating the size of each subset, we can calculate the number of elements of $\mathbf{P}({H}(x))$. The size of $\{0,\ldots,ts+\theta t+z-2\}$ is equal to $ts+\theta t+z-1$. The size of $\bigcup_{l=t}^{t+q-2} \{2ts+\theta l,\dots,ts+\theta(l+1)+z-2\}$ is equal to $(q-1)(\lambda +z-1)$. The size of $\{2ts+\theta(q+t-1),\dots ,(q+2)ts+\theta(t-1)+2z-2\}$ is equal to $-\lambda q+2z-1$. Therefore, $\mathbf{P}({H}(x))$ is equal to the sum of all these sizes, \ie $\mathbf{P}({H}(x))=ts+\theta t+z-1+(q-1)(\lambda +z-1)-\lambda q+2z-1=2ts+\theta(t-1)+(q+2)z-q-1$. This completes the proof. \hfill $\Box$
 \begin{lemma}\label{lemma:non-zero-coeff-AGE-z less than ts: 3}
For $\lambda+s-1 < z \leq ts, 0<\lambda <z, t\neq 1$ and $q\lambda < s$, we have
\begin{align}
    |\mathbf{P}({H}(x))|=\Upsilon_7(\lambda) 
    =& \theta(t+1)+q(z-1)-2\lambda +z+ts\nonumber\\
    &+\min\{0, z+s(1-t)-\lambda q-1\}
\end{align}
\end{lemma}
{\em Proof:} 
From (\ref{eq:phxforthe4lastcases}) and Lemmas \ref{lemma:d1234cupd'3AGE} and {\ref{lemma:d1234cupd''3AGE}}, we have
\begin{align}
    &\mathbf{P}({H}(x))=(\mathbf{\widehat{D}}_{123'4}\cup\mathbf{\widetilde{D}}'_3)\cup(\mathbf{\widehat{D}}_{123'4}\cup\mathbf{\widetilde{D}}''_3)\nonumber\\
    =& \{0,\ldots,ts+\theta t+z-2\}\nonumber
            \end{align}
    \begin{align}
    &\cup\bigcup_{l=t}^{t+q-2} \{2ts+\theta l,\dots,ts+\theta(l+1)+z-2\} \nonumber \\
   &\cup \{2ts+\theta(q+t-1),\dots ,(q+2)ts+\theta(t-1)+2z-2\}\nonumber\\
   &\cup \{0,\ldots,ts+\theta t+z-2\}\nonumber
            \end{align}
    \begin{align}
    &\cup\bigcup_{l=1}^{q-2} \{2ts+\theta(l+t-1),\dots,ts+\theta(l+t)+z-2\} \nonumber \\
    \cup & \{2ts+\theta(q+t-2),\dots,(q+1)ts+(t-1)\theta+s+z-2\}\nonumber\\
   &\cup \{2ts+\theta(q+t-1),\dots ,(q+2)ts+\theta(t-1)+2z-2\}\nonumber\\
   =& \{0,\ldots,ts+\theta t+z-2\}\nonumber\\
    &\cup\bigcup_{l=t}^{t+q-2} \{2ts+\theta l,\dots,ts+\theta(l+1)+z-2\} \nonumber \\
   &\cup \{2ts+\theta(q+t-1),\dots ,(q+2)ts+\theta(t-1)+2z-2\}\nonumber\\
    &\cup\bigcup_{l=t}^{t+q-3} \{2ts+\theta l,\dots,ts+\theta(l+1)+z-2\} \nonumber \\
    \cup & \{2ts+\theta(q+t-2),\dots,(q+1)ts+(t-1)\theta+s+z-2\}\nonumber\\
      =& \{0,\ldots,ts+\theta t+z-2\}\nonumber\\
    &\cup\bigcup_{l=t}^{t+q-2} \{2ts+\theta l,\dots,ts+\theta(l+1)+z-2\} \nonumber \\
    \cup & \{2ts+\theta(q+t-2),\dots,(q+1)ts+(t-1)\theta+s+z-2\}\nonumber\\
   &\cup \{2ts+\theta(q+t-1),\dots ,(q+2)ts+\theta(t-1)+2z-2\}\nonumber\\
   =& \{0,\ldots,ts+\theta t+z-2\}\nonumber\\
    &\cup\bigcup_{l=t}^{t+q-3} \{2ts+\theta l,\dots,ts+\theta(l+1)+z-2\} \nonumber \\
    \cup & \{2ts+\theta(q+t-2),\dots,(q+1)ts+(t-1)\theta+s+z-2\}\nonumber\\
   &\cup \{2ts+\theta(q+t-1),\dots ,(q+2)ts+\theta(t-1)+2z-2\}\label{eq:phxAGEtheLastFourb}
\end{align}
Next, we show that the subsets shown in (\ref{eq:phxAGEtheLastFourb}), do not have overlap except for the last two subsets of $\{2ts+\theta(q+t-2),\dots,(q+1)ts+(t-1)\theta+s+z-2\}$ and $\{2ts+\theta(q+t-1),\dots ,(q+2)ts+\theta(t-1)+2z-2\}$
\begin{align}
    & z \leq ts \nonumber \\
    \Rightarrow & z-2 < ts \nonumber \\
    \Rightarrow & ts+\theta(l+1)+z-2 < 2ts+\theta(l+1)
\end{align}
Therefore, by calculating the size of each subset, we can calculate the number of elements of $\mathbf{P}({H}(x))$. 
The size of $\{0,\ldots,ts+\theta t+z-2\}$ is equal to $ts+\theta t+z-1$.
The size of $\bigcup_{l=t}^{t+q-3} \{2ts+\theta l,\dots,ts+\theta(l+1)+z-2\}$ is equal to $(q-2)(\lambda+z-1)$. 
The size of $\{2ts+\theta(q+t-2),\dots,(q+1)ts+(t-1)\theta+s+z-2\}\cup \{2ts+\theta(q+t-1),\dots ,(q+2)ts+\theta(t-1)+2z-2\}$ is equal to $\min\{\lambda(1-2q)+3z+s-2, \theta -\lambda q+2z-1\}$. Therefore, $|\mathbf{P}({H}(x))|$ is equal to the sum of all these sizes, \ie $|\mathbf{P}({H}(x))|=\theta(t+1)+q(z-1)-2\lambda +z+ts+\min\{0, z+s(1-t)-\lambda q-1\}$. This completes the proof.\hfill $\Box$ 

\begin{lemma}\label{lemma:non-zero-coeff-AGE-z less than ts: 4}
For $z \leq \lambda+s-1 \leq ts, 0<\lambda <z, t\neq 1$ and $q\lambda \geq s$, we have
\begin{equation}
    |\mathbf{P}({H}(x))|=\Upsilon_8(\lambda)= 2ts+\theta(t-1)+3z+(\lambda+s-1)q-\lambda-s-1
\end{equation}
\end{lemma}
{\em Proof:} 
From (\ref{eq:phxforthe4lastcases}) and Lemmas \ref{lemma:d1234cupd'3AGE} and {\ref{lemma:d1234cupd''3AGE}}, we have
\begin{align}
    &\mathbf{P}({H}(x))=(\mathbf{\widehat{D}}_{123'4}\cup\mathbf{\widetilde{D}}'_3)\cup(\mathbf{\widehat{D}}_{123'4}\cup\mathbf{\widetilde{D}}''_3)\nonumber\\
    &=\{0,\ldots,\theta(t+1)+s-2\}\nonumber\\
    &\cup\bigcup_{l=t}^{t+q-3} \{2ts+\theta  l,\dots,\theta(l+2)+s-2\} \nonumber\\
    &\cup \{2ts+\theta (t+q-2),\dots,ts+\theta(t+q-1)+z-2\} \nonumber \\
   &\cup \{2ts+\theta(q+t-1),\dots ,(q+2)ts+\theta(t-1)+2z-2\}\nonumber\\
   &\cup \{0,\ldots,ts+\theta t+z-2\}\nonumber\\
    &\cup\bigcup_{l=1}^{q-2} \{2ts+\theta(l+t-1),\dots,ts+\theta(l+t)+z-2\} \nonumber \\
    &\cup \{2ts+\theta(q+t-2),\dots,ts+\theta(q+t-1)+z-2\}\nonumber\\
   &\cup \{2ts+\theta(q+t-1),\dots ,(q+2)ts+\theta(t-1)+2z-2\}\nonumber 
            \end{align}
    \begin{align}
   &=\{0,\ldots,\theta(t+1)+s-2\}\nonumber\\
    &\cup\bigcup_{l=t}^{t+q-3} \{2ts+\theta  l,\dots,\theta(l+2)+s-2\} \nonumber\\
    &\cup \{2ts+\theta (t+q-2),\dots,ts+\theta(t+q-1)+z-2\} \nonumber \\
   &\cup \{2ts+\theta(q+t-1),\dots ,(q+2)ts+\theta(t-1)+2z-2\}\nonumber\\
    &\cup\bigcup_{l=t}^{t+q-3} \{2ts+\theta l,\dots,ts+\theta(l+1)+z-2\} \nonumber \\
   &=\{0,\ldots,\theta(t+1)+s-2\}\nonumber\\
    &\cup\bigcup_{l=t}^{t+q-3} \{2ts+\theta  l,\dots,\theta(l+2)+s-2\} \nonumber\\
    &\cup \{2ts+\theta (t+q-2),\dots,ts+\theta(t+q-1)+z-2\} \nonumber \\
   &\cup \{2ts+\theta(q+t-1),\dots ,(q+2)ts+\theta(t-1)+2z-2\}\label{eq:phxAGEtheLastFourc}
\end{align}
Next, we show that the subsets shown in (\ref{eq:phxAGEtheLastFourc}) do not have overlap
\begin{align}
    & \lambda+s-1 \leq ts \nonumber \\
    \Rightarrow & \lambda+s-2 < ts \nonumber \\
    \Rightarrow & \theta(l+1)+s-2 < 2ts+\theta l
\end{align}
and
\begin{align}
    & z \leq ts \nonumber \\
    \Rightarrow & z-2 < ts \nonumber \\
    \Rightarrow & ts+\theta(t+q-1)+z-2 < 2ts+\theta(q+t-1)
\end{align}
Therefore, by calculating the size of each subset, we can calculate the number of elements of $\mathbf{P}({H}(x))$. 
The size of $\{0,\ldots,\theta(t+1)+s-2\}$ is equal to $\theta(t+1)+s-1$. 
The size of $\bigcup_{l=t}^{t+q-3} \{2ts+\theta  l,\dots,\theta(l+2)+s-2\}$ is equal to $(q-2)(2\lambda +s-1)$. 
The size of $\{2ts+\theta (t+q-2),\dots,ts+\theta(t+q-1)+z-2\}$ is equal to $z+\lambda-1$. 
The size of $\{2ts+\theta(q+t-1),\dots ,(q+2)ts+\theta(t-1)+2z-2\}$ is equal to $2z-q\lambda -1$. 
Therefore, $|\mathbf{P}({H}(x))|$ is equal to the sum of all these sizes, \ie $|\mathbf{P}({H}(x))|=2ts+\theta(t-1)+3z+(\lambda+s-1)q-\lambda-s-1$. This completes the proof.\hfill $\Box$

\begin{lemma}\label{lemma:non-zero-coeff-AGE-z less than ts: 5}
For $z \leq \lambda+s-1 \leq ts, 0<\lambda <z, t\neq 1$ and $q\lambda < s$:
\begin{align}
    |\mathbf{P}({H}(x))|=\Upsilon_9(\lambda)=&\theta(t+1)+q(s-1)-3\lambda+3z-1\nonumber\\
    &+\min\{0,ts-z+1+\lambda q-s\}
\end{align}
\end{lemma}
{\em Proof:} 
From (\ref{eq:phxforthe4lastcases}) and Lemmas \ref{lemma:d1234cupd'3AGE} and {\ref{lemma:d1234cupd''3AGE}}, we have
\begin{align}
    &\mathbf{P}({H}(x))=(\mathbf{\widehat{D}}_{123'4}\cup\mathbf{\widetilde{D}}'_3)\cup(\mathbf{\widehat{D}}_{123'4}\cup\mathbf{\widetilde{D}}''_3)\nonumber
    \end{align}
    \begin{align}
    &=\{0,\ldots,\theta(t+1)+s-2\}\nonumber\\
    &\cup\bigcup_{l=t}^{t+q-3} \{2ts+\theta  l,\dots,\theta(l+2)+s-2\} \nonumber\\
    &\cup \{2ts+\theta (t+q-2),\dots,ts+\theta(t+q-1)+z-2\} \nonumber \\
   &\cup \{2ts+\theta(q+t-1),\dots ,(q+2)ts+\theta(t-1)+2z-2\}\nonumber\\
   &\cup\{0,\ldots,ts+\theta t+z-2\}\nonumber\\
    &\cup\bigcup_{l=1}^{q-2} \{2ts+\theta(l+t-1),\dots,ts+\theta(l+t)+z-2\} \nonumber\\
    &\cup \{2ts+\theta(q+t-2),\dots,(q+1)ts+(t-1)\theta+s+z-2\}
             \end{align}
    \begin{align}
   &\cup \{2ts+\theta(q+t-1),\dots ,(q+2)ts+\theta(t-1)+2z-2\}\nonumber\\
   &=\{0,\ldots,\theta(t+1)+s-2\}\nonumber\\
    &\cup\bigcup_{l=t}^{t+q-3} \{2ts+\theta  l,\dots,\theta(l+2)+s-2\} \nonumber
            \end{align}
    \begin{align}
   &\cup \{2ts+\theta(q+t-1),\dots ,(q+2)ts+\theta(t-1)+2z-2\}\nonumber\\
    &\cup\bigcup_{l=t}^{t+q-3} \{2ts+\theta l,\dots,ts+\theta(l+1)+z-2\} \nonumber \\
    &\cup \{2ts+\theta(q+t-2),\dots,(q+1)ts+(t-1)\theta+s+z-2\}\nonumber
            \end{align}
    \begin{align}
   &=\{0,\ldots,\theta(t+1)+s-2\}\nonumber\\
    &\cup\bigcup_{l=t}^{t+q-3} \{2ts+\theta  l,\dots,\theta(l+2)+s-2\} \nonumber\\
    &\cup \{2ts+\theta(q+t-2),\dots,(q+1)ts+(t-1)\theta+s+z-2\}\nonumber\\
    &\cup \{2ts+\theta(q+t-1),\dots ,(q+2)ts+\theta(t-1)+2z-2\}\label{eq:phxAGEtheLastFourd}
\end{align}

Next, we show that the subsets shown in (\ref{eq:phxAGEtheLastFourd}), do not have overlap except for the last two subsets of $\{2ts+\theta(q+t-2),\dots,(q+1)ts+(t-1)\theta+s+z-2\}$ and $\{2ts+\theta(q+t-1),\dots ,(q+2)ts+\theta(t-1)+2z-2\}$
\begin{align}
    & \lambda+s-1 \leq ts \nonumber \\
    \Rightarrow & \lambda+s-2 < ts \nonumber \\
    \Rightarrow & \theta(l+1)+s-2 < 2ts+\theta l
\end{align}
Therefore, by calculating the size of each subset, we can calculate the number of elements of $\mathbf{P}({H}(x))$.
The size of $\{0,\ldots,\theta(t+1)+s-2\}$ is equal to $\theta(t+1)+s-1$.
The size of $\bigcup_{l=t}^{t+q-3} \{2ts+\theta  l,\dots,\theta(l+2)+s-2\}$ is equal to $(q-2)(2\lambda+s-1)$.
The size of $\{2ts+\theta(q+t-2),\dots,(q+1)ts+(t-1)\theta+s+z-2\}\cup \{2ts+\theta(q+t-1),\dots ,(q+2)ts+\theta(t-1)+2z-2\}$ is equal to $\min\{\lambda(1-2q)+3z+s-2, \theta -\lambda q+2z-1\}$. Therefore, $|\mathbf{P}({H}(x))|$ is equal to the sum of all these sizes, \ie $|\mathbf{P}({H}(x))|=\theta(t+1)+q(s-1)-3\lambda+3z-1+\min\{0,ts-z+1+\lambda q-s\}$. This completes the proof.\hfill $\Box$

From Lemmas \ref{lemma:t=1NAGECMPC}, \ref{lemma:non-zero-coeff-AGE-lambda=0 and z greater ts-t}, \ref{lemma:non-zero-coeff-AGE-lambda=0 and z less ts-t}, \ref{lemma:non-zero-coeff-AGE-lambda=z}, \ref{lemma:non-zero-coeff-AGE-z greater than ts}, \ref{lemma:non-zero-coeff-AGE-z less than ts: 1}, \ref{lemma:non-zero-coeff-AGE-z less than ts: 2}, \ref{lemma:non-zero-coeff-AGE-z less than ts: 3}, \ref{lemma:non-zero-coeff-AGE-z less than ts: 4} and \ref{lemma:non-zero-coeff-AGE-z less than ts: 5}, Theorem \ref{th:N_AGE} is proved.

\section*{Appendix D: Proof of Lemmas \ref{corol:AGE-Vs-Entang}, \ref{corol:AGE-Vs-SSMM}, \ref{corol:AGE-Vs-GCSANA}, and \ref{corol:AGE-Vs-polydot-1}} 
\setcounter{section}{0}

\subsection{Proof of Lemma \ref{corol:AGE-Vs-Entang} (AGE-CMPC Versus Entangled-CMPC)} 
$N_{\text{AGE-CMPC}}=2s+2z-1$ when $t=1$ using (\ref{eq:eq:N-AGE-CMPC-optimization}). On the other hand,  $N_{\text{Entangled-CMPC}}=2s+2z-1$ from \cite{8613446}. Thus, $N_{\text{AGE-CMPC}}=N_{\text{Entangled-CMPC}}$ when $t=1$.


$N_{\text{AGE-CMPC}}$ is expressed as in the following when $t \neq 1$ using (\ref{eq:eq:N-AGE-CMPC-optimization}) and (\ref{eq:N-AGE-CMPC}). 
\begin{align}
    N_{\text{AGE-CMPC}}&=\displaystyle\min_{\lambda} \Gamma\nonumber\\
    \leq& \Gamma \text{ for } \lambda=0\nonumber\\
    =&\begin{cases}
   2st^2+2z-1, & z>ts-s,\\\
   st^2+3st-2s+t(z-1)+1, & z \leq ts-s,\
   \end{cases}\nonumber\\
   =&N_{\text{Entangled-CMPC}},
\end{align}
where the last equality comes from \cite{8613446}.

From the above discussion, we conclude that $N_{\text{AGE-CMPC}}<N_{\text{Entangled-CMPC}}$ when $0 < \lambda^* \leq z$. For the case of $\lambda^*=0$, $N_{\text{AGE-CMPC}}=N_{\text{Entangled-CMPC}}$. This completes the proof of Lemma \ref{corol:AGE-Vs-Entang}.

\subsection{Proof of Lemma \ref{corol:AGE-Vs-SSMM} (AGE-CMPC Versus SSMM)}
$N_{\text{AGE-CMPC}}=2s+2z-1$ when $t=1$ using (\ref{eq:eq:N-AGE-CMPC-optimization}). On the other hand, $N_{\text{SSMM}}=2s+2z-1$ from \cite{Zhu2021ImprovedCF}. Thus, $N_{\text{AGE-CMPC}}=N_{\text{SSMM}}$ when $t=1$. Next, we consider the case of $t\neq 1$ and compare $N_{\text{AGE-CMPC}}$ with $N_{\text{SSMM}}$.

$N_{\text{AGE-CMPC}}$ is expressed as the following when $t \neq 1$ using (\ref{eq:eq:N-AGE-CMPC-optimization}) and (\ref{eq:N-AGE-CMPC}).
\begin{align}
    N_{\text{AGE-CMPC}}=&\displaystyle\min_{\lambda} \Gamma\nonumber\\
    \leq& \Gamma \text{ for } \lambda=z\nonumber\\
    =&2ts+(ts+z)(t-1)+2z-1\nonumber\\
    =&(t+1)(ts+z)-1\nonumber\\
    =&N_{\text{SSMM}},
\end{align}
where the last equality comes from Theorem 1 in \cite{Zhu2021ImprovedCF}. 

From the above discussion, we conclude that $N_{\text{AGE-CMPC}}<N_{\text{SSMM}}$ when $0 \leq \lambda^* < z$. For the case of $\lambda^*=z$, $N_{\text{AGE-CMPC}}=N_{\text{SSMM}}$. This completes the proof of Lemma \ref{corol:AGE-Vs-SSMM}.

\subsection{Proof of Lemma \ref{corol:AGE-Vs-GCSANA} (AGE-CMPC Versus GCSA-NA)}
$N_{\text{AGE-CMPC}}=2s+2z-1$ when $t=1$ using (\ref{eq:eq:N-AGE-CMPC-optimization}). On the other hand, $N_{\text{GCSA-NA}}=2s+2z-1$ from \cite{9333639}. Thus, $N_{\text{AGE-CMPC}}=N_{\text{GCSA-NA}}$ when $t=1$. Next, we consider the case of $t\neq 1$ and compare $N_{\text{AGE-CMPC}}$ with $N_{\text{GCSA-NA}}$.

$N_{\text{AGE-CMPC}}$ is expressed as the following when $t \neq 1$ using (\ref{eq:eq:N-AGE-CMPC-optimization}) and (\ref{eq:N-AGE-CMPC}).
\begin{align}
    N_{\text{AGE-CMPC}}&=\displaystyle\min_{\lambda} \Gamma\nonumber\\
    \leq& \Gamma \text{ for } \lambda=0\nonumber\\
    =&\begin{cases}
   2st^2+2z-1, & z>ts-s,\\\
   st^2+3st-2s+t(z-1)+1, & z \leq ts-s,\
   \end{cases}\nonumber\\
   &\begin{cases}
   =2st^2+2z-1, & z>ts-s,\\\
   \leq 2st^2+2z-1
   , & z \leq ts-s,\
   \end{cases}\label{eq:AGEvsGCSANA1}\\
       &\begin{cases}
   =N_{\text{GCSA-NA}}, & z>ts-s\\\
   \leq N_{\text{GCSA-NA}}
   , & z \leq ts-s\label{eq:AGEvsGCSANA2}\
   \end{cases},
\end{align}
where (\ref{eq:AGEvsGCSANA1}) comes from the condition of $z\leq ts-s$ as described in the following:
\begin{align}
    st^2+&3st-2s+t(z-1)+1\nonumber\\
    &=st^2+3st-2s+tz-t-2z+2z+1\nonumber\\
    &= st^2+3st-2s+(t-2)(z)-t+2z+1\nonumber\\
    &\leq st^2+3st-2s+(t-2)(ts-s)-t+2z+1\nonumber\\
    &=2st^2+2z-1-t+2\nonumber\\
    &\leq 2st^2+2z-1
\end{align}
and (\ref{eq:AGEvsGCSANA2}) comes from Theorem 1 in \cite{9333639}.

From the above discussion, we conclude that $N_{\text{AGE-CMPC}}<N_{\text{GCSA-NA}}$ when $0 < \lambda^* \leq z$. For the case of $\lambda^*=0$, $N_{\text{AGE-CMPC}}\leq N_{\text{GCSA-NA}}$. This completes the proof of Lemma \ref{corol:AGE-Vs-GCSANA}. 

\subsection{Proof of Lemma \ref{corol:AGE-Vs-polydot-1} (AGE-CMPC Versus PolyDot-CMPC)}
To prove this lemma, we consider different regions for the value of $z$, and prove that in all of the regions, the inequality of $N_{\text{AGE-CMPC}} \leq N_{\text{PolyDot-CMPC}}$ is valid.

(i) $z>ts, t\neq 1$: For this region, We consider the two cases of (a) $s\neq 1$ and (b) $s=1$.

(a) $s\neq 1$: From (\ref{eq:eq:N-AGE-CMPC-optimization}) and (\ref{eq:N-AGE-CMPC}), we have
\begin{align}
    N_{\text{AGE-CMPC}}=&\displaystyle\min_{\lambda} \Gamma\nonumber\\
    \leq& \Gamma \text{ for } 0<\lambda=ts-t<z\nonumber\\
    =&(q+2)ts+\theta(t-1)+2z-1\text{ for } \lambda=ts-t\nonumber\\
    =&(\min\{\floor{\frac{z-1}{2ts-t-ts}},t-1\}+2)ts+\nonumber\\
    &\quad \quad \quad \quad (2ts-t)(t-1)+2z-1\nonumber\\
    =&N_{\text{PolyDot-CMPC}},
\end{align}
where the last equality comes from \cite{PolyDot-CMPC}.

(b) $s=1$: From (\ref{eq:eq:N-AGE-CMPC-optimization}) and (\ref{eq:N-AGE-CMPC}), we have
\begin{align}
    N_{\text{AGE-CMPC}}=&\displaystyle\min_{\lambda} \Gamma\nonumber\\
    \leq& \Gamma \text{ for } \lambda=0\nonumber\\
    =&2t^2+2z-1\nonumber\\
    =&N_{\text{PolyDot-CMPC}},
\end{align}
where the last equality comes from $N_{\text{PolyDot-CMPC}}$ defined in \cite{PolyDot-CMPC} for $s=1$ and $z>ts$.

(ii) $\frac{t-1}{t-2}(ts-t)<z\leq ts, t\neq 1$: This condition exists only if the constraint of $\frac{t-1}{t-2}(ts-t)< ts$ is satisfied. This constraint is satisfied when
\begin{equation}\label{eq:temp1}
    s+1<t.
\end{equation} 

Next, we show that, for $0<\lambda=ts-t<z$, $\Gamma$ is equal to one of $\Upsilon_i(\lambda)$'s where $i=6,7,8,9$. Then, we show that each $\Upsilon_i(\lambda), i=6,7,8,9$ for $\lambda=ts-t$ is less than $N_{\text{PolyDot-CMPC}}$. For this purpose, we first assert that the conditions for this case, \ie $\lambda=ts-t<\frac{t-1}{t-2}(ts-t)<z\leq ts, t\neq 1$, do not satisfy the conditions for $\Upsilon_i(\lambda), i=1,2,3,4,5$.
The reason is that $0<t(s-1)=\lambda$ does not satisfy the condition for $\Upsilon_i(\lambda), i=1,2,3$. On the other hand, $z\leq ts$ does not satisfy the condition for $\Upsilon_4(\lambda)$. In addition, from (\ref{eq:temp1}), $s-1<t$ and thus $ts-s+1>ts-t=\lambda$, which does not satisfy the condition for $\Upsilon_5(\lambda)$.

We consider the following cases;  $q=0$ and $q=1$.

(a) $q=0$: For this case, based on the definition of $q$, we should have either (1) $\frac{z-1}{\lambda}=\frac{z-1}{ts-t}<1$, which is not possible as this contradicts the condition of (ii) that requires $ts-t< \frac{t-1}{t-2}ts-t<z$, so $ts-t\leq z-1$, or (2) $t=1$, which is not possible as (\ref{eq:temp1}) results in $s<0$, which is not a valid inequality. 

(b) $q=1$: From (\ref{eq:N-AGE-CMPC}), this falls under the condition of $\Upsilon_6(\lambda)$ and $\Upsilon_8(\lambda)$ as $q\lambda=\lambda=s(t-1)\ge s$. For this case, either the condition of $\lambda+s-1<z$ (condition of $\Upsilon_6(\lambda)$) or $z\leq \lambda+s-1$ (condition of $\Upsilon_8(\lambda)$) is satisfied. Both $\Upsilon_6(\lambda)$ and $\Upsilon_8(\lambda)$ are less than $N_{\text{PolyDot-CMPC}}$ as shown below.

For $s\neq 1$, we have 
\begin{align}\label{eq:gamma6vspolydotsneq1}
   \Upsilon_6(\lambda) &= 2ts+\theta(t-1)+(q+2)z-q-1 \nonumber \\
  & =  2ts+(ts+ts-t)(t-1)+3z-2 \nonumber\\
   & <  2ts+(2ts-t)(t-1)+3z-1\nonumber\\
   &= N_{\text{PolyDot-CMPC}},
\end{align}
where the last equality comes from $N_{\text{PolyDot-CMPC}}$ defined in \cite{PolyDot-CMPC} for $ts-t<z\leq ts, s\neq 1$. Next, we consider the case  $s=1$.

For $s=1$, from (\ref{eq:temp1}), we have $t>2$ and from the condition of (ii), we have $z\leq t$. Therefore, we have
\begin{align}
   \Upsilon_6(\lambda) &= 2ts+\theta(t-1)+(q+2)z-q-1 \nonumber \\
  & = 2t+(2t-t)(t-1)+3z-2 \nonumber\\
  & = t^2+t+3z-2\nonumber\\
   & < t^2+t+2z+z-1\nonumber\\
   & < t^2+t+tz+t-1\label{eq:tempp2}\\
   & = t^2+2t+tz-1\nonumber\\
   &= N_{\text{PolyDot-CMPC}}, 
   \label{eq:gamma6vspolydots=1}
\end{align}
where (\ref{eq:tempp2}) comes from $t>2$ and $z\leq t$. The last equality comes from $N_{\text{PolyDot-CMPC}}$ defined in \cite{PolyDot-CMPC} for $s=1, z\leq t$.

From (\ref{eq:gamma6vspolydotsneq1}) and (\ref{eq:gamma6vspolydots=1}), we conclude $\Upsilon_6(\lambda)<N_{\text{PolyDot-CMPC}}$.

For $s\neq 1$, we have
\begin{align}\label{eq:gamma8vspolydotsneq1}
   \Upsilon_8(\lambda) &= 2ts+\theta(t-1)+3z+(\lambda+s-1)q-\lambda-s-1 \nonumber \\
  & =  2ts+(2ts-t)(t-1)+3z+\lambda+s-1-\lambda-s-1 \nonumber \\
  & = 2ts+(2ts-t)(t-1)+3z-2 \nonumber \\
  & <  2ts+(2ts-t)(t-1)+3z-1\nonumber\\
   &= N_{\text{PolyDot-CMPC}},
\end{align}
where the last equality comes from $N_{\text{PolyDot-CMPC}}$ defined in \cite{PolyDot-CMPC} for $ts-t<z\leq ts, s\neq 1$. Next, we consider the case of $s=1$.

For $s=1$, from (\ref{eq:temp1}), we have $t>2$ and from the condition of (ii), we have $z\leq t$. Therefore, similar to (\ref{eq:tempp2}), we have
\begin{align}\label{eq:gamma8vspolydots=1}
   \Upsilon_8(\lambda) &= 2ts+\theta(t-1)+3z+(\lambda+s-1)q-\lambda-s-1 \nonumber \\
  & =  2t+(2t-t)(t-1)+3z+\lambda+1-1-\lambda-1-1 \nonumber \\
  & = t^2+t+3z-2 \nonumber \\
  & <  t^2+2t+tz-1\nonumber\\
   &= N_{\text{PolyDot-CMPC}},
\end{align}
where the last inequality comes from $t>2$ and $z\leq t$ and the last equality comes from $N_{\text{PolyDot-CMPC}}$ defined in \cite{PolyDot-CMPC} for $s=1, z\leq t$.

From (\ref{eq:gamma8vspolydotsneq1}) and (\ref{eq:gamma8vspolydots=1}), we conclude $\Upsilon_8(\lambda)<N_{\text{PolyDot-CMPC}}$.

From the above discussion, $\Gamma$ for $\lambda=ts-t$ is less than $N_{\text{PolyDot-CMPC}}$ for the condition of (ii). Therefore, we have:
\begin{align}
    N_{\text{AGE-CMPC}}=&\displaystyle\min_{\lambda} \Gamma\nonumber\\
    \leq& \Gamma \text{ for } 0<\lambda=ts-t<z\nonumber\\
    <&N_{\text{PolyDot-CMPC}},
\end{align}

(iii) $z \leq \frac{t-1}{t-2}(ts-t), s,t \neq 1$\footnote{Note that for this case, we have $s \neq 1$ as $z\leq \frac{t-1}{t-2}t(s-1)$.}: It is proved in \cite{PolyDot-CMPC} that for the condition of (iii), $N_{\text{SSMM}} \leq N_{\text{PolyDot-CMPC}}$. On the other hand, from Lemma (\ref{corol:AGE-Vs-SSMM}), $N_{\text{AGE-CMPC}} \leq N_{\text{SSMM}}$. Therefore, for this region, $N_{\text{AGE-CMPC}} \leq N_{\text{PolyDot-CMPC}}$.
 
(iv) $t=1$: For this region, from (\ref{eq:eq:N-AGE-CMPC-optimization}), we have:
\begin{align}
 N_{\text{AGE-CMPC}}&=2s+2z-1\nonumber\\
 &=N_{\text{PolyDot-CMPC}},
\end{align}
where the last equality comes from $N_{\text{PolyDot-CMPC}}$ defined in \cite{PolyDot-CMPC} for $t=1$.

From (i), (ii), (iii), and (iv), the number of workers required by AGE-CMPC method is always less than or equal to the number of workers required by PolyDot-CMPC \cite{PolyDot-CMPC}.
This completes the proof of Lemma \ref{corol:AGE-Vs-polydot-1}.

\end{document}